\documentstyle[epsfig,12pt]{article}
\advance\textheight by \topskip
\def\lsim{\mathrel{\lower2.5pt\vbox{\lineskip=0pt\baselineskip=0pt
\hbox{$<$}\hbox{$\sim$}}}}
\def\gsim{\mathrel{\lower2.5pt\vbox{\lineskip=0pt\baselineskip=0pt
\hbox{$>$}\hbox{$\sim$}}}}
\begin{document}
\def\gs{SU(2)_{\rm L} \times U(1)_{\rm Y}}
\def\ql{{\tilde{q}_{{i}_{\scriptscriptstyle L}}}}
\def\qr{{\tilde{q}_{{i}_{\scriptscriptstyle R}}}}
\def\tl{{\tilde{t}_{{\scriptscriptstyle L}}}}
\def\tr{{\tilde{t}_{{\scriptscriptstyle R}}}}
\def\bl{{\tilde{b}_{{\scriptscriptstyle L}}}}
\def\br{{\tilde{b}_{{\scriptscriptstyle R}}}}
\def\mt{{\tilde{m}^2}_{{\scriptscriptstyle t}}}
\def\mb{{\tilde{m}^2}_{{\scriptscriptstyle b}}}
\def\mz{{m}_{\scriptscriptstyle Z}^{2}}
\def\at{{A_{\scriptscriptstyle t}}}
\def\ab{{A_{\scriptscriptstyle b}}}
\def\qt{{Q_{\scriptscriptstyle t}}}
\def\qb{{Q_{\scriptscriptstyle b}}}
\def\tu{{\tilde{t}_{{\scriptscriptstyle 1}}}}
\def\td{{\tilde{t}_{{\scriptscriptstyle 2}}}}
\def\bu{{\tilde{b}_{{\scriptscriptstyle 1}}}}
\def\bd{{\tilde{b}_{{\scriptscriptstyle 2}}}}
\def\taul{{\tilde{\tau}_{{\scriptscriptstyle L}}}}
\def\taur{{\tilde{\tau}_{{\scriptscriptstyle R}}}}
\def\tauu{{\tilde{\tau}_{{\scriptscriptstyle 1}}}}
\def\taud{{\tilde{\tau}_{{\scriptscriptstyle 2}}}}
\def\mnu{{\tilde{m}^2_{{\scriptscriptstyle \nu}}}}
\def\mtau{{\tilde{m}^2}_{{\scriptscriptstyle \tau}_{\scriptscriptstyle 1}}}
\def\mtad{{\tilde{m}^2}_{{\scriptscriptstyle \tau}_{\scriptscriptstyle 2}}}
\def\w{{\tilde{W}^{\pm}}}
\def\h{{\tilde{H}^{\pm}}}
\def\chiu{{\tilde{\chi}^{+}}_{{\scriptscriptstyle 1}}}
\def\chid{{\tilde{\chi}^{+}}_{{\scriptscriptstyle 2}}}
\def\sw{s_{{\scriptscriptstyle W}}}
\def\cw{c_{{\scriptscriptstyle W}}}
\def\mfa{{\tilde{m}^2}_{{\scriptscriptstyle fa}}}
\def\mfb{{\tilde{m}^2}_{{\scriptscriptstyle fb}}}
\def\mfc{{\tilde{m}^2}_{{\scriptscriptstyle fc}}}
\def\mfd{{\tilde{m}^2}_{{\scriptscriptstyle fd}}}
\def\mi{{\tilde{M}^2}_{{\scriptscriptstyle i}}}
\def\mj{{\tilde{M}^2}_{{\scriptscriptstyle j}}}
\def\mcu{{\tilde{M}^{+\,2}}_{{\scriptscriptstyle 1}}}
\def\mcd{{\tilde{M}^{+\,2}}_{{\scriptscriptstyle 2}}}
\def\mnu{{\tilde{M}^{o\,2}}_{{\scriptscriptstyle 1}}}
\def\mnd{{\tilde{M}^{o\,2}}_{{\scriptscriptstyle 2}}}
\def\mnt{{\tilde{M}^{o\,2}}_{{\scriptscriptstyle 3}}}
\def\mnc{{\tilde{M}^{o\,2}}_{{\scriptscriptstyle 4}}}
\def\G1{G_{\alpha\mu \beta\nu\gamma\sigma}}
\newcommand{\sfe}{{\tilde f}}
\newcommand{\sg}{{\tilde \chi}}
\newcommand{\sne}{{\tilde \chi^o}}
\newcommand{\scp}{{\tilde \chi^+}}
\newcommand{\dx}{dx}  
\newcommand{\dy}{dy}  
\newcommand{\slas}[1]{\rlap/ #1}
\newcommand{\diag}{{\rm diag}}
\newcommand{\Tr}{{\rm Tr}}
\vspace*{-1cm}
\begin{flushright}
{\large FTUAM 98/8}\\
{Feb.-1999}\\
{hep-ph/9903211}
\end{flushright}
\begin{center}
\begin{large}
\begin{bf}THE SM AS THE QUANTUM LOW ENERGY EFFECTIVE
 THEORY OF THE MSSM \\
\end{bf}
\end{large}
\vspace{0.2cm}
ANTONIO DOBADO\\
\vspace{0.05cm} 
{\em  Departamento de F{\'\i}sica Te{\'o}rica\\
  Universidad Complutense de Madrid\\
 28040-- Madrid,\ \ Spain} \\
\vspace{0.3cm} 
 MARIA J. HERRERO \\
\vspace{0.05cm}
and\\
\vspace{0.1cm}
SIANNAH PE{\~N}ARANDA\\
\vspace{0.06cm}
{\em  Departamento de F{\'\i}sica Te{\'o}rica\\
  Universidad Aut{\'o}noma de Madrid\\
  Cantoblanco,\ \ 28049-- Madrid,\ \ Spain}
\end{center}
\begin{center}
{\bf ABSTRACT}
\end{center}
\begin{quotation}
\noindent
In the framework of the Minimal Supersymmetric Standard Model we compute
 the one-loop effective action for the electroweak bosons  obtained after
 integrating out the different sleptons, squarks, neutralinos and
 charginos, and present the result in terms of the physical sparticle
masses. In addition we study the asymptotic behaviour of the two, three and
four point Green's functions with external electroweak bosons in the limit
where the physical sparticle masses are very large in comparison with the
electroweak scale. We find that in this limit all the effects produced
 by the supersymmetric particles can be, either absorbed in the Standard Model
 parameters and gauge bosons wave functions, 
or else they are supressed by inverse powers of the supersymmetric  particle masses.
This work, therefore, completes the proof of decoupling of the heavy
 supersymmetric particles
 from the standard ones in the electroweak bosons effective action and
in the sense of the Appelquist-Carazzone Theorem that we started in a previous
work. From the point of view of effective field theories this work can be
seen as a (partial) proof that the SM can be obtained indeed from the MSSM
as the quantum low energy effective theory of this latter when the SUSY spectra
is much heavier than the electroweak scale. 
\end{quotation}
${\star}${\em e-mail: dobado@eucmax.sim.ucm.es}\\
${\dagger}$
{\em e-mail: herrero@delta.ft.uam.es}\,\,,\,\,{\em siannah@daniel.ft.uam.es}\\
\newpage
\section{Introduction}
In spite of the enormous amount of experimental evidence in favour of the
 Standard Model (SM), most of the physicists consider it just as a low
 energy manifestation of a more fundamental theory. Among the possible extensions of the SM one of the most popular
 nowdays is the so called Minimal Supersymmetric Standard Model (MSSM)~\cite{IS, HAMSSM}, which is
 the simplest theory that can be built from a supersymmetric version of the SM
 after the introduction of a minimal set of soft breaking terms~\cite{GG}. Those terms
 break the supersymmetry (SUSY) of the original Supersymmetric Standard Model 
 and give rise
 to contributions to the Higgs potential that finally produce the appropriate
 spontaneous breaking of the $SU(2)_L\times U(1)_Y$ electroweak gauge symmetry. 
Considering the MSSM as an interesting possibility motivated by many theoretical
 reasons, it is a quite natural question to ask in what sense, if any,  can
 the SM model be considered as a low energy  effective theory of the MSSM in the
 case where the SUSY partners of the standard particles are very heavy. In fact there
 are many partial indications that the SM is the low energy limit of the 
 MSSM~\cite{CHA}-\cite{QUI}. 
However, most of them are based on numerical estimates and are obtained after taking
  some of the mass parameters appearing in the soft breaking terms numerically very large.

In this work we would like to address the question of getting the SM from the MSSM 
 from a more formal and
 field theory
 complete point of view and, in addition, we will work  
 directly with the physical sparticles masses instead of using the soft breaking 
parameters. In order to do that we will pay special attention to two 
esential points: First we will define in a very precise way what we understand
 by a low energy effective theory. The definition that we will adopt here is 
the one corresponding to the so called decoupling or Appelquist-Carazzone 
Theorem~\cite{OJO}. Namely, a 
theory with just light fields $\phi$, is considered as the low energy effective
 theory of a larger theory with both heavy $\tilde \phi$ and light 
 $\phi$ fields if the effects of integrating out the 
heavy fields $\tilde \phi$, on the
Green's functions, can be reduced to renormalizations of the parameters 
of the effective theory, or produce extra terms which are supressed by inverse
 powers of the heavy $\tilde \phi$ masses~\cite{ANT}. The second important point
 to be taken into account is the precise way in which the large sparticle
 masses limit is taken. This is essential since, due to the divergences appearing
  in the loop integrals, large mass limits and momentum integrations do not commute 
 and even the large mass limit for the various particles
 may not commute among themselves. In this work we have
 choosen to take the limit where the sparticle  masses  $\tilde m_i$ are much
 larger than the electroweak boson masses and the external momenta and, at
 the same time, we will assume that the differences among the sparticles masses
 are much smaller than the sparticles masses themselves. These conditions give rise to  a precise definition  of the large
 sparticle masses regime and make it possible to define, in an unambiguous
 way, the resulting  low energy theory for the standard particles. If
 this low energy theory corresponds to the SM according to the
 Appelquist-Carazzone definition, we will say that the SUSY particles
 decouple from the SM. Notice that the decoupling of  the SUSY
 particles, in case it occurs, is neither inmediate nor trivial at all. This is 
 because the Appelquist-Carazzone
Theorem does not always apply~\cite{VE}-\cite{MJ}. For example it does not apply whenever we
 have spontaneous symmetry breaking or chiral fermions~\cite{ANT}. This is just
 the case of the MSSM. Therefore the decoupling of the SUSY particles in the
 Appelquist-Carazzone sense must be showed explicitly in this case. 

Thus, our program is the following. We will start with the MSSM sector
 involving the electroweak bosons for which we want to study the possible
 decoupling of the SUSY particles. Then we compute the Green's functions 
 for the electroweak gauge bosons that are 
obtained by integrating out the sleptons, squarks, neutralinos and charginos
at the one-loop level
 (the Higgs sector of the MSSM will be considered elsewhere). Next step is
 the analytical study  of the  behaviour of these Green's functions in the
 asymptotic regime of the large sparticle masses defined above. This task will be much easier by using
 the so called m-theorem~\cite{GMR} as it will be explained bellow. Finally by 
comparison of the obtained results with the tree level SM Green's functions
 for the electroweak bosons, we will be able to show the decoupling of the
 considered  SUSY particles according to the  Appelquist-Carazzone definition. 

The above program was started by the authors in~\cite{GEISHA}, where the two
point functions for electroweak gauge bosons and the $S$,$T$ and
 $U$ observables were considered. Here we continue that program and consider the 
three and four point functions for electroweak gauge bosons. We use the same notations 
and conventions for the MSSM as in our previous work that we do not repeat here
for brevity. We also refer the reader to that work for more
details and, in particular, for a broad discussion on the large sparticle 
masses limit. 
The present paper is organized as follows: In section two we review the definition
of the low-energy action for
 the electroweak bosons that we presented in ref.\cite{GEISHA}. 
The results for the two-point functions
are summarized in section three. The three and four
 point functions are obtained and discussed in sections four and five
 respectively. These and our previous results are analized together in order 
to establish the applicability criterion for the 
Appelquist-Carazzone Theorem in the studied case. Finally in section six we 
report the main conclusions of our work. In appendix A we define the one-loop integrals appearing
 in our computations by using the standard scalar and tensor integrals~\cite{PAS} and give the
  asymptotic forms of the last ones. Appendix B contains some operators and 
functions which are used in this article to present the results for the three and four 
functions. Appendix C is devoted to summarize the exact results to one loop for the
three and four-point sfermion contributions and for the three-point {\it inos}
contributions.

\section{The low-energy effective action for the electroweak bosons}
In this section we describe our computation of the effective action for
the electroweak bosons. It contains the two, three and four point Green's functions
and is obtained after the integration of the sfermions and {\it inos}, viz. charginos 
 and neutralinos of the MSSM~\cite{GEISHA}. 
 
In more generic words, our aim is to compute the effective action
 $\Gamma_{eff}[\phi]$ for the standard particles $\phi$   that is
 defined through functional integration of all the sparticles of
 the MSSM $\tilde \phi$. In short notation it is defined by,
\begin{equation}
\label{eq:gammaeff}{\rm e}^{i\Gamma_{eff}[\phi]}=\int [{\rm d}\tilde\phi]\,{\rm e}^{i \Gamma_{\rm MSSM}   [\phi,\tilde\phi]}
\end{equation}
with  
\begin{equation}\label{eq:gammaeffmssm}\Gamma_{\rm MSSM}[\phi,\tilde\phi] \equiv \int \dx{\cal L}_{\rm  MSSM}(\phi,\tilde \phi)\,\,;\,\,{\rm d}x\equiv{\rm d}^4x
\end{equation}
and ${\cal L}_{\rm  MSSM}$ is the MSSM Lagrangian.
The computation of the effective action will be performed at the one loop level by 
using dimensional regularization, in an arbitrary $R_\xi$ gauge and will include 
the integration of all the sfermions $\sfe$ (squarks $\tilde{q}$ and sleptons
$\tilde{l}$), neutralinos $\sne$ and  
charginos $\scp$. Our program starts, in particular, with the computation of the 
electroweak gauge boson effective action $\Gamma_{\rm MSSM}[V]$ ($V=A, Z $ and 
$W^{\pm}$) given by,
\begin{equation}
\label{eq:gammaeffV}{\rm e}^{i\Gamma_{eff}[V]}=\int [{\rm d}\sfe]\, [{\rm d}\sfe^*]\, [{\rm d}\scp]\, [{\rm d}\bar{\sg}^+]\, [{\rm d}\sne]\,{\rm e}^{i \Gamma_{\rm MSSM}[V,\sfe, \scp,\sne]}
\end{equation}
where,
\begin{eqnarray}
\label{eq:gammaMSSM}
\Gamma_{\rm MSSM}[V,\sfe, \scp,\sne]&\equiv&  \int\dx {\cal L}_{\rm MSSM}(V,\sfe, \scp,\sne) \nonumber\\ &=& \int\dx {\cal L}^{(0)} (V)+ \int\dx{\cal L}_{\sfe}(V,\sfe)+\int\dx{\cal L}_{\sg}(V,\sg) \nonumber\\&\equiv&\Gamma_0[V]+\Gamma_{\sfe}[V,\sfe]+\Gamma_{\sg}[V,\sg]
\end{eqnarray}
and ${\cal L}^{(0)}$, ${\cal L}_{\sfe}$, ${\cal L}_{\sg}$ 
are the free Lagrangian and the interaction Lagrangian of gauge bosons 
 with sfermions and {\it inos} respectively. 
From now on we will follow closely the definitions, notations and conventions introduced 
in~\cite{GEISHA}. In particular we will use the compact notation
:
\begin{eqnarray}
\label{eq:compact}
&&\phi(x)\equiv\phi_x\,,\,\delta(x-y)\equiv\delta_{xy}\,,\,A(x,y)
\equiv A_{xy}\nonumber\\&&\Tr A={\rm tr} \int\dx A_{xx}=\sum_a 
\int \dx A_{xx}^{aa}\,,
\end{eqnarray}
and 
\begin{equation}
\label{eq:matf}
\begin{array}{l}
\displaystyle 
\tilde{f} \equiv \left(
\begin{array}{l}
\tilde{t}_{1}\\  \tilde{t}_{2} \\
\tilde{b}_{1}\\  \tilde{b}_{2}
\end{array}
\right)
\hspace*{0.2cm} if \hspace*{0.2cm} \tilde{f} = \tilde{q} \,\,;
\hspace*{0.3cm} 
\displaystyle 
\tilde{f} \equiv \left(
\begin{array}{l}
\tilde{\nu}\\ 0 \\
\tilde{\tau}_{1}\\ \tilde{\tau}_{2}
\end{array}
\right) 
\hspace*{0.2cm} if \hspace*{0.2cm} \tilde{f} = \tilde{l}\,\,,\\
\end{array}
\label{eq:matcn}
\begin{array}{l}
\displaystyle 
\scp \equiv \left(
\begin{array}{l}
\scp_{1}\\  \scp_{2} 
\end{array}
\right)\,\,,
\hspace*{0.2cm} 
\displaystyle 
\sne \equiv \left(
\begin{array}{l}
\sne_{1} \\ \sne_{2} \\
\sne_{3}\\ \sne_{4}
\end{array}
\right)\,. 
\end{array}
\end{equation}
The actions appearing in eq.~(\ref{eq:gammaMSSM}) can be written as
\begin{equation} 
\label{eq:gammasf}
\Gamma_{\sfe}[V,\sfe]=\langle \sfe^{+} A_{\sfe}\sfe\rangle 
\end{equation}
where,
\begin{eqnarray}
  \label{eq:gammasfdef1}
  A_{\sfe}&\equiv&A_{\sfe}^{(0)}+A_{\sfe}^{(1)}+A_{\sfe}^{(2)}\nonumber \\  \label{eq:gammasfdef2}  \langle \sfe^{+} A_{\sfe}^{(i)}\sfe\rangle &\equiv&\sum_{\sfe}\int\dx\dy \sfe_x^{+}  A_{\sfe x y}^{(i)} \sfe_y\,\,,\,\,i=0,1,2
\end{eqnarray}
and the operators are:
\begin{eqnarray}
  \label{eq:opersfdef}
  A_{\sfe xy}^{(0)}&\equiv&(-\Box-\tilde M_{f}^2)_x\delta_{xy}\nonumber\\
  A_{\sfe xy}^{(1)}&\equiv&-i\,e\left(\partial_\mu A^\mu \hat{Q}_f+2\,\hat{Q}_f
    A_\mu\partial^\mu\right)_x\delta_{xy}
-\frac{i\,g}{c_w}\left(\partial_\mu Z^\mu \hat{G}_f+2\,\hat{G}_f
    Z_\mu\partial^\mu\right)_x\delta_{xy}\nonumber\\
  &-&\frac{i\,g}{\sqrt{2}}\left(\partial_\mu W^{+\mu} \Sigma_f^{tb}+2\,
    \Sigma_f^{tb} W_\mu^+\partial^\mu\right)_x\delta_{xy}+ {\rm h.c.}\nonumber\\
  A_{\sfe xy}^{(2)}&\equiv&\left(e^2 \hat{Q}_f^2 A_\mu
  A^\mu+\frac{2\,g\,e}{\cw}A_\mu Z^\mu
  \hat{Q}_f\hat{G}_f\right.
+\frac{g^2}{\cw^2}\hat{G}_f^2 Z_\mu Z^\mu
+ \frac{1}{2}g^2 \Sigma_f W_\mu^+ W^{\mu-}\nonumber\\
&+& \frac{eg}{\sqrt{2}} y_{f}A_{\mu}W^{\mu_{+}} \Sigma_{f}^{tb}
+ \frac{eg}{\sqrt{2}} y_{f}A_{\mu}W^{\mu_{-}}\Sigma_{f}^{bt}\nonumber\\
&-&\frac{g^{2}}{\sqrt{2}}y_{f} \frac{s_{\scriptscriptstyle W}^{2}}{c_{\scriptscriptstyle W}}
Z_{\mu}W^{\mu_{+}}\Sigma_{f}^{tb}
-\left.
\frac{g^{2}}{\sqrt{2}}y_{f} \frac{s_{\scriptscriptstyle W}^{2}}{c_{\scriptscriptstyle W}}
Z_{\mu}W^{\mu_{-}} \Sigma_{f}^{bt}\right)_x \delta_{xy}\,\,,
\end{eqnarray}
where $\sw^{2}=sin^{2}{\theta}_{\scriptscriptstyle W}$, 
$\cw^{2}=cos^{2}{\theta}_{\scriptscriptstyle W}$ and $y_{f}=\frac{1}{3}$ if 
$\tilde{f}=\tilde{q}$ or $y_{f}=-1$ if $\tilde{f}=\tilde{l}$.

Analogously we have
\begin{equation}
  \label{eq:gammasg}  
  \Gamma_{\sg}[V,\sg]=\frac{1}{2}\langle\bar{\sg}^o(A_0^{(0)}+A_0^{(1)})  
  \sne\rangle + \langle\bar{\sg}^+(A_+^{(0)}+A_+^{(1)})\scp\rangle +  
   \langle\bar{\sg}^o A_{0+}^{(1)}\scp\rangle + 
   \langle\bar{\sg}^+ A_{+0}^{(1)}\sne\rangle
\end{equation}
where
\begin{eqnarray}  
\label{eq:gammsgdef}  
\langle\bar{\sg}^o A_0^{(i)}\sne\rangle&\equiv&\int \dx\dy \bar{\sg}^o_x  
A_{0\,xy}^{(i)} \sne_y\,\,, \hspace*{0.4cm}  
\langle\bar{\sg}^+ A_+^{(i)}\scp\rangle \equiv \int \dx\dy \bar{\sg}^+_x  
A_{+\,xy}^{(i)} \scp_y \,\,;\,\,i=0,1 \nonumber\\  
\langle\bar{\sg}^o A_{0+}^{(1)}\scp\rangle&\equiv&\int \dx\dy \bar{\sg}^o_x  
A_{0+\,xy}^{(1)} \scp_y\,\,, \hspace*{0.2cm}  \langle\bar{\sg}^+ A_{+0}^{(1)}\sne\rangle \equiv \int \dx\dy \bar{\sg}^+_x  A_{+0\,xy}^{(1)} \sne_y \end{eqnarray} 
and the operators are:
\begin{eqnarray} \label{eq:opersgdef}
  A_{0\,xy}^{(0)}&\equiv&\left(i \slas{\partial}-\tilde M^0\right)_x \delta_{xy}\,\,, 
  \hspace*{0.5cm}  A_{+\,xy}^{(0)} \equiv \left(i \slas{\partial}-\tilde M^+\right)_x 
  \delta_{xy}\nonumber\\  A_{0\,xy}^{(1)}&\equiv&\frac{g}{c_w}Z_\mu 
  \gamma^\mu\left(O_L^{\prime\prime}  P_L+O_R^{\prime\prime} P_R\right)_x 
  \delta_{xy}\,\,, \hspace*{0.2cm}   A_{+\,xy}^{(1)} \equiv \left[\frac{g}{c_w}Z_\mu 
  \gamma^\mu\left(O_L^{\prime}  P_L+O_R^{\prime} P_R\right)-e\,A_\mu 
  \gamma^\mu\right]_x \delta_{xy}\nonumber\\  A_{0+\,xy}^{(1)}&\equiv&
  \left[g\, W_\mu^- \gamma^\mu\left(O_L      P_L+O_R P_R\right)\right]_x 
  \delta_{xy}\,\,, \hspace*{0.3cm}   A_{+0\,xy}^{(1)} \equiv \left[g\, W_\mu^+ 
  \gamma^\mu\left(O_L^{+}      P_L+O_R^{+} P_R\right)\right]_x \delta_{xy}\,.
\end{eqnarray}
In the above expressions the coupling matrices $\hat{Q}_f \,, \hat{G}_f \,,
\Sigma_{f}^{tb} \,, \Sigma_{f}^{bt} \,, \Sigma_{f}\,,$ and  
$ O_L\,, O_R\,, O_L^{\prime}\,,O_R^{\prime} \,,
O_L^{\prime\prime}\,, O_R^{\prime\prime}$ as well as the mass matrices 
$\tilde M_{f}, \tilde M^0$ and 
$\tilde M^+$ are defined in~\cite{GEISHA}.
 
The effective action can be written as:\\
\begin{equation}
\begin{array}{l}
\displaystyle e^{i \Gamma_{eff} [{\scriptscriptstyle V}]} = 
e^{i \Gamma_{o} [{\scriptscriptstyle V}]} e^{i \Gamma_{eff}^{\tilde{f}} 
[{\scriptscriptstyle V}]}e^{i \Gamma_{eff}^{\tilde{\chi}} 
[{\scriptscriptstyle V}]} 
\end{array}
\end{equation}
where,
\begin{equation}
\begin{array}{l}\label{eq:gammaeffF}\displaystyle e^{i \Gamma_{eff}^{\sfe} 
[{\scriptscriptstyle V}]} = \int [d\tilde{f}] [d\tilde{f}^{*}] 
e^{i \Gamma_{\tilde{f}} [{\scriptscriptstyle V},\tilde{f}]} 
\end{array}
\end{equation}
\begin{equation}
\begin{array}{l}
\label{eq:gammaeffCN}
\displaystyle e^{i \Gamma_{eff}^{\tilde{\chi}} [{\scriptscriptstyle V}]} = 
\int [d\tilde{\chi}^{+}] [d\bar{\tilde{\chi}}^{+}] [d\tilde{\chi}^{o}]  
e^{i \Gamma_{\tilde{\chi}} [{\scriptscriptstyle V},\tilde{\chi}]} 
\end{array}
\end{equation}
\hspace*{0.5cm}
After a Gaussian integration on the complex sfermion fields we find,
$$\Gamma_{eff}^{\sfe} [V] = i \Tr \log A_{\tilde{f}} = i \Tr \log [A_{\tilde{f}}^{(o)} (1 + A_{\tilde{f}}^{(o)-1}(A_{\tilde{f}}^{(1)} + A_{\tilde{f}}^{(2)}) )]$$
 and by  making the standard manipulations we get
\\
\begin{equation}
\begin{array}{l}
\displaystyle 
\Gamma_{eff}^{\sfe} [V] = i \sum_{k=1}^{\infty} \frac{(-1)^{k+1}}{k} \Tr 
[G_{\tilde{f}} (A_{\tilde{f}}^{(1)} + A_{\tilde{f}}^{(2)})]^{k} \end{array}
\end{equation}
 where the free sfermion propagator matrix $\displaystyle G_{\tilde{f}} \equiv A_{\tilde{f}}^{(o)^{-1}}$ is given by
\begin{equation}
\label{eq:propG}\displaystyle G_{\tilde{f} x y}^{ij} \equiv \int \frac{d^{{\scriptscriptstyle D}} q}{(2 \pi)^{{\scriptscriptstyle D}}} 
{\mu}_{o}^{4 - {\scriptscriptstyle D}} e^{-i q (x-y)} (q^{2} - \tilde{M}_{f}^{2})_{ij}^{-1} 
\end{equation}
with,\\
\hspace*{1cm} $\displaystyle (q^{2} - \tilde{M}_{f}^{2})^{-1} = \diag(\frac{1}{q^{2} - \tilde{m}_{t_{1}}^{2}},\frac{1}{q^{2} - \tilde{m}_{t_{2}}^{2}}, \frac{1}{q^{2} - \tilde{m}_{b_{1}}^{2}},\frac{1}{q^{2} - \tilde{m}_{b_{2}}^{2}})$ \hspace*{0.4cm} if \hspace*{0.4cm} $\tilde{f}=\tilde{q}$ \hspace*{0.2cm} ; 
or\\ \\
\hspace*{1cm} $\displaystyle (q^{2} - \tilde{M}_{f}^{2})^{-1} = \diag(\frac{1}{q^{2} - \tilde{m}_{\nu}^{2}}, \frac{1}{q^{2}},\frac{1}{q^{2} - \tilde{m}_{\tau_{1}}^{2}},\frac{1}{q^{2} - \tilde{m}_{\tau_{2}}^{2}})$ \hspace*{0.5cm} if 
\hspace*{0.5cm} $\tilde{f}=\tilde{l}$\\
\\
and the sums over the three generations and  the $N_{c}$ squarks
colors are implicit. Finally, if we keep just the terms that contribute to the two, three and
four point functions, the effective action generated from sfermions integration
can be written as, 
\begin{eqnarray}
  \label{eq:efffermions}
\displaystyle 
\Gamma_{eff}^{\sfe} [V] &=& i \Tr (G_{\tilde{f}} A_{\tilde{f}}^{(2)}) - 
\frac{i}{2} \Tr (G_{\tilde{f}} A_{\tilde{f}}^{(1)})^{2} - 
i \Tr (G_{\tilde{f}} A_{\tilde{f}}^{(1)} G_{\tilde{f}} A_{\tilde{f}}^{(2)})+
\frac{i}{3} \Tr (G_{\tilde{f}} A_{\tilde{f}}^{(1)})^{3} \nonumber\\
&- & \frac{i}{2} \Tr (G_{\tilde{f}} A_{\tilde{f}}^{(2)})^{2}+
i \Tr (G_{\tilde{f}} A_{\tilde{f}}^{(1)}G_{\tilde{f}}  A_{\tilde{f}}^{(1)}
G_{\tilde{f}} A_{\tilde{f}}^{(2)})-
\frac{i}{4} \Tr (G_{\tilde{f}} A_{\tilde{f}}^{(1)})^{4}+O(V^{5})\,\,,
\end{eqnarray}
Clearly, we can identify the first and second terms in eq.~(\ref{eq:efffermions}) with the one-loop
contributions from sfermions to the two-point functions; the third and fourth terms with 
the contributions to the
three-point functions and the last three terms are the corresponding contributions to the 
four-point functions.
  
On the other hand the contributions to the electroweak gauge bosons effective 
action comming from the neutralinos and the charginos are given by:
\begin{equation}
\begin{array}{l}\displaystyle 
e^{i \Gamma_{eff}^{\sg} [{\scriptscriptstyle V}]} = 
\int [d\tilde{\chi}^{+}] [d\bar{\tilde{\chi}}^{+}] [d\tilde{\chi}^{o}]e^{i \{\frac{1}{2} <\bar{\tilde{\chi}}^{o} (A_{o}^{(o)} + A_{o}^{(1)}) \tilde{\chi}^{o}> + <\bar{\tilde{\chi}}^{+} (A_{+}^{(o)} + A_{+}^{(1)}) \tilde{\chi}^{+}> + <\bar{\tilde{\chi}}^{o} A_{o+}^{(1)} \tilde{\chi}^{+}> + <\bar{\tilde{\chi}}^{+} A_{o+}^{(1)} \tilde{\chi}^{o}> \}}
\end{array}
\end{equation}
By performing first a standard Grassmann integration on the chargino fields we find,
$$\displaystyle 
e^{i \Gamma_{eff}^{\sg } [{\scriptscriptstyle V}]} = 
det(A_{+}^{(o)} + A_{+}^{(1)}) \int [d\tilde{\chi}^{o}] e^{i \frac{1}{2} 
<\bar{\tilde{\chi}}^{o} [A_{o}^{(o)} + A_{o}^{(1)}-
2A_{o+}^{(1)} (A_{+}^{(o)} + A_{+}^{(1)})^{-1} A_{+o}^{(1)}] \tilde{\chi}^{o}>}$$
Next we integrate over the neutralinos 
which are Majorana fermion fields  and find,
$$\displaystyle 
e^{i \Gamma_{eff}^{\sg } [{\scriptscriptstyle V}]} = det(A_{+}^{(o)} + A_{+}^{(1)}) det[A_{o}^{(o)} + A_{o}^{(1)} - 2A_{o+}^{(1)} (A_{+}^{(o)} + A_{+}^{(1)})^{-1} A_{+o}^{(1)}]^{\frac{1}{2}}.
$$
So that the effective action can be written as,
\begin{equation}
\begin{array}{l}
\displaystyle 
\Gamma_{eff}^{\tilde{\sg}} [V] = -i \Tr \log(A_{+}^{(o)} + A_{+}^{(1)}) - 
\frac{i}{2} \Tr  \log[A_{o}^{(o)} + A_{o}^{(1)} - 2A_{o+}^{(1)} (A_{+}^{(o)} + 
A_{+}^{(1)})^{-1} A_{+o}^{(1)}]
\end{array}
\end{equation}
Now, by introducing the chargino propagator
$k_{+} \equiv A_{+}^{(o)^{-1}}$ which is given by the matrix
\\
\begin{equation}
\label{eq:propcK}
\begin{array}{l}
\displaystyle 
k_{+ x y}^{ij} \equiv \int \frac{d^{{\scriptscriptstyle D}} q}
{(2 \pi)^{{\scriptscriptstyle D}}} {\mu}_{o}^{4 - {\scriptscriptstyle D}} e^{-i q (x - y)} (q{\hspace{-6pt}\slash} -\tilde{M}_{+})_{ij}^{-1}, \hspace*{0.5cm} i,j=1,2.
\end{array}
\end{equation}
and the neutralino propagator
 $k_{o} \equiv A_{o}^{(o)^{-1}}$ which is given by the matrix\\
\begin{equation}
\label{eq:propnK}
\begin{array}{l}\displaystyle k_{o x y}^{ij} \equiv \int \frac{d^{{\scriptscriptstyle D}} q}{(2 \pi)^{{\scriptscriptstyle D}}} {\mu}_{o}^{4 - {\scriptscriptstyle D}} 
e^{-i q (x - y)} (q{\hspace{-5pt}\slash} -\tilde{M}_{o})_{ij}^{-1}, \hspace*{0.5cm}i, j=1,2,3,4 
\end{array}
\end{equation}
we can writte the total {\it inos} contribution to the effective action as
\begin{eqnarray}
\label{eq:effinos}
\displaystyle 
\Gamma_{eff}^{\tilde{\chi}} [V] &=& \frac{i}{2} \Tr (k_{+} A_{+}^{(1)})^{2} +
\frac{i}{4} \Tr (k_{o} A_{o}^{(1)})^{2} + i \Tr (k_{o} A_{o+}^{(1)} k_{+}
A_{+o}^{(1)})-\frac{i}{3} \Tr(k_{+} A_{+}^{(1)})^{3}\nonumber\\
&-&  i \Tr (k_{o} A_{o+}^{(1)}k_{+} A_{+}^{(1)}  k_{+}A_{+o}^{(1)})
-i \Tr(k_{o} A_{o}^{(1)}k_{o} A_{o+}^{(1)}k_{+}A_{+o}^{(1)})
- \frac{i}{6} \Tr (k_{o} A_{o}^{(1)})^{3} \nonumber\\
&+&\frac{i}{4} \Tr(k_{+} A_{+}^{(1)})^{4}+
i \Tr (k_{o} A_{o+}^{(1)}{(k_{+} A_{+}^{(1)})}^{2}  k_{+}A_{+o}^{(1)})+
i \Tr {(k_{o} A_{o+}^{(1)}k_{+}A_{+o}^{(1)})}^{2}\nonumber\\
&+& i \Tr(k_{o} A_{o}^{(1)}k_{o} A_{o+}^{(1)}k_{+} A_{+}^{(1)}
k_{+}A_{+o}^{(1)})+i \Tr((k_{o} A_{o}^{(1)})^{2}k_{o}
A_{o+}^{(1)}k_{+}A_{+o}^{(1)})\nonumber\\
&+& \frac{i}{8} \Tr(k_{o} A_{o}^{(1)})^{4}+
O(V^{5})\,.
\end{eqnarray}
In the above formula the three first terms correspond with the one-loop contributions to
the two-point functions in the {\it inos} sector; the following four terms 
with the contributions to the three-point 
functions and the last six terms are the corresponding contributions to the four-point functions.

Thus the total effective action for the two, three, and four point Green's functions is
given by:
\begin{equation}
\label{eq:gammaop}
\displaystyle \Gamma_{eff} [V] = \Gamma_{o} [V]+\Gamma_{eff}^{\tilde{f}} [V]+
\Gamma_{eff}^{\tilde{\chi}} [V]\,, 
\end{equation}
where $\Gamma_{o} [V]$ is the effective action at tree level and 
 $\Gamma_{eff}^{\tilde{f}} [V]$ and $\Gamma_{eff}^{\tilde{\chi}} [V]$ are the 
effective actions generated from sfermions and {\it inos} respectively, which have been given in 
formulae  (\ref{eq:efffermions}) and (\ref{eq:effinos}).

The Feynman diagrams corresponding to the diferent terms appearing in the above 
equations~(\ref{eq:efffermions}) and (\ref{eq:effinos}) can be found in figure 1.

Finally, the effective action can be written as a function of the n-point Green's functions,
$\Gamma_{\mu \, \nu ... \, \rho}^{V_{1} \, V_{2} ... \, V_{n}}$, generically as:
\begin{equation}
\begin{array}{l}
\label{eq:effx}
\displaystyle \Gamma_{eff} [V] = \sum_{n}
\frac{1}{C_{\scriptstyle {V_{1}  V_{2}...\,V_{n}}}}
\int {\rm d}x_{1} ... {\rm d}x_{n} \,
\Gamma_{\mu \, \nu ... \, \rho}^{V_{1} \, V_{2} ... \, V_{n}} 
 (x_{1} \, x_{2} ... \,x_{n}) V_1^{\mu}(x_{1}) 
 \,V_2^{\nu}(x_{2}) \,... V_n^{\rho}(x_{n})\,\,, \\
\end{array}
\end{equation}
where $C_{\scriptstyle {V_{1} V_{2} ... V_{n}}}$ are the appropriate combinatorial
factors. For practical purposes, it is usefull to work in the momentum space where
the effective action is given by:
\begin{equation}
\label{eq:effp} 
\displaystyle
\Gamma_{eff} [V] = \sum_{n} \frac{1}{C_{\scriptstyle {V_{1} V_{2} ... V_{n}}}} 
\int  {\rm d}\tilde k_{1} ... {\rm d}\tilde k_{n} (2\pi)^{4}\,
\delta(\,{{\Sigma}_{i=1}^{n}k_{i}})
\Gamma_{\mu \, \nu ... \, \rho}^{V_{1} V_{2} ...  V_{n}} 
 (k_{1} \, k_{2} ... \,k_{n})\, V_1^{\mu}(k_{1})\,V_2^{\nu}(k_{2}) 
 \,... V_n^{\rho}(k_{n})\,, 
\end{equation}
where ${\rm d}\tilde k \equiv  {\rm d}^4 k /(2\pi)^4$ and the momentum-space Green's 
functions $\Gamma_{\mu \, \nu ... \, \rho}^{V_{1} \, V_{2} ... \, V_{n}}
(k_{1} \, k_{2} ... \,k_{n})$ are the Fourier transform of the ordinary space-time 
Green's functions $\Gamma_{\,\mu \, \nu ... \, \rho}^{V_{1} \, V_{2} ... \, V_{n}} 
(x_{1} \, x_{2} ... \,x_{n})$,
$$(2 \pi)^{4} \delta(\,{{\Sigma}_{i=1}^{n}k_{i}}) 
\Gamma_{\,\mu \,\nu... \, \rho}^{V_{1}\, V_{2} ...V_{n}}
(k_{1},k_{2}, ... ,k_{n}) \equiv 
\int dx_{1} dx_{2}...dx_{n} e^{i {\Sigma}_{i=1}^{n}k_{i}x_{i}} 
\Gamma_{\,\mu \,\nu... \, \rho}^{V_{1}\, V_{2} ... V_{n}}
(x_{1},x_{2}, ... ,x_{n}).$$
Our convention for the Fourier transform of the 
gauge bosons fields $V^{\mu}(k)$ is,
$$V^{\mu}(k)=\int dx\, e^{-ikx}V^{\mu}(x).$$ Finally, we
remind that in extracting the Green's functions from the effective action, the
proper symmetrization over the indexes and momenta corresponding to the
identical external fields must be performed. 

\section{Decoupling in the two-point functions}

In this section and in the following we study the asymptotic behaviour of the above 
effective action and the corresponding Green's functions in the regime where the 
sparticle masses are large. By large sparticle masses limit we mean generically,
$\tilde{m}_{\scriptscriptstyle i}^{2} \gg 
M_{\scriptscriptstyle EW}^{2}, k^{2}$, where $\tilde{m}_{\scriptscriptstyle i}$ denotes any of 
the physical sparticle masses, $M _{\scriptscriptstyle EW}$ any of the electroweak masses  
$(M_{\scriptscriptstyle Z}, M_{\scriptscriptstyle W}, m_{\scriptscriptstyle t},
\ldots)$ and $k$ denotes any of the external momenta. As for the analytical computation, whenever we refer 
to the large sparticle masses limit of a given one-loop
Feynman integral, we mean the asymptotic limit $\tilde{m}_{\scriptscriptstyle i}
\rightarrow \infty$ for all sparticle masses that are involved in that integral.
However, we would like to emphasize that this asymptotic limit is not fully defined 
unless one specifies in addition
the relative sizes of the involved masses.
In other words, the result may depend, in general, on the particular way this 
asymptotic limit is taken. Here we consider the asymptotic limit 
$\tilde{m}_{\scriptscriptstyle i,j}^{2}
\rightarrow \infty$, while keeping 
$|\frac{\tilde{m}_{\scriptscriptstyle i}^{2}-\tilde{m}_{\scriptscriptstyle j}^{2}}
{\tilde{m}_{\scriptscriptstyle i}^{2}+\tilde{m}_{\scriptscriptstyle j}^{2}}| \ll 1$
for all $i\neq j$. That is, we consider the
plaussible situation where there is a big gap between the SUSY particles and their standard
partners, but the differences among the SUSY masses are not large. 
First, we  
concentrate on the two-point functions. Details of this analysis can be found in 
\cite{GEISHA}. We just summarize here the main results. 

By working in the momentum space and by following the standard techniques 
it is possible to compute the two point functios coming from the integration of the 
sfermions and the {\it inos} according to the discussion introduced in the previous 
section. The corresponding part of the effective action can be written as
\begin{equation}
\Gamma_{eff} [V]_{[2]} = \frac{1}{C_{\scriptstyle V_{1}V_{2}}} \int 
 {\rm d}\tilde{p} \,\, {\rm d}\tilde{k} \,\,  \delta(p+k) (2\pi)^4
\Gamma_{\mu\,\nu}^{V_{1}V_{2}}(k) V_{1}^{\mu}(p)V_{2}^{\nu}(k)\,,
\end{equation}
where $C_{\scriptstyle V_{1}V_{2}}=n$ and $n$ denotes the number of external 
gauge bosons that are identical.

The exact results for each contribution to the two-point Green's functions in momentum
space and in a $R_{\xi}$ covariant gauge, $\Gamma^{A\, A}_{\mu\, \nu}(k),\,  \Gamma^{Z\, Z}_{\mu\, \nu}(k),\,$
$\Gamma^{A\, Z}_{\mu\, \nu}(k)$ and $\Gamma^{W\, W}_{\mu\, \nu}(k)$, can be found in
\cite{GEISHA}. 

As it was explained in the introduction and was mentioned at the begining of this 
section, we are interested in the asymptotic behavior 
of the Green's functions for very heavy SUSY masses. Thus we need to compute not 
just the exact results to one-loop of the Green's functions 
but their asymptotic expressions  valid in that limit. In order to get them we 
have analized the integrals by means of the so-called m-Theorem~\cite{GMR}. 
This theorem provides a powerful technique to study the asymptotic behaviour of
Feynman 
integrals in the limit where some of the masses are large. Notice that this is non
trivial since some of these integrals are divergent, in which case, the interchange of the 
integral with the large mass limit is not allowed. Thus, one should first compute the 
integrals with some regularization procedure as, for instance, 
dimensional regularization, and at the end take the large mass limit. Instead of this direct
way it is also possible to proceed as follows: First, one rearranges the integrand through
algebraic manipulations up to se\-parate the Feynman integral into a divergent part, which 
can be evaluated exactly using the standard techniques of dimensional regularization, and a
 convergent part that satisfies
the requirements demanded by the m-Theorem and therefore, goes to zero in the infinite mass
limit. By means of this procedure the correct asymptotic behaviour
of the integrals is guaranteed. This is the method we will follow in this work. Some examples of the computation of the Feynman integrals
 by means of the m-Theorem as well
as details of this theorem are given in~\cite{GEISHA}. The results for the one loop integrals
in the large masses limit that appear in the two-point functions are also
presented in that paper.

By following the above described method we have obtained the asymptotic
 behaviour of the two point functions in the large sparticle masses limit,
 which for the sfermions and {\it inos} sectors read respectivelly as follows:
\begin{eqnarray}
\label{eq:condition1}
\tilde{m}_{f_{i}}^{2} &\gg& M_{\scriptstyle EW}^{2}, \,k^{2}\,\,;\nonumber\\
|\tilde{m}_{f_{i}}^{2} - \tilde{m}_{f_{j}}^{2}| &\ll&
|\tilde{m}_{f_{i}}^{2} + \tilde{m}_{f_{j}}^{2}|\hspace*{0.2cm} \forall \,i,j\,, 
\end{eqnarray}
and
\begin{eqnarray}
\label{eq:condition2}
\tilde{M}_{i}^{2} &\gg& M_{\scriptstyle EW}^{2}, \,k^{2}\,,\nonumber\\
|\tilde{M}_{i}^{2} - \tilde{M}_{j}^{2}| &\ll& 
|\tilde{M}_{i}^{2} + \tilde{M}_{j}^{2}|\hspace*{0.2cm} \forall \,i,j \,,
\end{eqnarray}
where $\tilde{m}_{f_{i}}$ denotes the mass of the sfermion $\tilde{f}_{i}\,$, 
$\, \tilde{M}_{i}$ the mass of the {\it ino} $i$, $M_{\scriptstyle EW}$ is any of 
the electroweak masses and $k$ is any of the external momenta.
The results of the two-point functions $\Gamma^{V_{1}\, V_{2}}_{\mu\, \nu} (k)$ to 
one loop are given by,
\begin{equation}
\displaystyle 
\Gamma_{\mu \,\nu}^{V_{1} \, V_{2}}=
{\Gamma_{0}}_{\mu \,\nu}^{V_{1} \, V_{2}}+
\Delta {\Gamma}_{\mu \,\nu}^{V_{1} \, V_{2}}\,\,,
\end{equation}
where the tree level functions ${\Gamma_{0}}_{\mu \,\nu}^{V_{1} \, V_{2}}$ in a
$R_{\xi}$ covariant gauge are,
\begin{eqnarray}
\label{eq:efftree}
{\Gamma_0}^{V\, V}_{\mu\, \nu} (k) &=& (M_{\scriptscriptstyle V}^{2}-k^{2})
 g_{\mu\, \nu} + \left(1 - \frac{1}{\xi_{{\scriptscriptstyle V}}}\right)
 k_{\mu} k_{\nu} \,\,\,(V=Z,W)\,\,,\nonumber\\
{\Gamma_0}^{A\, A}_{\mu\, \nu} &=& 
-k^{2}g_{\mu\, \nu} + \left(1 - \frac{1}{\xi_{{\scriptscriptstyle A}}}\right)
 k_{\mu} k_{\nu} \,\,, \,\,\,
{\Gamma_{0}}_{\mu \,\nu}^{V_{1} \, V_{2}}=0 \,\,{\mbox if}\,\, V_{1}\neq V_{2}\,. 
\end{eqnarray}
and the contributions from sfermions and {\it  inos},
$\Delta {\Gamma}_{\mu \,\nu}^{V_{1} \, V_{2}}$, can be written as,
\begin{equation}
\Delta {\Gamma}_{\mu \,\nu}^{V_{1} \, V_{2}} (k) = 
\Sigma^{V_{1} \, V_{2}} (k)  g_{\mu\, \nu} + 
R^{V_{1} \, V_{2}} (k) k_{\mu} k_{\nu}\,\,.
\end{equation}
\hspace*{0.5cm}
We have shown in~\cite{GEISHA} that the asymptotic results are of the generic form:
\begin{eqnarray}
\Sigma^{V_{1} V_{2}} (k)&=&\Sigma^{V_{1} V_{2}}_{(0)}+
\Sigma^{V_{1} V_{2}}_{(1)} k^{2}+
O\left(\frac{k^{2}}{\Sigma \tilde{m}^{2}},\,
\frac{\bigtriangleup \tilde{m}^{2}}
{\Sigma\tilde{m}^{2}}\right) ,\nonumber\\
\nonumber\\
R^{V_{1} V_{2}} (k)&=&R^{V_{1} V_{2}}_{(0)}+
O\left(\frac{k^{2}}{\Sigma \tilde{m}^{2}},\,
\frac{\bigtriangleup \tilde{m}^{2}}
{\Sigma\tilde{m}^{2}}\right)\,\,,
\end{eqnarray}
where $\Sigma^{V_{1}\, V_{2}}_{(1)}$ and $R^{V_{1}\, V_{2}}_{(0)}$ contain the divergent 
contributions, namely the $O(1/\epsilon)$ terms in dimensional regularization and are 
functions of the large SUSY masses but are $k$ independent. Furthermore, we find
$R^{V_{1} V_{2}}_{(0)}=-\Sigma^{V_{1}\, V_{2}}_{(1)}$ in this asymptotic regime.
On the other hand, the $\Sigma^{V_{1}\, V_{2}}_{(0)}$ functions turn out to be finite
and $k$ independent, and they vanish in the asymptotic limit of infinite sparticle masses.
Here and in the following the terms denoted by
$O\left(\frac{k^{2}}{\Sigma \tilde{m}^{2}},\,
\frac{\bigtriangleup \tilde{m}^{2}}
{\Sigma\tilde{m}^{2}}\right)$ are suppressed by inverse powers of the large SUSY 
masses and vanish in the asymptotic regime. The large mass parameter of the asymptotic
expansion in the two-point functions is always taken to be the sum of the various squared masses involved in the
corresponding loop diagram which we denote here generically by 
$\Sigma \tilde{m}^{2}$. On the other hand, $\bigtriangleup \tilde{m}^{2}$ represents the
various corresponding squared mass differences which in our asymptotic limit are always
smaller than the corresponding sum.

These results can alternatively be expressed through the tranverse and longitudinal parts of 
the two-point functions, $\Sigma^{V_{1}\, V_{2}}_{\scriptscriptstyle T}$ and
$\Sigma^{V_{1}\, V_{2}}_{\scriptscriptstyle L}$, which are defined by,
\begin{equation}
\label{eq:tl}
\Gamma^{V_{1}\, V_{2}}_{\mu\, \nu} (k) = {\Gamma_0}^{V_{1}\, V_{2}}_{\mu\, \nu} (k)
+\Sigma_{\scriptscriptstyle T}^{V_{1}\, V_{2}} (k) \left( g_{\mu\, \nu}-
\frac{k_{\mu} k_{\nu}}{k^2}\right)+ \Sigma_{\scriptscriptstyle L}^{V_{1}\, V_{2}} (k)\,
\frac{k_{\mu} k_{\nu}}{k^2}\,.
\end{equation}
According to this definition, the asymptotic results whose explicit expressions are
given in~\cite{GEISHA} can be written, in a generic 
form, as:
\begin{eqnarray}
\Sigma^{V_{1} V_{2}}_{\scriptscriptstyle T} (k)&=&\Sigma^{V_{1} V_{2}}_{(0)}+
\Sigma^{V_{1} V_{2}}_{(1)} k^{2}+O\left(
\frac{k^{2}}{\Sigma \tilde{m}^{2}},\,
\frac{\bigtriangleup \tilde{m}^{2}}
{\Sigma\tilde{m}^{2}}\right),\nonumber\\
\nonumber\\
\Sigma^{V_{1}\, V_{2}}_{\scriptscriptstyle L} (k)&=& \Sigma^{V_{1} V_{2}}_{(0)}
+O\left(\frac{k^{2}}{\Sigma \tilde{m}^{2}},\,
\frac{\bigtriangleup \tilde{m}^{2}}
{\Sigma\tilde{m}^{2}}\right)\,\,,
\end{eqnarray}
Notice that $(\Sigma^{V_{1}\, V_{2}}_{\scriptscriptstyle L} -
\Sigma^{V_{1} V_{2}}_{\scriptscriptstyle T}) \propto k^{2}$. This result together
with the explicit
form of the $\Sigma^{V_{1} V_{2}}_{(0)}$ and $\Sigma^{V_{1} V_{2}}_{(1)}$ functions
demostrate that the decoupling indeed occurs in the two-point functions.
 
In order to illustrate the above result with one particular example, we choose to
present here the explicit expressions for the 
$\Sigma^{Z Z}$ contributions. The transverse 
contributions are~\cite{GEISHA}:
\begin{eqnarray}
\label{eq:SumqZZ}
\displaystyle \Sigma_{{\scriptscriptstyle T}}^{{\scriptscriptstyle ZZ}}
 (k)_{\tilde{q}} &=& N_{c} \frac{e^{2}}{16 \pi^{2}} \frac{1}
{s_{{\scriptscriptstyle W}}^{2} c_{{\scriptscriptstyle W}}^{2}}
\sum_{\tilde{q}} \left\{ 
\frac{1}{2}\left[ c_{t}^{2} s_{t}^{2} h({\tilde{m}}_{t_{1}}^{2},
 {\tilde{m}}_{t_{2}}^{2})
+ c_{b}^{2} s_{b}^{2} h({\tilde{m}}_{b_{1}}^{2}, 
{\tilde{m}}_{b_{2}}^{2})  \right]\right. \nonumber \\
&-& \frac{1}{3} k^{2} \left[
\left(\frac{c_{t}^{2}}{2} - \frac{2 s_{{\scriptscriptstyle W}}^{2}}{3}\right)^{2} 
\left(\Delta_\epsilon - \log \frac{\tilde{m}_{t_{1}}^{2}}{\mu_{o}^{2}}\right) 
\right.
+\left(\frac{s_{t}^{2}}{2} - \frac{2 s_{{\scriptscriptstyle W}}^{2}}{3}\right)^{2} 
\left(\Delta_\epsilon - \log \frac{\tilde{m}_{t_{2}}^{2}}{\mu_{o}^{2}}\right) \nonumber \\
 &+& \displaystyle
\left(-\frac{c_{b}^{2}}{2} + \frac{s_{{\scriptscriptstyle W}}^{2}}{3}\right)^{2}
\left(\Delta_\epsilon - \log
 \frac{\tilde{m}_{b_{1}}^{2}}{\mu_{o}^{2}}\right)
+\left(-\frac{s_{b}^{2}}{2} + \frac{s_{{\scriptscriptstyle W}}^{2}}{3}\right)^{2}
\left(\Delta_\epsilon - \log \frac{\tilde{m}_{b_{2}}^{2}}{\mu_{o}^{2}}\right) \nonumber \\
 &+& \displaystyle \frac{1}{2} s_{t}^{2} c_{t}^{2} 
\left(\Delta_\epsilon - \log \frac{\tilde{m}_{t_{1}}^{2} + \tilde{m}_{t_{2}}^{2}}
{2 \mu_{o}^{2}}\right)
+\left.\left. \frac{1}{2} s_{b}^{2} c_{b}^{2} 
\left(\Delta_\epsilon - \log \frac{\tilde{m}_{b_{1}}^{2} + \tilde{m}_{b_{2}}^{2}}{2 
\mu_{o}^{2}}\right) \right] \right\},\\ 
\nonumber\\
\label{eq:SumlZZ}
\displaystyle \Sigma_{{\scriptscriptstyle T}}^{{\scriptscriptstyle ZZ}}
 (k)_{\tilde{\l}} &=& -\frac{e^{2}}{16 \pi^{2}} \frac{1}{s_{{\scriptscriptstyle W}}^{2} 
 c_{{\scriptscriptstyle W}}^{2}} \sum_{\tilde{\l}} \left\{ 
-\frac{1}{2} c_{\tau}^{2} s_{\tau}^{2} h({\tilde{m}}_{\tau_{1}}^{2}, 
{\tilde{m}}_{\tau_{2}}^{2}) \right.
+\frac{1}{3} k^{2} \left[ \,\frac{1}{4}
\left(\Delta_\epsilon - \log \frac{\tilde{m}_{\nu}^{2}}{\mu_{o}^{2}}\right) \right. \nonumber \\
\displaystyle &+& \left(\frac{-c_{\tau}^{2}}{2} + s_{{\scriptscriptstyle W}}^{2}\right)^{2} 
\left(\Delta_\epsilon - \log
 \frac{\tilde{m}_{\tau_{1}}^{2}}{\mu_{o}^{2}}\right) 
+\left(-\frac{s_{\tau}^{2}}{2} + s_{{\scriptscriptstyle W}}^{2}\right)^{2} 
\left(\Delta_\epsilon - \log \frac{\tilde{m}_{\tau_{2}}^{2}}{\mu_{o}^{2}}\right)\nonumber \\ 
\displaystyle &+& \left. \left. \frac{1}{2} s_{\tau}^{2} c_{\tau}^{2} 
\left(\Delta_\epsilon - \log \frac{\tilde{m}_{\tau_{1}}^{2} + 
\tilde{m}_{\tau_{2}}^{2}}{2 \mu_{o}^{2}}\right) \right] \right\}, \\
\nonumber \\
\label{eq:SumcnZZ}
\displaystyle \Sigma_{{\scriptscriptstyle T}}^{{\scriptscriptstyle ZZ}}(k)_{\tilde{\chi}} &=& 
-\frac{e^{2}}{16 \pi^{2}} \frac{1}
{s_{{\scriptscriptstyle W}}^{2} c_{{\scriptscriptstyle W}}^{2}} \left\{
-\frac{1}{2} {(\tilde{M}_{3}^{o} - \tilde{M}_{4}^{o})}^{2}\right.
 \left(\Delta_{\epsilon} - \log \frac{\tilde{M}_{3}^o{}^{2} 
+ \tilde{M}_{4}^o{}^{2}}{2 \mu_{o}^{2}}\right) \nonumber \\
\displaystyle &+& \frac{1}{3} k^{2} \left[ 4 \left(\sw^{2}-1\right)^{2}  
\left(\Delta_{\epsilon} - \log \frac{\tilde{M}_{1}^+{}^{2}}
{\mu_{o}^{2}}\right)\right.
+ 4\left(\sw^{2}-\frac{1}{2}\right)^{2} 
\left(\Delta_{\epsilon} - \log \frac{\tilde{M}_{2}^+{}^{2}}{\mu_{o}^{2}}\right)\nonumber \\
\displaystyle &+& \left. \left.\left({\Delta}_{\epsilon} - 
\log \frac{\tilde{M}_{3}^o{}^{2}+\tilde{M}_{4}^o{}^{2}}{2 \mu_{o}^{2}}\right)
\right] \right\}, 
\end{eqnarray}
The results for the corresponding longitudinal parts can, generically, be written as:
\begin{equation}
\label{eq:trampa}
\Sigma_{{\scriptscriptstyle L}}^{{\scriptscriptstyle ZZ}}(k)= \left[\hspace*{0.1cm} 
{\rm Term}\hspace*{0.1cm}{\rm in} \hspace*{0.1cm}
\Sigma_{{\scriptscriptstyle T}}^{{\scriptscriptstyle ZZ}} (k)\hspace*{0.1cm}{\rm that}
\hspace*{0.1cm}{\rm is} \hspace*{0.1cm}k\hspace*{0.1cm}{\rm independent}  \,
 \right]\equiv \Sigma_{(0)}^{{\scriptscriptstyle ZZ}} \,.
\end{equation}
In the above equations $c_{f}=cos\theta_{f}, 
s_{f}=sin\theta_{f}$, with $\theta_{f}$ being the mixing angle in the $f$-sector, and
the sum in $\tilde{q}$ and $\tilde{l}$ run over the three squarks and sleptons
generations respectively. Besides,
\begin{equation}
\label{eq:lastint}
\hspace*{0.6cm}\displaystyle {\Delta}_\epsilon=\frac{2}{\epsilon }-{\gamma }_{\epsilon} 
+\log (4\pi) \hspace*{0.2cm}, \hspace*{0.2cm} \epsilon = 4-D\,;
\end{equation}
$\mu_{o}$ is the usual mass scale appearing in dimensional regularization and the function 
$h({m}_{1}^{2}, {m}_{2}^{2})$ is given by,
\begin{equation}
\label{eq:h}
h({m}_{1}^{2}, {m}_{2}^{2}) \equiv {m}_{1}^{2} \log \frac{2{m}_{1}^{2}}{{m}_{1}^{2} + {m}_{2}^{2}} +
{m}_{2}^{2} \log \frac{2{m}_{2}^{2}}{{m}_{1}^{2} + {m}_{2}^{2}}\,,
\end{equation}
which behaves as:
\begin{eqnarray}
h({m}_{1}^{2}, {m}_{2}^{2}) &\rightarrow&
\frac{{m}_{1}^{2} - {m}_{2}^{2}}{2} \left[
\frac{({m}_{1}^{2} - {m}_{2}^{2})}{({m}_{1}^{2} + {m}_{2}^{2})} +
O{\left(\frac{{m}_{1}^{2} - {m}_{2}^{2}}{{m}_{1}^{2} + {m}_{2}^{2}}\right)}^{2}
\right]
\end{eqnarray}
in the asymptotic limit. The explicit expressions for the other two-point functions, 
$\Gamma^{{\scriptscriptstyle AA}}, \Gamma^{{\scriptscriptstyle AZ}}$, and
$\Gamma^{{\scriptscriptstyle WW}}$ can be found in~\cite{GEISHA}.
 
As can be seen from our total results~\cite{GEISHA}, all the non-vanishing 
contributions to the two-point
functions in the asymptotic region are contained in
$\Sigma^{V_{1}\, V_{2}}_{(1)}$ and $R^{V_{1}\, V_{2}}_{(0)}$ and, besides,
$R^{V_{1}\, V_{2}}_{(0)}=-\Sigma^{V_{1}\, V_{2}}_{(1)}$. Therefore, they
 can be 
absorbed into a redefinition of the SM relevant para\-meters, $M_{\scriptscriptstyle W},
M_{\scriptscriptstyle Z}$ and $e$ and the gauge bosons wave functions. 
In consequence, the
decoupling of squarks, sleptons, charginos and neutralinos in the two point 
functions do indeed occur. 

\section{The three-point functions}

In this section we present the three-point functions for the electroweak gauge bosons to
one loop and analize the large masses limit of the SUSY particles.

In order to get the explicit expressions for the these functions one must work 
out the co\-rres\-ponding functional traces in the formulae (\ref{eq:efffermions}) and (\ref{eq:effinos}).
For this purpose one must substitute all the operators 
and propagators in these formulae, and compute all the appearing Dirac traces. The functional traces also 
involve to perform the sum in the corresponding matrix indexes, the sum over the various 
types of sfermions and the sum in color indexes in the case of squarks. We would like to 
mention that, in this paper, we have chosen to work in the momentum space, which 
turns out to simplify considerably the calculation of the functional traces.

By following the same procedure as in section 3 we have obtained the result for the effective action of the 
three-point functions coming from the integration of sfermions and {\it inos}. 
Generically, the corresponding part of the effective action 
can be written as,
\begin{equation}
\Gamma_{eff} [V]_{[3]} = \frac{1}{C_{\scriptstyle V_{1}V_{2}V_{3}}} \int 
 {\rm d}\tilde{p} \,\, {\rm d}\tilde{k} \,\, {\rm d}\tilde{r} \,\,  
\delta(p+k+r) (2\pi)^4
\Gamma_{\,\mu\,\nu\,\sigma}^{{\scriptstyle V_{1}V_{2}V_{3}}}\, V_{\,1}^{\mu}(p)
V_{\,2}^{\nu}(k) V_{\,3}^{\sigma}(r)\,,
\end{equation}
where $C_{\scriptstyle V_{1}V_{2}V_{3}}=n!$ and $n$ is the number of external gauge 
bosons that are identical.

\subsection{Sfermions contributions}

For simplicity, we show here the results in a general and compact
form and leave the details for the appendices. Once the appropiate traces 
have been computed, the corresponding effective action for the three-point 
functions coming from the sfermions integration can be expressed as,
\begin{eqnarray}
\label{eq:eff3}
{\Gamma_{eff}^{\sfe} [V]}_{[3]} &=&-{\pi}^2 \int {\rm d}\tilde{p}
\,{\rm d}\tilde{k}\,{\rm d}\tilde{r}\,\, \delta(p+k+r)\,
 \sum_{\tilde{f}}\left(\sum_{a,b} (\hat{O}^{1\,\mu})_{ab} 
(\hat{O}^{2\,\nu\sigma})_{ba}
  T^{a\,b}_{\mu}(p,\tilde{m}_{f_a}, \tilde{m}_{f_b})\,g_{\nu \sigma}\right.\nonumber\\ 
&-&\left. \frac{1}{3} \sum_{a,b,c} (\hat{O}^{1\,\mu})_{ab}
(\hat{O}^{1\,\nu})_{bc} (\hat{O}^{1\,\sigma})_{ca} \,\,
  T^{a\,b\, c}_{\mu \,\nu \,\sigma}(p,k,\tilde{m}_{f_a}, \tilde{m}_{f_b},
\tilde{m}_{f_c})\right),
\end{eqnarray}  
where, similarly to the two-point functions, the sum in ${\tilde f}$ runs over the three
generations and over the $N_{c}$ colors in the case of squarks, the indexes $a, b$ and $c$ 
run from one to four corresponding to the four entries of the sfermions column 
matrix $\tilde{f}$. $\,T^{a\,b}_{\mu}$ and $T^{a\,b\, c}_{\mu \,\nu \,\sigma}$ 
are the one-loop integrals that are defined as functions of the 
standard integrals in Appendix A, and $\hat{O}^{1\,\mu}$ and
$ \hat{O}^{2\,\mu\nu}$ are the {\em operators}
collected in Appendix B. It is important to emphasize that this formula is exact to one loop.

By substituting the definition of the {\em operators} involved in the above equation, we have obtained 
all the contributions to the 3-point functions to one loop. In particular, the
exact results for $AW^{+}W^{-}$ and $ZW^{+}W^{-}$ are given in Appendix C.
 
Futhermore, as we are interested in the large mass limit of the SUSY particles, we 
need the asymptotic expressions for the integrals appearing in the 
formula~(\ref{eq:eff3}), which we have obtained by means of the  
m-Theorem. The results of the these integrals in that limit can be
easily read from eqs.~(\ref{eq:int31}) and (\ref{eq:int31n}) respectively and
by using the corresponding asymptotic expressions for the scalar and tensor integrals
that have been presented in Appendix A. By substituting these asymptotic results
into eq.~(\ref{eq:eff3}) we finally get,
\begin{eqnarray}
\label{eq:efflim3}
{\Gamma_{eff}^{\sfe} [V]}_{[3]} &=& \frac{\,{\pi}^2}{9} \int 
 {\rm d}\tilde{p} \,\, {\rm d}\tilde{k} \,\, {\rm d}\tilde{r} \,\, \delta(p+k+r) 
 \sum_{\tilde{f}}  \nonumber\\
 && \left\{ \sum_{a,b,c} (\hat{O}^{1\,\mu})_{ab}
(\hat{O}^{1\,\nu})_{bc} (\hat{O}^{1\,\sigma})_{ca} \right.\,
\left(\left.
{\Delta}_\epsilon-\log \frac{\mfa+\mfb+\mfc}{3\mu_{o}^{2}}\right) \,
{\L}_{\mu\,\nu\,\sigma}\right\} \,,
\end{eqnarray}
where ${\L}_{\mu\,\nu\,\sigma}$ denotes the tensor appearing in the tree
 level vertex defined by,
\begin{equation}
\label{eq:optree}
{\L}_{\mu\,\nu\,\sigma}\equiv\left[(k-p)_{\sigma}g_{\mu\, \nu}+
(r-k)_{\mu}g_{\nu\, \sigma}+(p-r)_{\nu}g_{\mu\, \sigma}\right]\,.
\end{equation}

Therefore, the asymptotic result in eq.~(\ref{eq:efflim3}) is proportional to 
the tree level tensor 
${\L}_{\mu\,\nu\,\sigma}$. Thus, we can already conclude at this point
that the sfermions decouple in the three-point functions since this correction
being proportional to ${\L}_{\mu\,\nu\,\sigma}$ can be absorbed into redefinitions
of the SM parameters and the external gauge bosons wave functions. Notice that 
the two kind of one-loop Feynman integrals that appear in the three-point 
functions, $T^{a\,b}_{\mu}$ and $T^{a\,b\, c}_{\mu \,\nu \,\sigma}$, involve generically
two and three different sparticle masses respectively, which in our limit are
considered to be large. However, in order to implement the large SUSY masses limit
one must choose a proper combination of masses such that there is just one large mass 
parameter while the others are kept small. Our choice for the large mass
parameter is always the sum of the various squared SUSY masses involved in the
loop integral. The rest of the mass parameters can be expressed in terms of the
sparticle squared mass differences which in our approximation are small as
compared to their sum. The result in eq.~(\ref{eq:efflim3}) has
corrections not explicitely shown which are suppresed by inverse powers of these
large SUSY mass sums and therefore they vanish in our asymptotic limit.
For completeness, we present also here the explicit contributions to the 
three-point Green's functions with specific external 
gauge bosons, $\Gamma_{\mu \,\nu \,\sigma}^{V_{1} V_{2} V_{3}}$. 
Our results are presented in the  form:
\begin{equation}
\displaystyle \Gamma_{\mu \,\nu \,\sigma}^{V_{1} \, V_{2} \, V_{3}}=
{\Gamma_{0}}_{\mu \,\nu \,\sigma}^{V_{1} \, V_{2} \, V_{3}}+
\Delta {\Gamma}_{\mu \,\nu \,\sigma}^{V_{1} \, V_{2} \, V_{3}}\,,
\end{equation}
where the outgoing momenta assignements are $V_{1}^{\mu}(p)$, $V_{2}^{\nu}(k)$ and
$V_{3}^{\sigma}(r)$ and the tree level contributions are:
\begin{equation}
\displaystyle {\Gamma_{0}}_{\mu \,\nu \,\sigma}^{A W^+ W^-}=-e\,{\L}_{\mu \nu \sigma}\,\,,
\,\,\, {\Gamma_{0}}_{\mu \,\nu \,\sigma}^{Z W^+ W^-}=-g\cw\,{\L}_{\mu \nu \sigma}\,.
\end{equation}
\hspace*{0.5cm}
In order to get the sfermions contributions, one must substitute all the
{\em operators} that appear in eq.(\ref{eq:efflim3}), perform the corresponding 
sums and after a rather lengthy
calculation, the following results are obtained:
\begin{eqnarray}
\label{eq:AWW}
\displaystyle {\Delta {\Gamma}_{\mu \,\nu \,\sigma\hspace*{0.4cm}\tilde{q}}^{A W^+ W^-}}&=&
\frac{eg^{2}}{16 \pi^{2}} \frac{N_{c}}{9}\, {\L}_{\mu\,\nu\,\sigma} \sum_{\tilde{q}}
\frac{1}{2} \left\{\Delta_{\epsilon}
+f_{1}(\tilde{m}_{t_{1}}^{2}, \tilde{m}_{t_{2}}^{2}, \tilde{m}_{b_{1}}^{2},
\tilde{m}_{b_{2}}^{2})\right\} \nonumber\\
&+& \displaystyle {F_{1}}_{\mu \,\nu \,\sigma}\left[
O\left(\frac{p^{2}}{\Sigma \tilde{m}^{2}},\,\frac{\bigtriangleup \tilde{m}^{2}}
{\Sigma\tilde{m}^{2}}\right)\right]\,\,,\\
\nonumber\\
\label{eq:ZWW}
\displaystyle {\Delta {\Gamma}_{\mu \,\nu \,\sigma\hspace*{0.4cm}\tilde{q}}^{Z W^+ W^-}}&=&
 -\frac{g^{3}}{16 \pi^{2}} \frac{N_{c}}{6\cw}\,{\L}_{\mu\,\nu\,\sigma} 
\sum_{\tilde{q}}\frac{1}{3}s_{{\scriptscriptstyle W}}^{2} 
\left\{\Delta_{\epsilon} 
+f_{2}(\tilde{m}_{t_{1}}^{2}, \tilde{m}_{t_{2}}^{2}, \tilde{m}_{b_{1}}^{2},
\tilde{m}_{b_{2}}^{2})\right\} \nonumber\\
&+& \displaystyle {F_{2}}_{\mu \,\nu \,\sigma}\left[
O\left(\frac{p^{2}}{\Sigma \tilde{m}^{2}},\,\frac{\bigtriangleup \tilde{m}^{2}}
{\Sigma\tilde{m}^{2}}\right)\right]\,,
\end{eqnarray}
where, generically, $p$ denotes any of the external momenta, 
$\bigtriangleup \tilde{m}^{2}$ denotes the various squared mass differences
and $\Sigma\tilde{m}^{2}$ denotes the corresponding large mass parameter 
which in our case is always a sum of squared SUSY masses. The functions 
${F_{i}}_{\mu \,\nu \,\sigma} \,(i=1,2)$ are finite and  they go to zero in the 
limit of $\tilde{m}_{i,j}\rightarrow\infty \,(\forall i,j)$ with 
$|\tilde{m}_{i}^{2} - \tilde{m}_{j}^{2}| \ll
|\tilde{m}_{i}^{2} + \tilde{m}_{j}^{2}|$.
The functions $f_{1,2}(\tilde{m}_{t_{1}}^{2}, \tilde{m}_{t_{2}}^{2},
\tilde{m}_{b_{1}}^{2}, \tilde{m}_{b_{2}}^{2})$ are given explicitely
in Appendix B. These functions are also finite but different from zero 
in the large masses limit and therefore, they contain all the potentially 
non-decoupling effects of the three-point functions. More specifically, these
effects are given by the logarithmic dependence on the large mass parameter
of these two functions. Generically, these can be written as, 
\begin{equation}
f_{1,2}\,(\tilde{m}_{t_{1}}^{2}, \tilde{m}_{t_{2}}^{2}, \tilde{m}_{b_{1}}^{2},
\tilde{m}_{b_{2}}^{2}) = O\left(\log 
\frac{\Sigma \tilde{m}^{2}}{\mu_{o}^{2}}\right)+
O\left(\frac{\bigtriangleup \tilde{m}^{2}}
{\Sigma\tilde{m}^{2}}\right)\,.
\end{equation}

As we have mentioned before, the corrections $\Delta {\Gamma}$ are proportional to the 
tree level vertex ${\L}_{\mu\,\nu\,\sigma}$, and therefore the potentially non-decoupling 
effects in the three-point functions can be absorbed into redefinitions of the 
coupling constants and wave functions renormalization. Therefore, this is an explicit
proof of decoupling of squarks in the 
${\Gamma}_{\mu \,\nu \,\sigma}^{A W^+ W^-}$ and
${\Gamma}_{\mu \,\nu \,\sigma}^{Z W^+ W^-}$ Green functions.  

We would like to point out that the other three-point Green's functions are
exactly zero in our limit, as it was expected. As a check of the previous functional
computation we have also calculated all these three point functions by 
diagramatic methods and we have got the same results.

Similar results are obtained for the sleptons sector doing the
corresponding replacements: $\tilde{q} \rightarrow \tilde{l},\,
N_{c} \rightarrow 1,\, \tilde{m}_{t_{1}} \rightarrow  \tilde{m}_{\nu},\,
\tilde{m}_{b_{1}} \rightarrow  \tilde{m}_{\tau_{1}},\, 
\tilde{m}_{b_{2}} \rightarrow  \tilde{m}_{\tau_{2}},\, 
c_{t} \rightarrow 1, s_{t} \rightarrow 0, c_{b} \rightarrow c_{\tau}$,
$s_{b} \rightarrow s_{\tau}$ and $y_{f}=\frac{1}{3}\rightarrow
y_{f}=-1$. 

\subsection{{\it Inos} contributions}

To compute the {\it inos} contributions to the three-point functions one must 
work out the functional traces given in eq.~(\ref{eq:effinos}).
This leads to an expression containing several combinations of momenta, operators and 
Dirac traces corresponding to specific external gauge bosons 
$V_{1} \, V_{2} \, V_{3}$ that we give explicitely in Appendix B.

The result for the effective action coming from the integration of {\it inos} in 
the three-point functions can be expressed in a compact form as,
\begin{eqnarray}
\label{eq:eff3inos}
&& \displaystyle {\Gamma_{eff}^{\sg} [V]}_{[3]} = -i \int
{\rm d}\tilde{p} \,\, {\rm d}\tilde{k} \,\, {\rm d}\tilde{r} \,\,
(2\pi)^{4} \,\delta(p+k+r)\,\times\nonumber\\
&&\displaystyle \int d\widehat{q}\, \left[\,
\frac{1}{3} \sum_{i,j,k=1}^{2} {\cal F}^{ijk}(\tilde{M}_{i}^{+}, \tilde{M}_{j}^{+},
\tilde{M}_{k}^{+}) \right. 
\left\{  \,q_{\,1}^{\,\alpha}\,q_{\,2}^{\,\beta}\, q_{\,3}^{\,\gamma}\,
{\left(\, G\cdot O \,\right)}^{\scriptstyle +++}_{\scriptstyle \,1\,2\,3}\right.\nonumber\\ 
&&\displaystyle + q_{\,1}^{\,\alpha}\,\tilde{M}_{j}^{+} \tilde{M}_{k}^{+}
{\left( \,G\cdot O \,\right)}^{\scriptstyle +++}_{\scriptstyle \,1}
+ q_{\,2}^{\,\alpha}\,\tilde{M}_{i}^{+} \tilde{M}_{k}^{+}
{\left( \,G\cdot O \,\right)}^{\scriptstyle +++}_{\scriptstyle \,2}
+ \left. q_{\,3}^{\,\alpha}\,\tilde{M}_{i}^{+} \tilde{M}_{j}^{+}
{\left( \,G\cdot O \,\right)}^{\scriptstyle +++}_{\scriptstyle \,3}\right\}\nonumber\\
&&\displaystyle +\sum_{i=1}^{4} \sum_{j,k=1}^{2} 
{\cal F}^{ijk}(\tilde{M}_{i}^{o}, \tilde{M}_{j}^{+},\tilde{M}_{k}^{+})\left\{\,
q_{\,1}^{\,\alpha}\, q_{\,2}^{\,\beta}\, q_{\,3}^{\,\gamma}\,
{\left(\, G\cdot O \,\right)}^{\scriptstyle \,o++}_{\scriptstyle \,1\,2\,3}\right.\nonumber\\ 
&&\displaystyle +q_{\,1}^{\,\alpha}\,\tilde{M}_{j}^{+} \tilde{M}_{k}^{+}
{\left( \,G\cdot O \,\right)}^{\scriptstyle \,o++}_{\scriptstyle \,1}+
q_{\,2}^{\,\alpha}\,\tilde{M}_{i}^{o} \tilde{M}_{k}^{+}
{\left( \,G\cdot O \,\right)}^{\scriptstyle \,o++}_{\scriptstyle \,2}+\left.
q_{\,3}^{\,\alpha}\,\tilde{M}_{i}^{o} \tilde{M}_{j}^{+}
{\left( \,G\cdot O \,\right)}^{\scriptstyle \,o++}_{\scriptstyle \,3}\right\}\nonumber\\
&&\displaystyle +\sum_{i,j=1}^{4} \sum_{k=1}^{2} 
{\cal F}^{ijk}(\tilde{M}_{i}^{o}, \tilde{M}_{j}^{o},\tilde{M}_{k}^{+})\left\{\,
q_{\,1}^{\,\alpha}\, q_{\,2}^{\,\beta}\, q_{\,3}^{\,\gamma}\,
{\left(\, G\cdot O \,\right)}^{\scriptstyle \,oo+}_{\scriptstyle \,1\,2\,3}\right.\nonumber\\ 
&&\displaystyle + q_{\,1}^{\,\alpha}\,\tilde{M}_{j}^{o} \tilde{M}_{k}^{+}
{\left( \,G\cdot O \,\right)}^{\scriptstyle \,oo+}_{\scriptstyle \,1}+
q_{\,2}^{\,\alpha} \,\tilde{M}_{i}^{o} \tilde{M}_{k}^{+}
{\left( \,G\cdot O \,\right)}^{\scriptstyle \,oo+}_{\scriptstyle \,2}
+\left. q_{\,3}^{\,\alpha}\,\tilde{M}_{i}^{o} \tilde{M}_{j}^{o}
{\left( \,G\cdot O \,\right)}^{\scriptstyle \,oo+}_{\scriptstyle\,3}
\right\}\nonumber\\
&&\displaystyle +\frac{1}{6}\sum_{i,j,k=1}^{4} 
{\cal F}^{ijk}(\tilde{M}_{i}^{o}, \tilde{M}_{j}^{o},\tilde{M}_{k}^{o})\left\{\,
q_{\,1}^{\,\alpha}\, q_{\,2}^{\,\beta}\, q_{\,3}^{\,\gamma}\,
{\left(\, G\cdot O \,\right)}^{\scriptstyle \,ooo}_{\scriptstyle \,1\,2\,3}\right.\nonumber\\ 
&&\displaystyle +\left. q_{\,1}^{\,\alpha}\,\tilde{M}_{j}^{o} \tilde{M}_{k}^{o}
{\left( \,G\cdot O \,\right)}^{\scriptstyle \,ooo}_{\scriptstyle \,1}+
q_{\,2}^{\,\alpha} \,\tilde{M}_{i}^{o} \tilde{M}_{k}^{o}
{\left( \,G\cdot O \,\right)}^{\scriptstyle \,ooo}_{\scriptstyle \,2}
+\left. q_{\,3}^{\,\alpha}\,\tilde{M}_{i}^{o} \tilde{M}_{j}^{o}
{\left( \,G\cdot O \,\right)}^{\scriptstyle \,ooo}_{\scriptstyle \,3}\,\right\}
{\phantom {\frac{!}{!}}}\right]\,,
\end{eqnarray}
where $\left( \,G\cdot O \,\right)$ denotes the various products of traces and 
operators that are collected in Appendix B. The superscripts in 
$\left( \,G\cdot O \,\right)$ correspond with the type of 
sparticles appearing in the loop or, equivalently, in the internal Feynman's 
propagators, and the subscripts denote the corresponding momenta to be contracted with the results of the traces in each case. 
For example, in 
${\left(\, G\cdot O \,\right)}^{\scriptstyle +++}_{\scriptstyle \,1\,2\,3}$,
the superscripts ${\scriptstyle +++}$ denote the three charginos in the 
loop and the subscripts $123$ mean that the traces must
be contracted with the $q_{1}, q_{2}$ and $q_{3}$ momenta. The indexes $i,j,k$ in the above formula 
vary as $i,j,k=1,2$ if they refer to
charginos and as $i,j,k=1,\ldots,4$ if they refer to neutralinos and the generic function 
${\cal F}^{ijk}(\tilde{M}_{i}, \tilde{M}_{j},\tilde{M}_{k})$ is given by,
$${\cal F}^{ijk}(\tilde{M}_{i}, \tilde{M}_{j},\tilde{M}_{k})= 
\frac{1}{\left[q_{\,1}^2 - \tilde{M}_{i}^{2} \right]\,
\left[{q_{\,2}}^2 - \tilde{M}_{j}^{2} \right]\,
\left[{q_{\,3}}^2 - \tilde{M}_{k}^{2} \right]}\,,$$
where, 
\begin{equation}
\label{eq:qs}
q_{\,1}\,\,\equiv q\,\,\,\,, \,q_{\,2}\,\equiv q+p\,\,\,\,,
\,q_{\,3}\equiv q+p+k\,.
\end{equation}

As we have explained above, the next step is to compute each Dirac trace 
appearing in the expression (\ref{eq:eff3inos}), substitute the 
{\em operators}, perform the corresponding traces and finally to extract the
various three-point functions with specific external legs
which we do not present entirely here for brevity. We have computed each 
contribution 
to these functions and have checked that the results for 
$\Delta\Gamma^{AAA}, \,\Delta\Gamma^{AAZ},\,\Delta\Gamma^{AZZ}$ and 
$\Delta\Gamma^{ZZZ}$ are finite as it was expected. 

The exact results to one loop for the $AW^+W^-$ and $ZW^+W^-$ three point
functions are collected in Appendix C.

In order to get the assymptotic limit of the Green's 
functions in eqs.~(\ref{eq:eff3inos}), (\ref{eq:effAWW}) and (\ref{eq:effZWW}) we use the results of the one loop integrals 
in the large masses limit that are collected in Appendix A and the 
values for the coupling matrices $O_{L,\,R}\,, O_{L,\,R}^{\prime}$ and
$O_{L,\,R}^{\prime\prime}$ in the limit of large neutralino and chargino masses
that can be found in~\cite{GEISHA}. By substituting all these results into
eq.~(\ref{eq:eff3inos}) we find the {\it inos} 
contributions to the three-point part of the effective action which can be 
written as:
\begin{eqnarray}
\label{eq:3inoslimit}
\displaystyle && {\Gamma_{eff}^{\sg} [V]}_{[3]}= -\frac{4}{3}\,\pi^{2} \,\int
{\rm d}\tilde{p} \,\, {\rm d}\tilde{k} \,\, {\rm d}\tilde{r}\,\,
\delta(p+k+r) 
\, \sum_{i,j,k}\left\{\frac{1}{3}\left(\hat{O}^{1}+\hat{O}^{2}+
\hat{O}^{4}+\hat{O}^{6}+\hat{O}^{8}\right)^{\mu \,\nu\,\sigma}_{\,i\,j\,k}
\right.\nonumber\\
&&+\frac{1}{6}\,\hat{O}^{12\, \mu \,\nu\,\sigma}_{\hspace*{0.4cm}i\,j\,k}
\left.+\left(\hat{O}^{16}+\hat{O}^{18}\right)^{\mu\,\nu\,\sigma}_{\,i\,j\,k}
+\hat{O}^{22\, \mu \,\nu\,\sigma}_{\hspace*{0.4cm}i\,j\,k}\right\}
\left({\Delta}_\epsilon-\log \frac{\tilde{M}_{i}^{2}+\tilde{M}_{j}^{2}+
\tilde{M}_{k}^{2}}{3\mu_{o}^{2}}\right){\L}_{\mu\,\nu\,\sigma},
\end{eqnarray}
where ${\L}_{\mu\,\nu\,\sigma}$ represents the tree level tensor defined in 
eq.~(\ref{eq:optree}) and the {\em operators}
$\hat{O}^{\mu \,\nu\,\sigma}_{\,i\,j\,k}$ can be found in Appendix B.
Notice that the indices $ijk$ vary in accordance with the {\it inos}
particles appearing in the loops, i.e, $i,j,k=1,2$ if they refer to charginos and
$i,j,k=1,...,4$ if they refer to neutralinos.

The fact that this result is proportional again to the tree level tensor ${\L}_{\mu\,\nu\,\sigma}$
enables us to conclude that the {\it inos} also decouple in the three-point
functions. For completeness we have worked out, in detail, the explicit
expressions for the three-point functions with specific external gauge
bosons that are different from zero in our limit. By using the same notation 
as in subsection 4.1 we have obtained,
\begin{eqnarray}
\label{eq:AWWinoslim}
\displaystyle {\Delta {\Gamma}_{\mu \,\nu \,\sigma\hspace*{0.4cm}\sg}^{A W^+ W^-}}&=&
\frac{eg^{2}}{16 \pi^{2}} \frac{4}{3}\, {\L}_{\mu\,\nu\,\sigma} 
 \left\{\frac{3}{2} \Delta_{\epsilon}+ f_{3}(\tilde{M}_{1}^{+},\tilde{M}_{2}^{+},
\tilde{M}_{1}^{0},\tilde{M}_{2}^{0},
\tilde{M}_{3}^{0},\tilde{M}_{4}^{0})\right\} \nonumber\\
&+& \displaystyle {F_{3}}_{\mu \,\nu \,\sigma}\left[
O\left(\frac{p^{2}}{\Sigma \tilde{m}^{2}},\,\frac{\bigtriangleup \tilde{m}^{2}}
{\Sigma\tilde{m}^{2}}\right)\right]\,\,,\\
\label{eq:ZWWinoslim}
\displaystyle {\Delta {\Gamma}_{\mu \,\nu \,\sigma\hspace*{0.4cm}\sg}^{Z W^+ W^-}}&=&
 \frac{g^{3}}{16 \pi^{2}} \frac{4}{3\cw} \,{\L}_{\mu\,\nu\,\sigma} 
\left\{\frac{3}{2}c_{{\scriptscriptstyle W}}^{2} 
\Delta_{\epsilon}+ f_{4}(\tilde{M}_{1}^{+},\tilde{M}_{2}^{+},
\tilde{M}_{1}^{0},\tilde{M}_{2}^{0},
\tilde{M}_{3}^{0},\tilde{M}_{4}^{0})\right\} \nonumber\\
&+& \displaystyle {F_{4}}_{\mu \,\nu \,\sigma}\left[
O\left(\frac{p^{2}}{\Sigma \tilde{m}^{2}},\,\frac{\bigtriangleup \tilde{m}^{2}}
{\Sigma\tilde{m}^{2}}\right)\right]\,,
\end{eqnarray}
where the functions ${F_{i}}_{\mu \,\nu \,\sigma} \,(i=3,4)$ are finite and  
we have proved explicitly that they go to zero in our asymptotic limit.
On the other hand, the functions 
$f_{i}(\tilde{M}_{1}^{+},\tilde{M}_{2}^{+},
\tilde{M}_{1}^{0},\tilde{M}_{2}^{0},
\tilde{M}_{3}^{0},\tilde{M}_{4}^{0}) \, (i=3,4)$ are finite and 
different from zero in the large masses limit and therefore, they contain all the potentially 
non-decoupling effects of the three-point functions. Their explicit expressions
can be found in Appendix B. However, as we have commented above, the corrections $\Delta\Gamma$ given in
(\ref{eq:AWWinoslim}) and (\ref{eq:ZWWinoslim}),
are also proportional to the tree level tensor ${\L}_{\mu\,\nu\,\sigma}$ and
therefore, the mentioned potentially non-decoupling effects can be absorbed into
redefenitions of the SM parameters and the gauge bosons wave functions. The
results in eqs.~(\ref{eq:AWWinoslim}) and (\ref{eq:ZWWinoslim}) demostrate
explicitely, therefore, the decoupling of {\it inos} in the 
${\Gamma}^{A W^+ W^-}$ and ${\Gamma}^{Z W^+ W^-}$ functions.

In addition, we have checked that after the proper symmetrization over the
identical external fields, the $\Delta{\Gamma}^{AAA}, \Delta{\Gamma}^{AAZ},
\Delta{\Gamma}^{AZZ}$ and $\Delta{\Gamma}^{ZZZ}$ contributions are exactly zero
in our limit as it was expected since there are no corresponding tree level
vertices. As a check of the previous functional
computation we have also calculated all these three point functions by 
diagramatic methods and we have got the same results. 

\section{The four-point functions and higher}

In this section we compute the four-point Green's functions with external 
gauge bosons, $A, Z, W^{+}, W^{-}$, at one loop level. At the end of this
 section we discuss also the case of higher 
point functions, which completes our analysis of decoupling of SUSY particles.

Let us begin by writting the expression of the corresponding part of the effective action as 
a function of the 4-point functions
$\Gamma_{\mu \,\nu \,\sigma\,\lambda}^{V_{1} \, V_{2} \, V_{3}\,V_{4}}$,
\begin{equation}
\begin{array}{l}
\displaystyle
\Gamma_{eff} [V]_{[4]} = \frac{1}{C_{\scriptstyle {V_{1} V_{2} V_{3} V_{4}}}} 
\int{\rm d}\tilde{p} \,\, {\rm d}\tilde{k} \,\, {\rm d}\tilde{r} \,\, 
 {\rm d}\tilde{t} \,\,\delta(p+k+r+t) (2\pi)^4 
 \Gamma_{\,\mu \, \nu \,\sigma\,\lambda}^{{\scriptstyle V_{1}V_{2}V_{3}V_{4}}} 
 V_{\,1}^{\mu}(p) V_{\,2}^{\nu}(k) V_{\,3}^{\sigma}(r) V_{\,4}^{\lambda}(t) \,,
\end{array}
\end{equation}
where $C_{\scriptstyle V_{1}V_{2}V_{3}V_{4}}$ is the appropriate combinatorial
factor for the number of identical external gauge bosons.

By working in the momentum space and by following the same techniques described
in the previous sections one computes the four-point functions coming from the integrations 
of sfermions and {\it inos}. Clearly, this computation involves to work
out again the corresponding functional traces 
given in eqs.~(\ref{eq:efffermions}) and (\ref{eq:effinos}).

\subsection{Sfermions contributions} 

The resulting effective action for 4-point functions that are generated
from sfermions integration can be 
summarized in the following expression,
\begin{eqnarray}  
 \label{eq:eff4} 
{\Gamma_{eff}^{\tilde{f}}[V]}_{[4]} &=&{\pi}^2 \int {\rm d}\tilde{p} \,
{\rm d}\tilde{k} \,{\rm d}\tilde{r} \,{\rm d}\tilde{t} \,\,\delta(p+k+r+t) 
 \sum_{\tilde{f}}\times \nonumber\\ 
&& \left( \frac{1}{2}\sum_{a,b} (\hat{O}^{2\,\mu\nu})_{\,ab} 
 (\hat{O}^{2\,\sigma\lambda})_{\,ba} \,g_{\mu\,\nu}g_{\sigma\lambda}\,
 J^{a\,b}_{p+k}(p+k,\tilde{m}_{f_a}, \tilde{m}_{f_b})\right.  \nonumber\\ 
&& -\sum_{a,b,c}(\hat{O}^{1\,\mu})_{ab} (\hat{O}^{1\,\nu})_{bc}
(\hat{O}^{2\,\sigma\lambda})_{ca}\,g_{\sigma \lambda} \,\,
J^{a\,b\, c}_{\mu \,\nu}(p,k,\tilde{m}_{f_a}, \tilde{m}_{f_b},\tilde{m}_{f_c})
\nonumber\\ 
&&\left.+\frac{1}{4} \sum_{a,b,c,d}(\hat{O}^{1\,\mu})_{ab}
(\hat{O}^{1\,\nu})_{bc} (\hat{O}^{1\,\sigma})_{cd}
(\hat{O}^{1\,\lambda})_{da} \,\, J^{a\,b\,c\,d}_{\mu \,\nu\, \sigma\,\lambda} 
(p,k,r,\tilde{m}_{f_a}, \tilde{m}_{f_b},\tilde{m}_{f_c})\right),
\end{eqnarray}
where the indexes $a, b, c$ and $d$ run from one to four corresponding
to the four entries of the sfermion matrix in eq.(\ref{eq:matf}) and the integrals 
and {\em operators} appearing in this formula are the ones given in Appendix A and B respectively. 

From this formula we have obtained the sfermion contributions to the 4-point
functions,
${\Gamma}_{\mu \,\nu \,\sigma\,\lambda}^{V_{1} \, V_{2} \, V_{3}\, V_{4}}$. 
In the case of the squarks we have presented the exact results for 
the ${\Delta {\Gamma}_{\mu \,\nu \,\sigma\,\lambda
\hspace*{0.4cm}\tilde{q}}^{A A W^+ W^-}}$, 
${\Delta {\Gamma}_{\mu \,\nu \,\sigma\,\lambda
\hspace*{0.4cm}\tilde{q}}^{Z Z  W^+ W^-}}$,
${\Delta {\Gamma}_{\mu \,\nu \,\sigma\,\lambda
\hspace*{0.4cm}\tilde{q}}^{A Z W^+ W^-}}$ and
${\Delta {\Gamma}_{\mu \,\nu \,\sigma\,\lambda
\hspace*{0.4cm}\tilde{q}}^{W^+ W^- W^+ W^-}}$ in Appendix C.
We have checked explicitely in addition that the other 4-point functions
not shown in this appendix are finite as it corresponds to the Green's
functions that do not have tree level contributions.

In order to get the asymptotic expressions for the effective action given 
in~(\ref{eq:eff4}), one proceeds as in the previous sections. Notice that
for the  sfermions
contributions to the four point part of the effective action it is not possible to 
write, directly, an 
expression equivalent to the one given in~(\ref{eq:efflim3}). In the
first step after substituting just the asymptotic
results of the integrals in eq.(\ref{eq:eff4}), one does not obtain yet a
 result proportional to the tree level
vertex for the effective action and, one could think that it may be
some non-decoupling effect in the Appelquist-Carazzone sense.
However, this is not the case, and in order to conclude anything about the
 decoupling of sfermions in the 4-point
functions one needs to go a step further and to compute the different
contributions to the four-point Green's functions, which 
involve at the same time to perform the sums in the corresponding matrix
indices and over the various types
of sfermions. Finally, after performing these sums one gets the results for the sfermions contributions to the four-point
functions that indeed show decoupling since they turn out to be proportional
to the corresponding tree level contribution.

Analogously to the previous section, we write our results as: 
\begin{equation}
\label{eq:not4}
\displaystyle \Gamma_{\mu \,\nu \,\sigma\,\lambda}^{V_{1} \, V_{2} \, V_{3}\, V_{4}}=
{\Gamma_{0}}_{\mu \,\nu \,\sigma\,\lambda}^{V_{1} \, V_{2} \, V_{3}\, V_{4}}+
\Delta {\Gamma}_{\mu \,\nu \,\sigma\,\lambda}^{V_{1} \, V_{2} \, V_{3}\, V_{4}}\,,
\end{equation}
where the outgoing momenta assignements are 
$V_{1}^{\mu}(p)$, $V_{2}^{\nu}(k)$, $V_{3}^{\sigma}(r)$ and
$V_{4}^{\lambda}(t)$ and the different contributions to the effective action 
at tree level are defined by,
\begin{eqnarray}
\displaystyle \Gamma_{0\,\,\mu \,\nu \,\sigma\,\lambda}^{AAW^+ W^-}=-e^{2}
{\ss}_{\mu \nu \sigma \lambda}&,&
\Gamma_{0\,\,\mu \,\nu \,\sigma\,\lambda}^{AZW^+W^-}=
-g^{2}\sw\cw{\ss}_{\mu \nu \sigma \lambda}\,, \nonumber\\
\nonumber\\
\displaystyle \Gamma_{0\,\,\mu \,\nu \,\sigma\,\lambda}^{ZZW^+ W^-}=
-g^{2}\cw^{2}{\ss}_{\mu \nu \sigma \lambda}&,&
\Gamma_{0\,\,\mu \,\nu \,\sigma\,\lambda}^{W^+ W^-W^+W^-}=
g^{2}{\ss}_{\mu \sigma \nu \lambda}\,\,,
\end{eqnarray}
with,
\begin{equation}
\label{eq:optree4}
{\ss}_{\mu\,\nu\,\sigma\,\lambda}\equiv\left[ 2g_{\mu \,\nu}g_{\sigma \,\lambda}-
g_{\mu \,\sigma}g_{\nu \,\lambda}-g_{\mu \,\lambda}g_{\nu \,\sigma}\right]\,.
\end{equation}

The results for the squark contributions to the four-point functions that are
different from zero are the following,
\begin{eqnarray}
\label{eq:AAWW}
\displaystyle {\Delta {\Gamma}_{\mu \,\nu \,\sigma\,\lambda\hspace*{0.4cm}\tilde{q}}
^{A A W^+ W^-}}&=&
-\frac{N_{c}}{6} \frac{e^{2}g^{2}}{16 \pi^{2}}\,\sum_{\tilde{q}}\left\{ 
\,{\ss}_{\mu \nu \sigma \lambda} \Delta_{\epsilon}+
g_{\mu \,\nu}\,g_{\sigma \,\lambda}\,
g_{1}(\tilde{m}_{t_{1}}^{2}, \tilde{m}_{t_{2}}^{2}, \tilde{m}_{b_{1}}^{2},
\tilde{m}_{b_{2}}^{2})\right.\nonumber\\
&+& \left.(g_{\mu \,\sigma}\,g_{\nu \,\lambda}+g_{\mu \,\lambda}g_{\nu \,\sigma})
\,g_{2}(\tilde{m}_{t_{1}}^{2}, \tilde{m}_{t_{2}}^{2}, \tilde{m}_{b_{1}}^{2},
\tilde{m}_{b_{2}}^{2})\right\} \nonumber\\
&=&-\frac{N_{c}}{6} \frac{e^{2}g^{2}}{16 \pi^{2}} \,{\ss}_{\mu \nu \sigma \lambda} 
\,\sum_{\tilde{q}}\left\{\Delta_{\epsilon}-\log \frac{\hat{M}^{2}}{\mu_{o}^{2}}\right\} 
+{G_{1}}_{\mu \,\nu \,\sigma\,\lambda}\left[
O\left(\frac{p^{2}}{\Sigma \tilde{m}^{2}},\,\frac{\bigtriangleup \tilde{m}^{2}}
{\Sigma\tilde{m}^{2}}\right)\right]\,\,,\\
\nonumber\\
\label{eq:AZWW}
\displaystyle {\Delta {\Gamma}_{\mu \,\nu \,\sigma\,\lambda\hspace*{0.4cm}\tilde{q}}
^{A Z W^+ W^-}}&=&-\frac{N_{c}}{6} 
\frac{eg^{3}}{16 \pi^{2}} \,\sum_{\tilde{q}}\left\{ 
\,c_{\scriptscriptstyle W}\,{\ss}_{\mu \nu \sigma \lambda} \Delta_{\epsilon}+
\frac{1}{c_{\scriptscriptstyle W}}\, g_{\mu \,\nu}\,g_{\sigma \,\lambda}\,
g_{3}(\tilde{m}_{t_{1}}^{2}, \tilde{m}_{t_{2}}^{2}, \tilde{m}_{b_{1}}^{2},
\tilde{m}_{b_{2}}^{2})\right.\nonumber\\
&+& \left.\frac{1}{c_{\scriptscriptstyle W}}\, (g_{\mu \,\sigma}\,g_{\nu \,\lambda}+g_{\mu \,\lambda}g_{\nu \,\sigma})
\,g_{4}(\tilde{m}_{t_{1}}^{2}, \tilde{m}_{t_{2}}^{2}, \tilde{m}_{b_{1}}^{2},
\tilde{m}_{b_{2}}^{2})\right\} \nonumber\\
&=&-\frac{N_{c}}{6} 
\frac{eg^{3}}{16 \pi^{2}}c_{\scriptscriptstyle W} \,{\ss}_{\mu \nu \sigma \lambda} 
\sum_{\tilde{q}}\left\{\Delta_{\epsilon}-\log \frac{\hat{M}^{2}}{\mu_{o}^{2}}\right\}
+{G_{2}}_{\mu \,\nu \,\sigma\,\lambda}\left[
O\left(\frac{p^{2}}{\Sigma \tilde{m}^{2}},\,\frac{\bigtriangleup \tilde{m}^{2}}
{\Sigma\tilde{m}^{2}}\right)\right]\,\,,\nonumber\\
\\
\label{eq:ZZWW}
\displaystyle {\Delta {\Gamma}_{\mu \,\nu \,\sigma\,\lambda\hspace*{0.4cm}\tilde{q}}
^{Z Z W^+ W^-}}&=&-\frac{N_{c}}{6} \frac{g^{4}}{16 \pi^{2}}
\,\sum_{\tilde{q}}\left\{c_{\scriptscriptstyle W}^{2}
 \,{\ss}_{\mu \nu \sigma \lambda} \Delta_{\epsilon}+
\frac{1}{c_{\scriptscriptstyle W}^{2}}\,g_{\mu \,\nu}\,g_{\sigma \,\lambda}\,
g_{5}(\tilde{m}_{t_{1}}^{2}, \tilde{m}_{t_{2}}^{2}, \tilde{m}_{b_{1}}^{2},
\tilde{m}_{b_{2}}^{2})\right.\nonumber\\
&+& \left.\frac{1}{c_{\scriptscriptstyle W}^{2}}\, (g_{\mu \,\sigma}\,g_{\nu \,\lambda}+g_{\mu \,\lambda}g_{\nu \,\sigma})
\,g_{6}(\tilde{m}_{t_{1}}^{2}, \tilde{m}_{t_{2}}^{2}, \tilde{m}_{b_{1}}^{2},
\tilde{m}_{b_{2}}^{2})\right\} \nonumber\\
&=&-\frac{N_{c}}{6} \frac{g^{4}}{16 \pi^{2}}c_{\scriptscriptstyle W}^{2} 
\,{\ss}_{\mu \nu \sigma \lambda}\sum_{\tilde{q}}
\left\{\Delta_{\epsilon}-\log \frac{\hat{M}^{2}}{\mu_{o}^{2}}\right\}
+{G_{3}}_{\mu \,\nu \,\sigma\,\lambda}\left[
O\left(\frac{p^{2}}{\Sigma \tilde{m}^{2}},\,\frac{\bigtriangleup \tilde{m}^{2}}
{\Sigma\tilde{m}^{2}}\right)\right]\,\,,\nonumber\\
\\
\label{eq:WWWW}
\displaystyle {\Delta {\Gamma}_{\mu \,\nu \,\sigma\,\lambda\hspace*{0.9cm}\tilde{q}}
^{W^+ W^- W^+ W^-}}&=&-\frac{N_{c}}{3}\frac{g^{4}}{16 \pi^{2}}
\,\sum_{\tilde{q}}\left\{ \,{\ss}_{\mu \sigma\nu  \lambda}\, \Delta_{\epsilon}+
g_{\mu \,\nu}\,g_{\sigma \,\lambda}\,
g_{7}(\tilde{m}_{t_{1}}^{2}, \tilde{m}_{t_{2}}^{2}, \tilde{m}_{b_{1}}^{2},
\tilde{m}_{b_{2}}^{2})\right.\nonumber\\
&+& \left. (g_{\mu \,\sigma}\,g_{\nu \,\lambda}+g_{\mu \,\lambda}g_{\nu \,\sigma})
\,g_{8}(\tilde{m}_{t_{1}}^{2}, \tilde{m}_{t_{2}}^{2}, \tilde{m}_{b_{1}}^{2},
\tilde{m}_{b_{2}}^{2})\right\} \nonumber\\
&=&-\frac{N_{c}}{3}\frac{g^{4}}{16 \pi^{2}}\,{\ss}_{\mu \sigma \nu \lambda} 
\sum_{\tilde{q}}\left\{\Delta_{\epsilon}-\log \frac{\hat{M}^{2}}{\mu_{o}^{2}}\right\}
+{G_{4}}_{\mu \,\nu \,\sigma\,\lambda}\left[
O\left(\frac{p^{2}}{\Sigma \tilde{m}^{2}},\,\frac{\bigtriangleup \tilde{m}^{2}}
{\Sigma\tilde{m}^{2}}\right)\right]\,.
\end{eqnarray}
\hspace*{0.5cm}
The functions ${G_{k}}_{\mu \,\nu \,\sigma\,\lambda}$
and $g_{k}(\tilde{m}_{t_{1}}^{2}, \tilde{m}_{t_{2}}^{2}, \tilde{m}_{b_{1}}^{2},
\tilde{m}_{b_{2}}^{2}) \,(k=1,\ldots4)$ are both finite, but the first ones vanish in our asymptotic
limit whereas the second ones are different from zero in this limit. Therefore the
latter contain all the potentially non-decoupling effects of the four-point functions. 
The explicit formulae of the $g_{k}$ functions $(k=1,\ldots8)$ are collected in 
Appendix B. As a check of the previous functional
computation we have also calculated all these four point functions by 
diagramatic methods and we have got the same results.

Notice that, if one takes the sum of the corresponding SUSY squared masses
involved as the large parameter in the asymptotic expansion,
$\Sigma\tilde{m}^{2}$, one gets that the dominant contributions to these
$g_{k}$ functions are logarithmic. Generically, we can write,
\begin{equation}
g_{k}(\tilde{m}_{t_{1}}^{2}, \tilde{m}_{t_{2}}^{2}, \tilde{m}_{b_{1}}^{2},
\tilde{m}_{b_{2}}^{2}) \,=O\left(\log \frac{\Sigma \tilde{m}^{2}}{\mu_{o}^{2}}\right)+
O\left(\frac{\bigtriangleup \tilde{m}^{2}}{\Sigma \tilde{m}^{2}}\right)\,,
\end{equation}
where $\bigtriangleup \tilde{m}^{2}$ denotes the various squared mass differences and,
as in the previous cases, all the contributions of the type
$O\left(\frac{\bigtriangleup \tilde{m}^{2}}{\Sigma \tilde{m}^{2}}\right)$
vanish in our asymptotic limit. Notice also that in the previous expressions of the
four point functions that are given in terms of these $g_{k}$ functions, the
decoupling is not manifest yet since the Lorentz tensorial structure is apparently
not proportional to the tree level one. However, after rewritting these results in
terms of the proper variable which in this case is given by,
$$\hat{M}^{2} \equiv \frac{1}{4}(\tilde{m}_{t_{1}}^{2}+\tilde{m}_{t_{2}}^{2}+
\tilde{m}_{b_{1}}^{2}+\tilde{m}_{b_{2}}^{2})$$
one finds out that the one loop corrections to the four point functions,
$\Delta {\Gamma }$, in the asymptotic limit of large $\hat{M}^{2}$ are indeed
proportional to the tree level contribution. This can be seen in the last lines of
eqs.~(\ref{eq:AAWW}) to (\ref{eq:WWWW}) respectively. Therefore the potentially 
non-decoupling effects in the four-point functions can also be absorbed into redefinitions
of the coupling constants and wave functions. 
 
Similar expressions are obtained for the sleptons sector doing the
corresponding replacements mentioned at the end of subsection 4.1.

In summary, the results in this subsection show explicitely the
decoupling of squarks and sleptons in the four-point functions.

\subsection{{\it Inos} contributions}

Here we consider the effective action for 4-point functions generated 
from the integration of charginos and neutralinos, which results after
computing the corresponding last functional
traces given in eq. (\ref{eq:effinos}). By inserting the operators and
propagators of eqs. (\ref{eq:opersgdef}), (\ref{eq:propcK}) and (\ref{eq:propnK})
into equation (\ref{eq:effinos}) and after a lengthy calculation, that we
do not present here for brevity, the 
{\it inos} contributions to the 4-point part of the effective action can be
summarized as follows,
\begin{eqnarray}
\label{eq:eff4inos}
&& \displaystyle  {\Gamma_{eff}^{\sg} [V]}_{[4]} = i \int
{\rm d}\tilde{p} \,{\rm d}\tilde{k} \, {\rm d}\tilde{r} \,{\rm d}\tilde{t}\,\,
(2\pi)^{4} \,\delta (p+k+r+t)\,\times\nonumber\\
&& \int d\widehat{q}\, \left[\frac{1}{4} \sum_{i,j,k,l=1}^{2} {\cal G}^{ijkl}(\tilde{M}_{i}^{+}, 
\tilde{M}_{j}^{+},\tilde{M}_{k}^{+},\tilde{M}_{l}^{+})\right.
\left\{ \,q_{\,1}^{\,\alpha}\,q_{\,2}^{\,\beta}\, 
q_{\,3}^{\,\gamma}\,q_{\,4}^{\,\rho}
{\left( \,G\cdot O \,\right)}^{\scriptstyle ++++}_{\scriptstyle \,1\,2\,3\,4}
\right.\nonumber\\
&& \displaystyle + q_{\,1}^{\,\alpha}\,q_{\,2}^{\,\beta}\,\tilde{M}_{k}^{+} 
\tilde{M}_{l}^{+}
{\left(\,G\cdot O \,\right)}^{\scriptstyle ++++}_{\scriptstyle \,1\,2}
+ q_{\,1}^{\,\alpha}\,q_{\,3}^{\,\gamma}\,\tilde{M}_{j}^{+} 
\tilde{M}_{l}^{+} 
{\left(\,G\cdot O \,\right)}^{\scriptstyle ++++}_{\scriptstyle \,1\,3}
+ q_{\,1}^{\,\alpha}\,q_{\,4}^{\,\rho}\,\tilde{M}_{j}^{+} \tilde{M}_{k}^{+}
{\left(\,G\cdot O \,\right)}^{\scriptstyle ++++}_{\scriptstyle \,1\,4}\nonumber\\
&& \displaystyle \left.+ q_{\,2}^{\,\beta}\,q_{\,3}^{\,\gamma}\,\tilde{M}_{l}^{+} 
\tilde{M}_{i}^{+}
{\left(\,G\cdot O \,\right)}^{\scriptstyle ++++}_{\scriptstyle \,2\,3}
+ q_{\,2}^{\,\beta}\,q_{\,4}^{\,\rho}\,\tilde{M}_{k}^{+} \tilde{M}_{i}^{+}
{\left(\,G\cdot O \,\right)}^{\scriptstyle ++++}_{\scriptstyle \,2\,4}
+ q_{\,3}^{\,\gamma}\,q_{\,4}^{\,\rho}\,\tilde{M}_{i}^{+} \tilde{M}_{j}^{+}
{\left(\,G\cdot O \,\right)}^{\scriptstyle ++++}_{\scriptstyle \,3\,4}\right.\nonumber\\
&& \displaystyle \left.+\tilde{M}_{i}^{+} \tilde{M}_{j}^{+} \tilde{M}_{k}^{+}
\tilde{M}_{l}^{+}\,{\left(\,G\cdot O \,\right)}^{\scriptstyle ++++}
\right\}\nonumber\\
&& +\sum_{i=1}^{4} \sum_{j,k,l=1}^{2} {\cal G}^{ijkl}(\tilde{M}_{i}^{o}, \tilde{M}_{j}^{+},
\tilde{M}_{k}^{+}, \tilde{M}_{l}^{+})\,
\left\{ q_{\,1}^{\,\alpha}\,q_{\,2}^{\,\beta}\, q_{\,3}^{\,\gamma}\,q_{\,4}^{\,\rho}\,
{\left( \,G\cdot O \,\right)}^{\scriptstyle \,o+++}_{\scriptstyle \,1\,2\,3\,4}
\right.\nonumber\\
&& +q_{\,1}^{\,\alpha}\,q_{\,2}^{\,\beta}\,\tilde{M}_{k}^{+} \tilde{M}_{l}^{+}
{\left(\,G\cdot O \,\right)}^{\scriptstyle o+++}_{\scriptstyle \,1\,2}
+q_{\,1}^{\,\alpha}\,q_{\,3}^{\,\gamma}\,\tilde{M}_{j}^{+} \tilde{M}_{l}^{+}
{\left(\,G\cdot O \,\right)}^{\scriptstyle o+++}_{\scriptstyle \,1\,3}
+q_{\,1}^{\,\alpha}\,q_{\,4}^{\,\rho}\,\tilde{M}_{j}^{+} \tilde{M}_{k}^{+}
{\left(\,G\cdot O \,\right)}^{\scriptstyle o+++}_{\scriptstyle \,1\,4}\nonumber\\
&& \left. +q_{\,2}^{\,\beta}\,q_{\,3}^{\,\gamma}\,\tilde{M}_{i}^{o} \tilde{M}_{l}^{+}
{\left(\,G\cdot O \,\right)}^{\scriptstyle o+++}_{\scriptstyle \,2\,3}
 +q_{\,2}^{\,\beta}\,q_{\,4}^{\,\rho}\,\tilde{M}_{i}^{o} \tilde{M}_{k}^{+}
{\left(\,G\cdot O \,\right)}^{\scriptstyle o+++}_{\scriptstyle \,2\,4}
+q_{\,3}^{\,\gamma}\,q_{\,4}^{\,\rho}\,\tilde{M}_{i}^{o} \tilde{M}_{j}^{+}
{\left(\,G\cdot O \,\right)}^{\scriptstyle o+++}_{\scriptstyle
  \,3\,4}\right.\nonumber\\
&& \displaystyle \left.+\tilde{M}_{i}^{o} \tilde{M}_{j}^{+} \tilde{M}_{k}^{+}
\tilde{M}_{l}^{+}\,{\left(\,G\cdot O \,\right)}^{\scriptstyle o+++}\right\}\nonumber\\  
&& +\sum_{i,j=1}^{4} \sum_{k,l=1}^{2} {\cal G}^{ijkl}(\tilde{M}_{i}^{o}, \tilde{M}_{j}^{o},
\tilde{M}_{k}^{+}, \tilde{M}_{l}^{+})\,
\left\{ q_{\,1}^{\,\alpha}\,q_{\,2}^{\,\beta}\, q_{\,3}^{\,\gamma}\,q_{\,4}^{\,\rho}\,
{\left( \,G\cdot O \,\right)}^{\scriptstyle \,oo++}_{\scriptstyle \,1\,2\,3\,4}
\right.\nonumber\\
&& +q_{\,1}^{\,\alpha}\,q_{\,2}^{\,\beta}\,\tilde{M}_{k}^{+} \tilde{M}_{l}^{+}
{\left(\,G\cdot O \,\right)}^{\scriptstyle oo++}_{\scriptstyle \,1\,2}
+q_{\,1}^{\,\alpha}\,q_{\,3}^{\,\gamma}\,\tilde{M}_{j}^{o} \tilde{M}_{l}^{+}
{\left(\,G\cdot O \,\right)}^{\scriptstyle oo++}_{\scriptstyle \,1\,3}
+q_{\,1}^{\,\alpha}\,q_{\,4}^{\,\rho}\,\tilde{M}_{j}^{o} \tilde{M}_{k}^{+}
{\left(\,G\cdot O \,\right)}^{\scriptstyle oo++}_{\scriptstyle \,1\,4}\nonumber\\
&& \left. +q_{\,2}^{\,\beta}\,q_{\,3}^{\,\gamma}\,\tilde{M}_{i}^{o} \tilde{M}_{l}^{+}
{\left(\,G\cdot O \,\right)}^{\scriptstyle oo++}_{\scriptstyle \,2\,3}
 +q_{\,2}^{\,\beta}\,q_{\,4}^{\,\rho}\,\tilde{M}_{i}^{o} \tilde{M}_{k}^{+}
{\left(\,G\cdot O \,\right)}^{\scriptstyle oo++}_{\scriptstyle \,2\,4}
+q_{\,3}^{\,\gamma}\,q_{\,4}^{\,\rho}\,\tilde{M}_{i}^{o} \tilde{M}_{j}^{o}
{\left(\,G\cdot O \,\right)}^{\scriptstyle oo++}_{\scriptstyle \,3\,4}\right.\nonumber\\
&& \displaystyle \left.+\tilde{M}_{i}^{o} \tilde{M}_{j}^{o} \tilde{M}_{k}^{+}
\tilde{M}_{l}^{+}\,{\left(\,G\cdot O \,\right)}^{\scriptstyle oo++}
\right\}\nonumber\\
&& +\sum_{i,j,k=1}^{4} \sum_{l=1}^{2} {\cal G}^{ijkl}(\tilde{M}_{i}^{o}, \tilde{M}_{j}^{o},
\tilde{M}_{k}^{o}, \tilde{M}_{l}^{+})\,
\left\{ q_{\,1}^{\,\alpha}\,q_{\,2}^{\,\beta}\, q_{\,3}^{\,\gamma}\,q_{\,4}^{\,\rho}\,
{\left( \,G\cdot O \,\right)}^{\scriptstyle \,ooo+}_{\scriptstyle \,1\,2\,3\,4}
\right.\nonumber\\
&& +q_{\,1}^{\,\alpha}\,q_{\,2}^{\,\beta}\,\tilde{M}_{k}^{+} \tilde{M}_{l}^{+}
{\left(\,G\cdot O \,\right)}^{\scriptstyle ooo+}_{\scriptstyle \,1\,2}
+q_{\,1}^{\,\alpha}\,q_{\,3}^{\,\gamma}\,\tilde{M}_{j}^{o} \tilde{M}_{l}^{+}
{\left(\,G\cdot O \,\right)}^{\scriptstyle ooo+}_{\scriptstyle \,1\,3}
+q_{\,1}^{\,\alpha}\,q_{\,4}^{\,\rho}\,\tilde{M}_{j}^{o} \tilde{M}_{k}^{o}
{\left(\,G\cdot O \,\right)}^{\scriptstyle ooo+}_{\scriptstyle \,1\,4}\nonumber\\
&& +q_{\,2}^{\,\beta}\,q_{\,3}^{\,\gamma}\,\tilde{M}_{i}^{o} \tilde{M}_{l}^{+}
{\left(\,G\cdot O \,\right)}^{\scriptstyle ooo+}_{\scriptstyle \,2\,3}
 +q_{\,2}^{\,\beta}\,q_{\,4}^{\,\rho}\,\tilde{M}_{i}^{o} \tilde{M}_{k}^{o}
{\left(\,G\cdot O \,\right)}^{\scriptstyle ooo+}_{\scriptstyle \,2\,4}
+q_{\,3}^{\,\gamma}\,q_{\,4}^{\,\rho}\,\tilde{M}_{i}^{o} \tilde{M}_{j}^{o}
{\left(\,G\cdot O \,\right)}^{\scriptstyle ooo+}_{\scriptstyle \,3\,4}\nonumber\\
&&\displaystyle \left.+\tilde{M}_{i}^{o} \tilde{M}_{j}^{o} \tilde{M}_{k}^{o}
\tilde{M}_{l}^{+}\,{\left(\,G\cdot O \,\right)}^{\scriptstyle ooo+}
\right\}\nonumber\\
&& +\frac{1}{8} \sum_{i,j,k,l=1}^{4} {\cal G}^{ijkl}(\tilde{M}_{i}^{o}, \tilde{M}_{j}^{o},
\tilde{M}_{k}^{o}, \tilde{M}_{l}^{o})\,
\left\{ q_{\,1}^{\,\alpha}\,q_{\,2}^{\,\beta}\, q_{\,3}^{\,\gamma}\,q_{\,4}^{\,\rho}\,
{\left( \,G\cdot O \,\right)}^{\scriptstyle \,oooo}_{\scriptstyle \,1\,2\,3\,4}
\right.\nonumber\\
&& +q_{\,1}^{\,\alpha}\,q_{\,2}^{\,\beta}\,\tilde{M}_{k}^{o} \tilde{M}_{l}^{o}
{\left(\,G\cdot O \,\right)}^{\scriptstyle oooo}_{\scriptstyle \,1\,2}
+q_{\,1}^{\,\alpha}\,q_{\,3}^{\,\gamma}\,\tilde{M}_{j}^{o} \tilde{M}_{l}^{o}
{\left(\,G\cdot O \,\right)}^{\scriptstyle oooo}_{\scriptstyle \,1\,3}
+q_{\,1}^{\,\alpha}\,q_{\,4}^{\,\rho}\,\tilde{M}_{j}^{o} \tilde{M}_{k}^{o}
{\left(\,G\cdot O \,\right)}^{\scriptstyle oooo}_{\scriptstyle \,1\,4}\nonumber\\
&& \left.\left. +q_{\,2}^{\,\beta}\,q_{\,3}^{\,\gamma}\,\tilde{M}_{i}^{o} 
\tilde{M}_{l}^{o}
{\left(\,G\cdot O \,\right)}^{\scriptstyle oooo}_{\scriptstyle \,2\,3}
 +q_{\,2}^{\,\beta}\,q_{\,4}^{\,\rho}\,\tilde{M}_{i}^{o} \tilde{M}_{k}^{o}
{\left(\,G\cdot O \,\right)}^{\scriptstyle oooo}_{\scriptstyle \,2\,4}
+q_{\,3}^{\,\gamma}\,q_{\,4}^{\,\rho}\,\tilde{M}_{i}^{o} \tilde{M}_{j}^{o}
{\left(\,G\cdot O \,\right)}^{\scriptstyle oooo}_{\scriptstyle
\,3\,4}\right.\right.\nonumber\\
&& \displaystyle \left.\left.+\tilde{M}_{i}^{o} \tilde{M}_{j}^{o} 
\tilde{M}_{k}^{o}
\tilde{M}_{l}^{o}\,{\left(\,G\cdot O \,\right)}^{\scriptstyle oooo}\,
\right\}{\phantom {\frac{!}{!}}}\right]\,.
\end{eqnarray}
Analogously to eq.~(\ref{eq:eff3inos}) we have
used here the shorthand notation $\left(\,G\cdot O\,\right)$ for the various
products of traces and operators whose explicit expressions are collected
 in Appendix B. Notice that there are some terms without subscripts which means 
there is no momentum contracted with the results of the traces. For example, in
${\left(\,G\cdot O \,\right)}^{\scriptstyle ++++}$, the superscripts denote the four 
 charginos in the loop and the absence of subscripts indicates there is no 
 contraction with any momenta. The definitions of the 
$q_{1}, q_{2}, q_{3}$ and $q_{4}$ momenta, as well as the 
generic function ${\cal G}^{ijkl}(\tilde{M}_{i},
\tilde{M}_{j},\tilde{M}_{k},\tilde{M}_{l})$ are given by,
\begin{equation}
\label{eq:q4}
q_{\,1}\,\,\equiv q\,\,\,, \,q_{\,2}\,\equiv q+p\,\,\,,
\,q_{\,3}\equiv q+p+k\,\,\,,q_{4}\equiv q+p+k+r\,,
\end{equation}
and
$${\cal G}^{ijkl}(\tilde{M}_{i}, \tilde{M}_{j},\tilde{M}_{k},\tilde{M}_{l})= 
\frac{1}{\left[q_{1}^2 - \tilde{M}_{i}^{2} \right]\,
\left[q_{2}^2 - \tilde{M}_{j}^{2} \right]\,
\left[q_{3}^2 - \tilde{M}_{k}^{2} \right]\,
\left[q_{4}^2 - \tilde{M}_{l}^{2}\right]}\,.$$
 
In order to obtain the exact contributions to one loop level from the
{\it inos} sector, one must work out the corresponding Dirac 
traces in  (\ref{eq:eff4inos}) and then write down the results in terms
of the standard one-loop Feynman integrals. We have performed such computation
but due to the length of the final expressions we prefer not to 
present these exact results here and to restrict ourselves to the
presentation and discussion of just the corresponding asymptotic results.

By starting with the exact result given in formula (\ref{eq:eff4inos}) and
by inserting the 
asymptotic results of the corresponding integrals and coupling matrices
we have derived the 4-point Green's
functions from {\it inos} sector in the large masses limit. The results of these integrals have been given in 
Appendix A. After a lengthy calculation 
we can summarize the result for the four point part of the effective
action in the asymptotic limit by the following expression:
\begin{eqnarray}
\label{eq:4inoslimit}
\displaystyle &&{\Gamma_{eff}^{\sg} [V]}_{[4]} = \frac{4}{3}\,{\pi}^{2}\, \int
{\rm d}\tilde{p} \,\, {\rm d}\tilde{k} \,\, {\rm d}\tilde{r}\,\,
{\rm d}\tilde{t}\,\,\delta(p+k+r+t)\times \nonumber\\
&&\,\sum_{i,j,k,l}\left\{\frac{1}{4}\left(\check{O}^{1}+\check{O}^{2}+
\check{O}^{4}+\check{O}^{6}+\check{O}^{8}+\check{O}^{10}+\check{O}^{12}+
\check{O}^{14}+\check{O}^{18}+
\check{O}^{22}+\check{O}^{26}+\check{O}^{30}
\right)^{\mu \,\nu\,\sigma\,\lambda}_{\,i\,j\,k\,l}\right.\nonumber\\
&&+\left.\frac{1}{8}\,\check{O}^{38\, \mu \,\nu\,\sigma\,\lambda}_
{\hspace*{0.4cm}i\,j\,k\,l}+\left(
\check{O}^{46}+\check{O}^{50}+\check{O}^{54}+\check{O}^{58}+\check{O}^{60}
+\check{O}^{68}+\check{O}^{76}+\check{O}^{84}\right)
^{\mu \,\nu\,\sigma\,\lambda}_{\,i\,j\,k\,l}\,\right\}\times\nonumber\\
&&\,\,\left({\Delta}_\epsilon-\log \frac{\tilde{M}_{i}^{2}+\tilde{M}_{j}^{2}+
\tilde{M}_{k}^{2}+\tilde{M}_{l}^{2}}
{4\mu_{o}^{2}}\right){\ss}_{\sigma\,\mu\,\nu\,\lambda}\,,
\end{eqnarray}
where ${\ss}_{\sigma\,\mu\,\nu\,\lambda}$ is the 
tree level tensor defined in eq.~(\ref{eq:optree4}) but with the Lorentz indices
being interchanged, and the operators $\check{O}^{\mu \,\nu\,\sigma\,\lambda}
_{\,i\,j\,k\,l}$ are given in Appendix B.

At this point, we can already see that the asymptotic result from the {\it inos}
 sector is proportional to the tree level tensor after the proper
 symmetrization over the identical external fields and therefore, we can 
 conclude that the {\it inos} decouple in the 4-point functions.

For completeness we present in the following the corresponding asymptotic results
for the 4-point functions with specific external gauge bosons. After a
lengthy computation we get,
\begin{eqnarray}
\label{eq:AAWWinos}
\displaystyle {\Delta {\Gamma}_{\mu \,\nu \,\sigma\,\lambda
\hspace*{0.4cm}{\sg}}^{A A W^+ W^-}}&=&
 \frac{e^{2}g^{2}}{12 \pi^{2}} \,{\ss}_{\mu \nu \sigma \lambda} 
\,\left\{-\frac{3}{2}\Delta_{\epsilon} +
g_{9}(\tilde{M}_{1}^{+},\tilde{M}_{2}^{+},
\tilde{M}_{1}^{0},\tilde{M}_{2}^{0},
\tilde{M}_{3}^{0},\tilde{M}_{4}^{0})\right\} \nonumber\\
&+& \displaystyle {G_{5}}_{\mu \,\nu \,\sigma\,\lambda}\left[
O\left(\frac{p^{2}}{\Sigma \tilde{m}^{2}},\,\frac{\bigtriangleup \tilde{m}^{2}}
{\Sigma\tilde{m}^{2}}\right)\right]\,\,,\\
\nonumber\\
\label{eq:AZWWinos}
\displaystyle {\Delta {\Gamma}_{\mu \,\nu \,\sigma\,\lambda
\hspace*{0.4cm}{\sg}}^{A Z W^+ W^-}}&=&
\frac{eg^{3}}{12 \pi^{2}}\frac{1}{c_{\scriptscriptstyle W}} \,{\ss}_{\mu \nu \sigma \lambda} 
\left\{-\frac{3}{2} \cw^{2}\Delta_{\epsilon}
+g_{10}(\tilde{M}_{1}^{+},\tilde{M}_{2}^{+},
\tilde{M}_{1}^{0},\tilde{M}_{2}^{0},
\tilde{M}_{3}^{0},\tilde{M}_{4}^{0})\right\} \nonumber\\
&+& \displaystyle {G_{6}}_{\mu \,\nu \,\sigma\,\lambda}\left[
O\left(\frac{p^{2}}{\Sigma \tilde{m}^{2}},\,\frac{\bigtriangleup \tilde{m}^{2}}
{\Sigma\tilde{m}^{2}}\right)\right]\,\,,\\
\nonumber\\
\label{eq:ZZWWinos}
\displaystyle {\Delta {\Gamma}_{\mu \,\nu \,\sigma\,\lambda
\hspace*{0.4cm}{\sg}}^{Z Z W^+ W^-}}&=&-
\frac{g^{4}}{48 \pi^{2}} \frac{1}{c_{\scriptscriptstyle W}^{2}} 
\,{\ss}_{\mu \nu \sigma \lambda}\, \left\{6\,c_{\scriptscriptstyle W}^{4}
\Delta_{\epsilon}+g_{11}(\tilde{M}_{1}^{+},\tilde{M}_{2}^{+},
\tilde{M}_{1}^{0},\tilde{M}_{2}^{0},
\tilde{M}_{3}^{0},\tilde{M}_{4}^{0})\right\}\nonumber\\
&+&{G_{7}}_{\mu \,\nu \,\sigma\,\lambda}\left[
O\left(\frac{p^{2}}{\Sigma \tilde{m}^{2}},\,\frac{\bigtriangleup \tilde{m}^{2}}
{\Sigma\tilde{m}^{2}}\right)\right]\,\,,\\
\nonumber\\
\label{eq:WWWWinos}
\displaystyle {\Delta {\Gamma}_{\mu \,\nu \,\sigma\,\lambda
\hspace*{0.9cm}{\sg}}^{W^+ W^- W^+ W^-}}&=&
\frac{g^{4}}{96 \pi^{2}}\,{\ss}_{\mu \sigma\nu  \lambda} 
\left\{10\,\Delta_{\epsilon}
+g_{12}(\tilde{M}_{1}^{+},\tilde{M}_{2}^{+},
\tilde{M}_{1}^{0},\tilde{M}_{2}^{0},
\tilde{M}_{3}^{0},\tilde{M}_{4}^{0})\right\} \nonumber\\
&+& \displaystyle {G_{8}}_{\mu \,\nu \,\sigma\,\lambda}\left[
O\left(\frac{p^{2}}{\Sigma \tilde{m}^{2}},\,\frac{\bigtriangleup \tilde{m}^{2}}
{\Sigma\tilde{m}^{2}}\right)\right]\,,
\end{eqnarray}
where the functions ${G_{k}}_{\mu \,\nu \,\sigma\,\lambda}$
and $g_{k}(\tilde{m}_{t_{1}}^{2}, \tilde{m}_{t_{2}}^{2}, \tilde{m}_{b_{1}}^{2},
\tilde{m}_{b_{2}}^{2}) \,(k=5,\ldots8)$ are both finite, but the first ones vanish in our asymptotic
limit whereas the second ones are different from zero in this limit. The explicit form of the
latter can be found in Appendix B. In principle, their contain all the potentially non-decoupling 
effects of these four-point functions. As a check of the previous functional
computation we have also calculated all these four point functions by 
diagramatic methods and we have got the same results. 

As in the previous n-point Green's functions the one-loop corrections in
eqs. (\ref{eq:AAWWinos}) to (\ref{eq:WWWWinos}) are 
also proportional to the tree level vertex and at the end, we can conclude that those 
potentially non-decoupling effects in the four-point functions can be reabsorbed into redefinitions
of the various SM parameters. Therefore, we can guarantee that the
decoupling of the {\it inos} in the
4-point Green's functions take place as well.

In addition, we have checked that after the proper symmetrization over the
indices and momenta of the identical external fields, the $\Delta{\Gamma}^{AAAA},
 \Delta{\Gamma}^{AAAZ},
\Delta{\Gamma}^{AAZZ}, \Delta{\Gamma}^{AZZZ}$ and $\Delta{\Gamma}^{ZZZZ}$
contributions are exactly zero
in our limit as it was expected since there are no corresponding tree level
vertices. This is a rather non trivial check of our computation. 

As can be seen from all the results in the present article and those
obtained and discussed 
in~\cite{GEISHA}, we have proved explicitely that the decoupling of sfermions, charginos and 
neutralinos in the two, three and four point functions with external gauge
bosons do indeed occur and this decoupling proceeds by assuming that all
the sparticle masses 
are large as compared to the electroweak scale but close to each other.

Once we have shown the decoupling of SUSY particles in the two, three
and four point functions we can ask about the decoupling in the n-point 
functions, with $n>4$. In this case two important observations
are in order. First, due to the renormalizability of the MSSM there are no
divergent contributions to the five or higher point functions since
those functions vanish at the tree level and we are working in renormalizable
gauges. Thus, those Green's functions are finite and so are the sum
of the Feynman integrals corresponding to each given Green's function. In this case
their asymptotic behaviour in the above defined 
region can trivially be obtained. Then it is inmediate to check
that the decoupling of the SUSY particles also takes place.  

\section{Conclusions}

In this work we have studied the decoupling properties of the SUSY particles
appearing in the MSSM. In particular we have shown that the SM can be
considered as the low energy effective theory of the MSSM in the limit
where the sparticle masses are large. Our proof of decoupling in the Green's
functions with external gauge bosons is quite general and does not
depend on the
particular form of the soft breaking terms since it is performed completly in terms
of the SUSY masses.  The decoupling is shown in the sense of the
Appelquist-Carazzone Theorem. By this we mean that in the appropriate asymptotic
region of large SUSY masses considered in this work, the effect of the SUSY
particles 
on the gauge boson Green's functions can be absorbed in the SM parameters
and gauge boson wave functions or else they 
correspond to new terms which are suppressed by negative powers of the SUSY
masses.

However, in addition to the SUSY particles, the MSSM has also other particles 
which are not present in the SM. These are the extra Higgs scalar doublets that
must be added to the MSSM in order to produce fermion masses and that through
their SUSY partners give rise to an anomaly free theory. In order to 
provide a complete proof that the SM is really the low energy effective
theory of the MSSM one should show that these extra scalars also decouple in the
above mentioned sense. There are some indications that this could be the
case. In addition, one must also study not just the Green's functions with
external gauge bosons but also with
all the possible SM particles in the external legs. Particularly interesting
in the context of decoupling could be the Green's functions with external
heavy fermions where due to the enhacement effect of the heavy fermion masses
the decoupling of the SUSY particles could either not occur or to proceed much
slowly. Work in progress in these directions is being done but the results will be presented
elsewhere.

\section*{Acknowledgements.}

This work has 
been partially supported by the Spanish Mi\-nisterio de Educaci{\'o}n y Cultura 
under projects CICYT AEN97-1678 and AEN93-0776, and the fellowship AP95 00503301.

\section*{Appendix A.}
\vspace{0.4cm}
\setcounter{equation}{0}
\renewcommand{\theequation}{A.\arabic{equation}}

In this appendix we give the definition of the one-loop integrals 
that have been used in the computation of the three and
four-point functions and their results in the large masses limit. The
one-loop integrals contributing to the two-point functions were presented
in our previous work \cite{GEISHA} to which we refer the reader for
completeness. As all these integrals can be written in terms of  
the standard scalar and tensor integrals~\cite{PAS}, we  
start by reviewing the definition of these standard 2-, 3- and 4-point integrals in the following.
From now on, we denote:
$$\int d\widehat{q} \equiv \int \frac{d^{{\scriptscriptstyle D}} q}
{(2 \pi)^{{\scriptscriptstyle D}}} {\mu}_{o}^{4 - {\scriptscriptstyle D}}.$$

$\bullet$  Standard integrals.
\begin{eqnarray}
\label{eq:stanint}
A_{0}(m_1) &\equiv& -i 16 \pi^{2} \int d\widehat{q} \,\frac{1}{D_1} \,,\nonumber\\
B_{0,\,\mu,\,\mu\nu}(p,m_1,m_2) &\equiv& -i 16 \pi^{2} \int d\widehat{q} \,
\frac{\{1,\,q_{\mu},\,q_{\mu}q_{\nu}\}}{D_1\,D_2} \,,\nonumber\\
C_{0,\,\mu,\,\mu\nu,\,\mu\nu\sigma}(p,k,m_1,m_2,m_3) &\equiv& -i 16 \pi^{2} 
\int d\widehat{q} \,
\frac{\{1,\,q_{\mu},\,q_{\mu}q_{\nu},\,q_{\mu}q_{\nu}q_{\sigma}\}}{D_1\,D_2\,D_3} \,,\nonumber\\
D_{0,\,\mu,\,\mu\nu,\,\mu\nu\sigma,\,\mu\nu\sigma\lambda}(p,k,r,m_1,m_2,m_3,m_4) &\equiv& 
-i 16 \pi^{2} \int d\widehat{q} \,
\frac{\{1,\,q_{\mu},\,q_{\mu}q_{\nu},\,q_{\mu}q_{\nu}q_{\sigma},\,
q_{\mu}q_{\nu}q_{\sigma}q_{\lambda}\}}{D_1\,D_2\,D_3\,D_4} \,,\nonumber\\
\end{eqnarray}
with the denominators given by,
\begin{eqnarray}
\label{eq:den}
D_1 &=& \left[q^2 - m_1^2 \right] \nonumber\\
D_2 &=& \left[(q+p)^2 - m_2^2 \right] \nonumber\\
D_3 &=& \left[(q+p+k)^2 - m_3^2 \right] \nonumber\\
D_4 &=& \left[(q+p+k+r)^2 - m_4^2 \right] 
\end{eqnarray}

$\bullet$ One-loop integrals.

The 3-point integrals appearing in eqs. (\ref{eq:eff3}), (\ref{eq:effAWW})
and (\ref{eq:effZWW}) are given in terms of the standard integrals by,
\begin{eqnarray}
\label{eq:int31}
T^{a\,b}_{\mu}(p,\tilde{m}_{f_a}, \tilde{m}_{f_b}) &=& 
2 B_{\mu}(p,\tilde{m}_{f_a}, \tilde{m}_{f_b})+ p_{\mu}B_{0}(p,\tilde{m}_{f_a}, 
\tilde{m}_{f_b})\,\,,\\
\nonumber\\
\label{eq:int31n}
T^{a\,b\, c}_{\mu \,\nu \,\sigma} 
(p,k,\tilde{m}_{f_a}, \tilde{m}_{f_b},\tilde{m}_{f_c})&=&   
\left\{ 8 C_{\mu \,\nu \,\sigma}+
4 [(p+k)_{\sigma} C_{\mu \,\nu}+ (2p+k)_{\nu} C_{\mu \,\sigma}+
p_{\mu} C_{\nu \,\sigma}] \right.\nonumber\\
&+& 2 [(p+k)_{\sigma}(2p+k)_{\nu} C_{\mu}+p_{\mu}(p+k)_{\sigma} C_{\nu}+
p_{\mu}(2p+k)_{\nu} C_{\sigma}]\nonumber\\
&+& \left. p_{\mu}(2p+k)_{\nu}(p+k)_{\sigma}\,C_{0}\right\} \,
(p,k,\tilde{m}_{f_a}, \tilde{m}_{f_b},\tilde{m}_{f_c})\,\,,
\end{eqnarray}
\begin{eqnarray}
\label{eq:int32}
{\cal T}_{\,\mu \,\nu \,\sigma}^{\,i\,j\,k}
(p,k,\tilde{m}_{i}, \tilde{m}_{j},\tilde{m}_{k})&=& \left\{
4 C_{\mu \,\nu \,\sigma}+ 2 (k+p)_{\sigma}\,C_{\mu \,\nu}+2 (k+2p)_{\nu}\,C_{\mu \,\sigma}+
2 p_{\mu} C_{\nu \,\sigma}\right.\nonumber\\
&+& (k_{\nu}p_{\mu}+k_{\mu}p_{\nu}+2p_{\mu}p_{\nu})\,C_{\sigma} +
(k_{\nu}p_{\sigma}+k_{\sigma}p_{\nu}+2p_{\nu}p_{\sigma})\,C_{\mu}\nonumber\\
&+& (k_{\sigma}p_{\nu}-k_{\mu}p_{\sigma})\,C_{\nu}\nonumber\\
&-&g_{\alpha \beta}\left[\,g_{\mu \nu}\,C_{\alpha \,\beta \,\sigma}
+g_{\sigma \nu}\,C_{\alpha \,\beta \,\mu}
+g_{\sigma \mu}\,C_{\alpha \,\beta \,\nu}
+2g_{\sigma \nu}C_{\beta \,\mu}(k+p)_{\alpha}\right.\nonumber\\
&+& C_{\alpha \,\beta}\, (g_{\sigma \mu}(k+2p)_{\nu}+g_{\mu \nu}k_{\sigma}-
 g_{\sigma \nu}k_{\mu})
 +2g_{\mu \nu}\,C_{\beta \,\sigma}\,p_{\alpha}\nonumber\\
&+& C_{\beta}\,(g_{\sigma \mu}\,p_{\alpha}(k+2p)_{\nu}+
g_{\sigma \mu}\,k_{\alpha}\,p_{\nu}\nonumber\\
&+&g_{\mu \nu}(k_{\sigma}p_{\alpha}-k_{\alpha}p_{\sigma})
+ g_{\sigma \nu}(k_{\alpha}p_{\mu}-k_{\mu}p_{\alpha}))\nonumber\\
&+&\left.\left. p_{\beta}(k+p)_{\alpha}(g_{\mu \nu}\,C_{\sigma}+g_{\sigma \nu}\,
C_{\mu}-g_{\sigma \mu}\,C_{\nu})
\right]\right\}\,(p,k,\tilde{m}_{i}, \tilde{m}_{j},\tilde{m}_{k}),\\
\nonumber\\
{\cal I}_{\,\mu \,\nu \,\sigma}^{\,i\,j\,k}
(p,k,\tilde{m}_{i}, \tilde{m}_{j},\tilde{m}_{k})&=&
\left\{\,g_{\mu \nu}\,C_{\sigma}+g_{\sigma \nu}\,C_{\mu}
-g_{\sigma \mu}\,C_{\nu}\,\right\}
(p,k,\tilde{m}_{i}, \tilde{m}_{j},\tilde{m}_{k})\,\,,\\
\nonumber\\
{\cal P}_{\,\mu \,\nu \,\sigma}^{\,i\,j\,k}
(p,k,\tilde{m}_{i}, \tilde{m}_{j},\tilde{m}_{k})&=&
\left\{g_{\sigma \nu}\,(p_{\mu}\,C_{0}+C_{\mu})+
g_{\sigma \mu}\,(p_{\nu}\,C_{0}+C_{\nu})\right.\nonumber\\
&-& \left.
g_{\mu \nu}\,(p_{\sigma}\,C_{0}+C_{\sigma})\right\}\,
(p,k,\tilde{m}_{i}, \tilde{m}_{j},\tilde{m}_{k})\,\,,\\
\nonumber\\
{\cal J}_{\,\mu \,\nu \,\sigma}^{\,i\,j\,k}
(p,k,\tilde{m}_{i}, \tilde{m}_{j},\tilde{m}_{k})&=&\left\{\,
g_{\mu \nu}\,(C_{\sigma}+(k+p)_{\sigma}\,C_{0})
-g_{\sigma \nu}\,(C_{\mu}+(k+p)_{\mu}\,C_{0})\right.\nonumber\\
&+& \left.
g_{\sigma \mu}\,(C_{\nu}+(k+p)_{\nu}\,C_{0})\right\}\,
(p,k,\tilde{m}_{i}, \tilde{m}_{j},\tilde{m}_{k})\,\,,
\end{eqnarray}
where the variables within the last parentheses correspond to the arguments of the corresponding 
integrals.

Now, we present the 4-point integrals appearing in the computation 
of the four-point functions. Let us begin for those
involved in the computation of sfermions contributions that is, in eqs.
(\ref{eq:eff4}):
\begin{eqnarray}
\label{eq:int41}
J^{a\,b}_{p+k}(p+k,\tilde{m}_{f_a}, \tilde{m}_{f_b}) &=& 
B_{0}(p+k,\tilde{m}_{f_a}, \tilde{m}_{f_b})\,\,,\\
\nonumber\\
J^{a\,b\, c}_{\mu \,\nu} 
(p,k,\tilde{m}_{f_a}, \tilde{m}_{f_b},\tilde{m}_{f_c})&=&   
\left\{ 4 C_{\mu \,\nu}+ 2 (k+2p)_{\nu}\, C_{\mu}\right.\nonumber\\
&+& \left.
2 p_{\mu} C_{\nu}+p_{\mu}(k+2p)_{\nu}\, C_{0} 
\right\} \,(p,k,\tilde{m}_{f_a}, \tilde{m}_{f_b},\tilde{m}_{f_c})\,\,,\\
\nonumber\\
\label{eq:int4f}
J^{a\,b\,c\,d}_{\mu\,\nu\,\sigma \,\lambda}
(p,k,r,\tilde{m}_{f_a}, \tilde{m}_{f_b},\tilde{m}_{f_c},\tilde{m}_{f_d})&=&
\left\{ 16D_{\mu \,\nu\,\sigma \,\lambda} + 8 (k+p+r)_{\lambda}\,D_{\mu \,\nu\,\sigma}
+8\, p_{\mu}\,D_{\nu \,\sigma\,\lambda}\right.\nonumber\\
&+& 8 (2k+2p+r)_{\sigma}\,D_{\mu \,\nu\,\lambda}+
 8 (k+2p)_{\nu}\,D_{\mu \,\sigma\,\lambda} \nonumber\\
&+& 4(2k+2p+r)_{\sigma}\,(k+p+r)_{\lambda} \,(D_{\mu \,\nu}+p_{\mu}\,D_{\nu})\nonumber\\ 
&+& 2(2k+2p+r)_{\sigma}\,(k+2p)_{\nu}\, (2D_{\mu \,\lambda}+p_{\mu}\,D_{\lambda})\nonumber\\
&+& 2(k+2p)_{\nu}\,(k+p+r)_{\lambda}\,(2D_{\mu \,\sigma}+p_{\mu}\,D_{\sigma})\nonumber\\
&+& 2\,(2k+2p+r)_{\sigma}\,(k+2p)_{\nu}\,(k+p+r)_{\lambda}\,D_{\mu}\nonumber\\
&+& p_{\mu}\,(k+2p)_{\nu}\,(k+p+r)_{\lambda}\,(2k+2p+r)_{\sigma}\,
D_{0}\nonumber\\
&+& 4p_{\mu}\,(k+p+r)_{\lambda}\, D_{\nu \,\sigma}+
4p_{\mu}\,(k+2p)_{\nu}\,D_{\sigma \,\lambda}\nonumber\\ 
&+& \left. 4p_{\mu}\,(2k+2p+r)_{\sigma}\,D_{\nu \,\lambda}\right\} \,(p,k,\tilde{m}_{f_a}, \tilde{m}_{f_b},\tilde{m}_{f_c},
\tilde{m}_{f_d})\,.\nonumber\\
\end{eqnarray}

$\bullet$ Asymptotic results.

As we said before, we compute all the integrals in the large masses limit by
 using the
m-Theorem~\cite{GMR}. Some examples of the applicability of this theorem
to the present context of decoupling of SUSY particles can be found in
ref.~\cite{GEISHA}.

We present in the following the results for the standard one-loop
integrals in the limit of heavy SUSY particles. In taking this limit we
require in addition  
that the differences of masses be always smaller than their sums, i.e 
$\tilde{m}^2 \gg k^2$ and $|\tilde{m}_{i}^2-\tilde{m}_{j}^2| \ll|\tilde{m}_{i}^2+\tilde{m}_{j}^2| $.
The results of the standard integrals in our asymptotic limit are as follows,
\begin{eqnarray}
\label{eq:io}
A_{0}(m_1) &=& \left( {\Delta}_\epsilon+1-\log \frac{m_1^2}{\mu_{o}^{2}} \right){m_1^2}\,,\nonumber\\
\nonumber\\
B_{0}(p,m_1,m_2) &=& 
\left( {\Delta}_\epsilon-\log \frac{m_1^2+m_2^2}{2\mu_{o}^{2}} \right)\,,\nonumber\\
\nonumber\\
B_{\mu}(p,m_1,m_2) &=& -\frac{1}{2}\, p_{\mu} 
\left( {\Delta}_\epsilon-\log \frac{m_1^2+m_2^2}{2\mu_{o}^{2}} \right)\,,\nonumber\\
\nonumber\\
B_{\mu \nu}(p,m_1,m_2) &=& \frac{1}{4} \,(m_1^2+m_2^2) \left({\Delta}_\epsilon+1
-\log \frac{m_1^2+m_2^2}{2\mu_{o}^{2}}\right)g_{\mu\, \nu} \nonumber\\
&-&\frac{1}{12}\,p^{2} \left({\Delta}_\epsilon-\log 
\frac{m_1^2+m_2^2}{2\mu_{o}^{2}}\right)g_{\mu\, \nu} \nonumber\\
&+&\frac{1}{3} p_{\mu}p_{\nu}\left({\Delta}_\epsilon-
\log \frac{m_1^2+m_2^2}{2\mu_{o}^{2}}\right)\,,\nonumber \\
\nonumber\\
C_{0}(p,k,m_1,m_2,m_3) &=& 0\,\,\,, \,\,\,\,C_{\mu}(p,k,m_1,m_2,m_3) = 0\,,\nonumber\\
\nonumber\\
C_{\mu\nu}(p,k,m_1,m_2,m_3) &=& \frac{1}{4}\left({\Delta}_\epsilon-
\log \frac{m_1^2+m_2^2+m_3^2}{3\mu_{o}^{2}}\right)g_{\mu\, \nu}\,,\nonumber\\
C_{\mu\nu\sigma}(p,k,m_1,m_2,m_3) &=& -\frac{1}{12}
{(2p+k)}_{\rho}\left({\Delta}_\epsilon-\log \frac{m_1^2+m_2^2+m_3^2}{3\mu_{o}^{2}}\right)\times\nonumber\\
&&\hspace*{0.3cm} \left[g_{\mu\, \nu}\,g_{\sigma\, \rho}+g_{\mu\, \sigma}\,g_{\nu\, \rho}+
g_{\mu\, \rho}\,g_{\nu\, \sigma}\right]\,,\nonumber \\
\nonumber\\
D_{0}(p,k,r,m_1,m_2,m_3,m_4) &=& 0\,\,, \,\,\,D_{\mu}(p,k,r,m_1,m_2,m_3,m_4) = 0\,,\nonumber\\
\nonumber\\
D_{\mu\nu}(p,k,r,m_1,m_2,m_3,m_4) &=& 0\,\,, \,\,\,D_{\mu\nu\sigma}(p,k,r,m_1,m_2,m_3,m_4) = 0\,,\nonumber\\
\nonumber\\
\label{eq:last4}
D_{\mu\nu\sigma\lambda}(p,k,r,m_1,m_2,m_3,m_4)&=& \frac{1}{24}
\left({\Delta}_\epsilon-\log \frac{m_1^2+m_2^2+m_3^2+m_4^2}{4\mu_{o}^{2}}\right)\times\nonumber\\
&&\hspace*{0.3cm} \left[g_{\mu\, \nu}\,g_{\sigma\,\lambda}+g_{\mu\, \sigma}\,
g_{\nu\, \lambda}+
g_{\mu\, \lambda}\,g_{\nu\, \sigma}\right]\,.
\end{eqnarray}
The corrections to these formulae are suppressed by inverse powers of the sums of the 
corresponding squared masses and vanish in the asymptotic large masses limit.

Finally, notice that the results for the 
3- and 4-point integrals appearing in our calculations can be easily obtained 
from the above formulae
by substitution in eqs.~(\ref{eq:int31}) to (\ref{eq:int4f}) correspondingly. We will not present here these 
results for brevity.

This complete our analysis and results of the 3 and 4-point integrals that have 
appeared in the present work.

\section*{Appendix B.}
\vspace{0.4cm}
\setcounter{equation}{0}
\renewcommand{\theequation}{B.\arabic{equation}}

In this appendix we collect the definitions of all the {\em operators} that have been introduced in this work as well as the different functions, $f_{i}$ 
$\,(i=1\ldots 4)$, and $g_{i}$ $\,(i=1\ldots 12)$, appearing in the asymptotic results for the 3 and 4-point Green's
functions respectively. Since we work in the momentum space all these operators are
functions of the corresponding momenta. Thus, for instance, the three-point-function operator
given by $\hat{O}^{\mu\nu\sigma} \sim V_{1}^{\mu} V_{2}^{\nu} V_{3}^{\sigma}$ really
means $\hat{O}^{\mu\nu\sigma} \sim V_{1}^{\mu}(p) V_{2}^{\nu}(k)V_{3}^{\sigma}(r)$ and 
similarly for the other operators. In the following we omit this explicit momentum 
dependence for brevity.

The {\em operators} in eqs.~(\ref{eq:eff3}) are defined by,
\begin{eqnarray}
  \label{eq:operos}
 \hat{O}^{1\,\mu} &=& eA^{\mu} \hat{Q}_f +\frac{g}{c_w} Z^\mu \hat{G}_f+
  \frac{g}{\sqrt{2}} W^{+\mu} \Sigma_f^{tb}+
  \frac{g}{\sqrt{2}} W^{-\mu} \Sigma_f^{bt} \,\,, \nonumber\\
 \hat{O}^{2\,\mu\nu} &=&  e^2 \hat{Q}_f^2 A^{\mu }A^\nu+
 \frac{2\,g\,e}{c_w}A^{\mu }Z^\nu
  \hat{Q}_f\hat{G}_f+
  \frac{g^2}{c_w^2}\hat{G}_f^2 Z^{\mu}Z^\nu+\frac{g^2}{2}
  \Sigma_f W^{\mu+} W^{\nu-} \nonumber\\
 &+& \frac{eg}{\sqrt{2}} y_{\tilde{f}}A^{\mu}\left(W^{\nu+ } \Sigma_f^{tb}+
  W^{\nu- } \Sigma_f^{bt}\right)-
  \frac{g^2}{\sqrt{2}}y_{\tilde{f}}\frac{s_w^2}{c_w}
  Z^{\mu }\left(W^{\nu+} \Sigma_f^{tb}+W^{\nu- } \Sigma_f^{bt}\right)\,.
\end{eqnarray}

In order to write a general expression for the three and four-point functions from {\it inos} contributions
that have been presented in eqs.~(\ref{eq:eff3inos}) and (\ref{eq:eff4inos}), we have introduced the shorthand 
notation
$\left(\, G\cdot O \,\right)$, which we give explicitely in the following. For this purpose
we use the compact notation:
\begin{eqnarray}
\label{eq:traces}
G_{\,\mu \,\nu\,\alpha\,\sigma} &\equiv& \Tr \,[\,\gamma_{\mu} \,\gamma_{\nu} \,\gamma_{\alpha}\, 
\gamma_{\sigma} \,]\,,\nonumber\\
\G1 &\equiv& \Tr\, [\,\gamma_{\alpha}\, \gamma_{\mu} \,\gamma_{\beta}\, \gamma_{\nu} \,\gamma_{\gamma} \,
\gamma_{\sigma} \,] \,,\nonumber\\
\label{eq:trace1}
G_{\,\alpha\, \mu \,\beta\, \nu\, \gamma\, \sigma\, \rho\, \lambda} &\equiv&
\Tr \, [\,\gamma_{\alpha}\, \gamma_{\mu} \,\gamma_{\beta}\, \gamma_{\nu} \,\gamma_{\gamma} \,
\gamma_{\sigma} \,\gamma_{\rho}\, \gamma_{\lambda}\,]\,\,.
\end{eqnarray}

The expressions for each $\left(\, G\cdot O \,\right)$ term in
eq.~(\ref{eq:eff3inos}) for the three-point functions are:
\begin{eqnarray}
\label{eq:GOterm3}
\displaystyle {\left(\, G\cdot O \,\right)}^{\scriptstyle +++}_{\scriptstyle \,1\,2\,3} &=&
 \G1 \left( \hat{O}^{1}+\hat{O}^{2}+\hat{O}^{4}+\hat{O}^{6}+\hat{O}^{8}\right)
\,,\nonumber\\
\displaystyle {\left(\, G\cdot O \,\right)}^{\scriptstyle +++}_{\scriptstyle \,1} &=&
 G_{\,\alpha \,\mu \,\nu\,\sigma} \left(
{\hat{O}}^{1}+{\hat{O}}^{3}+{\hat{O}}^{5}+\hat{O}^{6}+\hat{O}^{9} \right)\,,\nonumber\\
\displaystyle {\left(\, G\cdot O \,\right)}^{\scriptstyle +++}_{\scriptstyle \,2} &=&
 G_{\,\mu \,\alpha \,\nu\,\sigma}\left(
\hat{O}^{1} +\hat{O}^{3}+\hat{O}^{4}+\hat{O}^{7}+\hat{O}^{10}\right)\,,\nonumber\\
\displaystyle {\left(\, G\cdot O \,\right)}^{\scriptstyle +++}_{\scriptstyle \,3} &=&
 G_{\,\mu \,\nu\,\alpha \,\sigma}\left(
{\hat{O}}^{1}+{\hat{O}}^{2}+{\hat{O}}^{5}+\hat{O}^{7}+\hat{O}^{11}\right)\,,\nonumber\\
\displaystyle {\left(\, G\cdot O \,\right)}^{\scriptstyle \,o++}_{\scriptstyle \,1\,2\,3} &=& 
\G1 \left(\,\hat{O}^{16}+\hat{O}^{18}\right)\,\,\,,\,\,
{\left(\, G\cdot O \,\right)}^{\scriptstyle \,o++}_{\scriptstyle \,1} = 
G_{\,\alpha \,\mu \,\nu\,\sigma}\left(\,
\hat{O}^{16}+\hat{O}^{19}\right)\,,\nonumber\\
\displaystyle {\left(\, G\cdot O \,\right)}^{\scriptstyle \,o++}_{\scriptstyle \,2} &=& 
G_{\,\mu \,\alpha\,\nu\,\sigma}\left(\,
\hat{O}^{17}+\hat{O}^{20}\right)\,\,\,,\,\,
{\left(\, G\cdot O \,\right)}^{\scriptstyle \,o++}_{\scriptstyle \,3} = 
G_{\,\mu \,\nu\,\alpha\,\sigma}\left(
\hat{O}^{17}+\hat{O}^{21}\right)\,,\nonumber\\
\displaystyle {\left(\, G\cdot O \,\right)}^{\scriptstyle \,oo+}_{\scriptstyle \,1\,2\,3} &=& 
G_{\,\alpha\,\mu \,\beta\,\nu\,\gamma\,\sigma}\,\hat{O}^{22}
\,\,,\,\,
{\left(\, G\cdot O \,\right)}^{\scriptstyle \,oo+}_{\scriptstyle \,1} = 
G_{\,\alpha \,\mu \,\nu\,\sigma}\,\hat{O}^{23}\,,\nonumber\\
\displaystyle {\left(\, G\cdot O \,\right)}^{\scriptstyle \,oo+}_{\scriptstyle \,2} &=& 
G_{\,\mu \,\alpha\,\nu\,\sigma}\,\hat{O}^{24}\,\,,\,\,
{\left(\, G\cdot O \,\right)}^{\scriptstyle \,oo+}_{\scriptstyle \,3} = 
G_{\,\mu \,\nu\,\alpha\,\sigma}\,\hat{O}^{25}\,\,,\nonumber\\
\displaystyle {\left(\, G\cdot O \,\right)}^{\scriptstyle \,ooo}_{\scriptstyle \,1\,2\,3} &=& 
G_{\,\alpha\,\mu \,\beta\,\nu\,\gamma\,\sigma}\,\hat{O}^{12}
\,\,,\,\,
{\left(\, G\cdot O \,\right)}^{\scriptstyle \,ooo}_{\scriptstyle \,1} = 
G_{\,\alpha \,\mu \,\nu\,\sigma}\,\hat{O}^{13}\,,\nonumber\\
\displaystyle {\left(\, G\cdot O \,\right)}^{\scriptstyle \,ooo}_{\scriptstyle \,2} &=& 
G_{\,\mu \,\alpha\,\nu\,\sigma}\,\hat{O}^{14}\,\,,\,\,
{\left(\, G\cdot O \,\right)}^{\scriptstyle \,ooo}_{\scriptstyle \,3} = 
G_{\,\mu \,\nu\,\alpha\,\sigma}\,\hat{O}^{15}\,\,,
\end{eqnarray}
where the traces are given in (\ref{eq:traces}) and the {\em operators} whose
indices have been omitted here for shortness are given by,
\begin{eqnarray}
\hat{O}^{1\, \mu \,\nu\,\sigma}_{\,\,\,i\,j\,k} &=& -e^{3} 
A_{\mu} \,A_{\nu}\,A_{\sigma}\,\delta_{ij}\,\delta_{jk}\,\delta_{ki}
+ e^{2}\frac{g}{2\cw}\left[\, A_{\mu} \,A_{\nu} \,Z_{\sigma}\, \delta_{ij}\,\delta_{jk} 
({O'}_{L}+{O'}_{R})_{k\,i}\right.\nonumber\\
& +&\left. A_{\mu} \,Z_{\nu} \,A_{\sigma} \,\delta_{ij}\, 
\delta_{ki}({O'}_{L}+{O'}_{R})_{j\,k}
+ Z_{\mu} \,A_{\nu} \,A_{\sigma}\,\delta_{jk}\, \delta_{ki} 
({O'}_{L}+{O'}_{R})_{i\,j}\,\right]\,,\nonumber\\
\hat{O}^{2\,(3)\, \mu \,\nu\,\sigma}_{\hspace*{0.6cm}i\,j\,k} &=& 
-e \frac{g^{2}}{2{\cw}^{2}} 
A_{\mu}\, Z_{\nu} \,Z_{\sigma}\,\delta_{ij} \left( {O'}_{L_{k\,i}} 
{O'}_{L\,(R)_{j\,k}}+
{O'}_{R_{k\,i}} {O'}_{R\,(L)_{j\,k}}\right)\,\,,\nonumber\\
\hat{O}^{4\,(5)\, \mu \,\nu\,\sigma}_{\hspace*{0.6cm}i\,j\,k} &=& 
-e \frac{g^{2}}{2{\cw}^{2}} 
A_{\sigma}\, Z_{\mu} \,Z_{\nu}\,\delta_{ki} \left( {O'}_{L_{i\,j}} {O'}_{L\,(R)_{j\,k}}+
{O'}_{R_{i\,j}} {O'}_{R\,(L)_{j\,k}}\right)\,\,,\nonumber\\
\hat{O}^{6\,(7)\, \mu \,\nu\,\sigma}_{\hspace*{0.6cm}i\,j\,k} &=& 
-e \frac{g^{2}}{2{\cw}^{2}} 
A_{\nu}\, Z_{\mu} \,Z_{\sigma}\,\delta_{jk} \left( {O'}_{L_{i\,j}} 
{O'}_{L\,(R)_{k\,i}}+
{O'}_{R_{i\,j}} {O'}_{R\,(L)_{k\,i}}\right)\,\,,\nonumber\\
\hat{O}^{8\,(9)\, \mu \,\nu\,\sigma}_{\hspace*{0.6cm}i\,j\,k} &=& 
\frac{g^{3}}{2{\cw}^{3}} 
Z_{\mu}\, Z_{\nu} \,Z_{\sigma}\,\left({O'}_{L_{i\,j}} 
{O'}_{L\,(R)_{j\,k}} {O'}_{L_{k\,i}}+{O'}_{R_{i\,j}}
{O'}_{R\,(L)_{j\,k}} {O'}_{R_{k\,i}}\right)\,\,,\nonumber\\
\hat{O}^{10\,(11)\, \mu \,\nu\,\sigma}_{\hspace*{1.0cm}i\,j\,k} &=& 
\frac{g^{3}}{2{\cw}^{3}} 
Z_{\mu}\, Z_{\nu} \,Z_{\sigma}\,\left( {O'}_{L_{i\,j}} 
{O'}_{L\,(R)_{j\,k}} {O'}_{R_{k\,i}}+{O'}_{R_{i\,j}}
{O'}_{R\,(L)_{j\,k}} {O'}_{L_{k\,i}} 
\right)\,\,,\nonumber\\
\hat{O}^{12\,(13)\, \mu \,\nu\,\sigma}_{\hspace*{1.0cm}i\,j\,k} &=& 
\frac{g^{3}}{2{\cw}^{3}} 
Z_{\mu}\, Z_{\nu} \,Z_{\sigma}\,\left({O''}_{L_{i\,j}} 
{O''}_{L\,(R)_{j\,k}} {O''}_{L_{k\,i}}+{O''}_{R_{i\,j}}
{O''}_{R\,(L)_{j\,k}} {O''}_{R_{k\,i}}\right)\,\,,\nonumber\\
\hat{O}^{14\,(15)\, \mu \,\nu\,\sigma}_{\hspace*{1.0cm}i\,j\,k} &=& 
\frac{g^{3}}{2{\cw}^{3}} 
Z_{\mu}\, Z_{\nu} \,Z_{\sigma}\,\left( {O''}_{L_{i\,j}} 
{O''}_{L\,(R)_{j\,k}} {O''}_{R_{k\,i}}+{O''}_{R_{i\,j}}
{O''}_{R\,(L)_{j\,k}} {O''}_{L_{k\,i}} 
\right)\,\,,\nonumber\\
\hat{O}^{16\,(17)\, \mu \,\nu\,\sigma}_{\hspace*{1.0cm}i\,j\,k} &=& 
-e \frac{g^{2}}{2} A_{\nu}\, W^{-}_{\mu} \,
W^{+}_{\sigma}\,\delta_{jk} \left( {O}_{L_{i\,j}} {O}^{+}_{L\,(R)_{k\,i}}+
{O}_{R_{i\,j}} {O}^{+}_{R\,(L)_{k\,i}}\right)\,\,,\nonumber\\ 
\hat{O}^{18\,(19)\, \mu \,\nu\,\sigma}_{\hspace*{1.0cm}i\,j\,k} &=& 
\frac{g^{3}}{2{\cw}} Z_{\nu}\, W^{-}_{\mu}\,W^{+}_{\sigma}\,
\left({O}_{L_{i\,j}} {O'}_{L\,(R)_{j\,k}} {O}^{+}_{L_{k\,i}}+{O}_{R_{i\,j}}
{O'}_{R\,(L)_{j\,k}} {O}^{+}_{R_{k\,i}}\right)\,\,,\nonumber\\ 
\hat{O}^{20\,(21)\, \mu \,\nu\,\sigma}_{\hspace*{1.0cm}i\,j\,k} &=& 
\frac{g^{3}}{2{\cw}} Z_{\nu}\, W^{-}_{\mu}\,W^{+}_{\sigma}\,
\left({O}_{L_{i\,j}} {O'}_{L\,(R)_{j\,k}} {O}^{+}_{R_{k\,i}}+{O}_{R_{i\,j}}
{O'}_{R\,(L)_{j\,k}} {O}^{+}_{L_{k\,i}}\right)\,\,,\nonumber\\
\hat{O}^{22\,(23)\, \mu \,\nu\,\sigma}_{\hspace*{1.0cm}i\,j\,k} &=& 
\frac{g^{3}}{2{\cw}} Z_{\mu}\, W^{-}_{\nu}\,W^{+}_{\sigma}\,
\left({O''}_{L_{i\,j}} {O}_{L\,(R)_{j\,k}} {O}^{+}_{L_{k\,i}}+{O''}_{R_{i\,j}}
{O}_{R\,(L)_{j\,k}} {O}^{+}_{R_{k\,i}}\right)\,\,,\nonumber\\ 
\hat{O}^{24\,(25)\, \mu \,\nu\,\sigma}_{\hspace*{1.0cm}i\,j\,k} &=& 
\frac{g^{3}}{2{\cw}} Z_{\mu}\, W^{-}_{\nu}\,W^{+}_{\sigma}\,
\left({O''}_{L_{i\,j}} {O}_{L\,(R)_{j\,k}} {O}^{+}_{R_{k\,i}}+{O''}_{R_{i\,j}}
{O}_{R\,(L)_{j\,k}} {O}^{+}_{L_{k\,i}}\right)\,\,.
\end{eqnarray}

The generic terms $\left(\, G\cdot O \,\right)$ in the {\it inos} 
contributions to the four-point functions given in eq.~(\ref{eq:eff4inos}) can be written as:
\begin{eqnarray}
\displaystyle {\left( \,G\cdot O \,\right)}^{\scriptstyle ++++}_
{\scriptstyle \,1\,2\,3\,4}&=&
G_{\alpha\mu \beta\nu\gamma\sigma\rho\lambda} 
\left( {\check{O}}^{1}+{\check{O}}^{2}+{\check{O}}^{4}+{\check{O}}^{6}+
{\check{O}}^{8}+{\check{O}^{10}}+{\check{O}^{12}}+{\check{O}^{14}}+
{\check{O}^{18}}\right.\nonumber\\
&+&\left.{\check{O}^{22}}+{\check{O}^{26}}+{\check{O}^{30}}\,\right)\,\,,\nonumber\\
\displaystyle {\left( \,G\cdot O \,\right)}^{\scriptstyle ++++}_
{\scriptstyle \,1\,4}&=&
G_{\alpha \mu \nu\sigma\rho\lambda} \left(
{\check{O}}^{1}+{\check{O}}^{2}+{\check{O}}^{5}+{\check{O}}^{6}+
{\check{O}}^{9}+{\check{O}}^{10}+\check{O}^{13}+{\check{O}}^{17}+
\check{O}^{18} \right.\nonumber\\
&+&\left.{\check{O}^{23}}+{\check{O}^{27}}+{\check{O}^{31}}\,\right) \,\,,\nonumber\\
\displaystyle {\left( \,G\cdot O \,\right)}^{\scriptstyle ++++}_
{\scriptstyle \,1\,3}&=&
G_{\alpha \mu \nu\gamma\sigma\lambda} \left(
{\check{O}}^{1}+{\check{O}}^{3}+\check{O}^{5}+\check{O}^{6}+\check{O}^{8}+
\check{O}^{11}+ \check{O}^{13}+{\check{O}}^{16}+{\check{O}}^{19}\right.\nonumber\\
&+&\left.{\check{O}}^{23}+{\check{O}}^{29}+{\check{O}}^{34}\,\right)\,\,,\nonumber\\
\displaystyle {\left( \,G\cdot O \,\right)}^{\scriptstyle ++++}_
{\scriptstyle \,1\,2}&=&
G_{\alpha \mu \beta\nu\sigma\lambda} \left(
{\check{O}}^{1}+{\check{O}}^{3}+\check{O}^{4}+\check{O}^{6}+\check{O}^{9}+
 \check{O}^{11}+\check{O}^{12}+{\check{O}}^{15}+{\check{O}}^{19}
 \right.\nonumber\\
&+&\left.{\check{O}}^{22}+{\check{O}}^{28}+{\check{O}}^{32}\,\right)\,\,,\nonumber\\
\displaystyle {\left( \,G\cdot O \,\right)}^{\scriptstyle ++++}_
{\scriptstyle \,3\,4}&=&
G_{\mu \nu\gamma\sigma\rho\lambda} \left(
{\check{O}}^{1}+{\check{O}}^{2}+\check{O}^{4}+\check{O}^{7}+
\check{O}^{8}+\check{O}^{11}+\check{O}^{13}+\check{O}^{14}+{\check{O}}^{21}
\right.\nonumber\\
&+&\left.{\check{O}}^{25}+{\check{O}}^{29}+{\check{O}}^{35}\,
\right)\,\,,\nonumber\\
\displaystyle {\left( \,G\cdot O \,\right)}^{\scriptstyle ++++}_
{\scriptstyle \,2\,4}&=&
G_{\mu \beta\nu\sigma\rho\lambda} \left(
{\check{O}}^{1}+{\check{O}}^{2}+\check{O}^{5}+\check{O}^{7}+\check{O}^{9}
+\check{O}^{11}+\check{O}^{12}+\check{O}^{17}+\check{O}^{21}\right.\nonumber\\
&+&\left.{\check{O}}^{24}+{\check{O}}^{28}
+{\check{O}}^{33}\,\right)\,\,,\nonumber\\
\displaystyle {\left( \,G\cdot O \,\right)}^{\scriptstyle ++++}_
{\scriptstyle \,2\,3}&=&
G_{\mu \beta\nu\gamma\sigma\lambda} \left(
{\check{O}}^{1}+{\check{O}}^{3}+\check{O}^{5}+\check{O}^{7}+\check{O}^{8}
+\check{O}^{10}+\check{O}^{12}+\check{O}^{16}+\check{O}^{20}\right.\nonumber\\
&+&\left.{\check{O}}^{24}+{\check{O}}^{26}+{\check{O}}^{36}\right)\,\,,\nonumber\\
\displaystyle {\left( \,G\cdot O \,\right)}^{\scriptstyle ++++}&=&
G_{\mu \nu\sigma\lambda} \left(
{\check{O}}^{1}+{\check{O}}^{3}+{\check{O}}^{4}+{\check{O}}^{7} +{\check{O}}^{9}+
{\check{O}}^{10}+{\check{O}}^{13}+{\check{O}}^{15}+{\check{O}}^{20}\right.\nonumber\\ 
&+&\left.{\check{O}}^{25}+{\check{O}}^{27}+{\check{O}}^{37}\right)
\,\,,\nonumber\\
\nonumber\\
\displaystyle {\left( \,G\cdot O \,\right)}^{\scriptstyle o+++}_
{\scriptstyle \,1\,2\,3\,4}&=&
G_{\alpha\mu \beta\nu\gamma\sigma\rho\lambda} 
\left( {\check{O}}^{46}+{\check{O}}^{50}+{\check{O}}^{58}+\check{O}^{60}
\right)\,\,,\nonumber\\
\vspace*{0.3cm}
\displaystyle {\left( \,G\cdot O \,\right)}^{\scriptstyle o+++}_{\scriptstyle \,1\,4}&=&
 G_{\alpha \mu\nu\sigma\rho\lambda} \left(
{\check{O}}^{46}+{\check{O}}^{51}+{\check{O}}^{58}+{\check{O}}^{61}\right)\, \,,\nonumber\\
\vspace*{0.3cm}
\displaystyle {\left( \,G\cdot O \,\right)}^{\scriptstyle o+++}_{\scriptstyle \,1\,3}&=&
G_{\alpha \mu \nu\gamma\sigma\lambda} \left(
{\check{O}}^{47}+{\check{O}}^{51}+{\check{O}}^{58}+{\check{O}}^{64}\right)\, \,,\nonumber\\
\vspace*{0.3cm}
\displaystyle {\left( \,G\cdot O \,\right)}^{\scriptstyle o+++}_{\scriptstyle \,1\,2}&=&
G_{\alpha \mu \beta\nu\sigma\lambda} \left(
{\check{O}}^{47}+{\check{O}}^{50}+{\check{O}}^{58}+{\check{O}}^{62}\right)\, \,,\nonumber\\
\vspace*{0.3cm}
\displaystyle {\left( \,G\cdot O \,\right)}^{\scriptstyle o+++}_{\scriptstyle \,3\,4}&=&
G_{\mu \nu\gamma\sigma\rho\lambda} \left(
{\check{O}}^{49}+{\check{O}}^{53}+{\check{O}}^{59}+{\check{O}}^{65}\right)
\,\,,\nonumber\\
\vspace*{0.3cm}
\displaystyle {\left( \,G\cdot O \,\right)}^{\scriptstyle o+++}_{\scriptstyle \,2\,4}&=&
G_{\mu \beta\nu\sigma\rho\lambda} \left(
{\check{O}}^{49}+{\check{O}}^{52}+{\check{O}}^{59}+{\check{O}}^{63}\right)
\,\,,\nonumber\\
\vspace*{0.3cm}
\displaystyle {\left( \,G\cdot O \,\right)}^{\scriptstyle o+++}_{\scriptstyle \,2\,3}&=&
 G_{\mu \beta\nu\gamma\sigma\lambda} \left(
{\check{O}}^{48}+{\check{O}}^{52}+{\check{O}}^{59}+{\check{O}}^{66}\right)
\,\,,\nonumber\\
\vspace*{0.3cm}
\displaystyle {\left( \,G\cdot O \,\right)}^{\scriptstyle o+++}&=&
 G_{\mu \nu\sigma\lambda} \left(
{\check{O}}^{48}+{\check{O}}^{53}+{\check{O}}^{59}+{\check{O}}^{67}\right)
\,\,,\nonumber\\
\nonumber\\
\displaystyle {\left( \,G\cdot O \,\right)}^{\scriptstyle oo++}_
{\scriptstyle \,1\,2\,3\,4}&=&G_{\alpha\mu \beta\nu\gamma\sigma\rho\lambda} 
\left({\check{O}}^{54}+{\check{O}}^{68}+{\check{O}}^{84}\,\right)\,\,,\nonumber\\
\vspace*{0.3cm}
\displaystyle {\left( \,G\cdot O \,\right)}^{\scriptstyle oo++}
_{\scriptstyle \,1\,4}&=&G_{\alpha\mu \nu\sigma\rho\lambda}\left(
{\check{O}}^{55}+{\check{O}}^{69}+{\check{O}}^{85}\,\right)\,\,,\nonumber\\
\vspace*{0.3cm}
\displaystyle {\left( \,G\cdot O \,\right)}^{\scriptstyle oo++}_
{\scriptstyle \,1\,3}&=&G_{\alpha\mu \nu\gamma\sigma\lambda}
\left({\check{O}}^{55}+{\check{O}}^{72}+
{\check{O}}^{88}\,\right)\,\,,\nonumber\\
\vspace*{0.3cm}
\displaystyle {\left( \,G\cdot O \,\right)}^{\scriptstyle oo++}_
{\scriptstyle \,1\,2}&=&G_{\alpha\mu \beta\nu\sigma\lambda}\left(
{\check{O}}^{54}+{\check{O}}^{70}+{\check{O}}^{86}\,\right)\,\,,\nonumber\\
\vspace*{0.3cm}
\displaystyle {\left( \,G\cdot O \,\right)}^{\scriptstyle oo++}_
{\scriptstyle \,3\,4}&=&G_{\mu \nu\gamma\sigma\rho\lambda}\left(
{\check{O}}^{57}+{\check{O}}^{73}+
{\check{O}}^{89}\,\right)\,\,,\nonumber\\
\vspace*{0.3cm}
\displaystyle {\left( \,G\cdot O \,\right)}^{\scriptstyle oo++}_
{\scriptstyle \,2\,4}&=&G_{\mu \beta\nu\sigma\rho\lambda}\left(
{\check{O}}^{56}+{\check{O}}^{71}+
{\check{O}}^{87}\,\right)\,\,,\nonumber\\
\vspace*{0.3cm}
\displaystyle {\left( \,G\cdot O \,\right)}^{\scriptstyle oo++}_
{\scriptstyle \,2\,3}&=&G_{\mu \beta\nu\gamma\sigma\lambda}\left(
{\check{O}}^{56}+{\check{O}}^{74}+
{\check{O}}^{90}\,\right)\,\,,\nonumber\\
\vspace*{0.3cm}
\displaystyle {\left( \,G\cdot O \,\right)}^{\scriptstyle oo++}&=&
G_{\mu \nu\sigma\lambda}\left({\check{O}}^{57}+
{\check{O}}^{75}+{\check{O}}^{91}\,\right)\,\,,\nonumber\\
\nonumber\\
\displaystyle {\left( \,G\cdot O \,\right)}^{\scriptstyle ooo+}_
{\scriptstyle \,1\,2\,3\,4}&=&G_{\alpha\mu \beta\nu\gamma\sigma\rho\lambda}
{\check{O}}^{76}\,\,\,\,\,,\,\,\,
{\left( \,G\cdot O \,\right)}^{\scriptstyle ooo+}_{\scriptstyle \,1\,4}=
G_{\alpha\mu \nu\sigma\rho\lambda}{\check{O}}^{77}\,,\nonumber\\
\vspace*{0.3cm}
\displaystyle {\left( \,G\cdot O \,\right)}^{\scriptstyle ooo+}_
{\scriptstyle \,1\,3}&=&G_{\alpha\mu \nu\gamma\sigma\lambda}
{\check{O}}^{80}\,\,\,\,\,\,,\,\,\,
{\left( \,G\cdot O \,\right)}^{\scriptstyle ooo+}_{\scriptstyle \,1\,2}=
G_{\alpha\mu \beta\nu\sigma\lambda}{\check{O}}^{78}\,,\nonumber\\
\vspace*{0.3cm}
\displaystyle {\left( \,G\cdot O \,\right)}^{\scriptstyle ooo+}_
{\scriptstyle \,3\,4}&=&G_{\mu \nu\gamma\sigma\rho\lambda}
{\check{O}}^{81}\,\,\,\,\,\,,\,\,\,
{\left( \,G\cdot O \,\right)}^{\scriptstyle ooo+}_{\scriptstyle \,2\,4}=
G_{\mu \beta\nu\sigma\rho\lambda}{\check{O}}^{79}\,,\nonumber\\
\vspace*{0.3cm}
\displaystyle {\left( \,G\cdot O \,\right)}^{\scriptstyle ooo+}_
{\scriptstyle \,2\,3}&=&G_{\mu \beta\nu\gamma\sigma\lambda}
{\check{O}}^{82}\,\,\,\,\,\,,\,\,\,
{\left( \,G\cdot O \,\right)}^{\scriptstyle ooo+}=
G_{\mu\nu\sigma \lambda}{\check{O}}^{83}\,,\nonumber\\
\nonumber\\
\displaystyle {\left( \,G\cdot O \,\right)}^{\scriptstyle oooo}_{\scriptstyle \,1\,2\,3\,4}&=&
G_{\alpha\mu\beta\nu\gamma\sigma\rho\lambda}{\check{O}}^{38}\,\,\,\,\,,\,\,\,
{\left( \,G\cdot O \,\right)}^{\scriptstyle oooo}_{\scriptstyle \,1\,4}=
G_{\alpha\mu \nu\sigma\rho\lambda}{\check{O}}^{39}\,,\nonumber\\
\vspace*{0.3cm}
\displaystyle {\left( \,G\cdot O \,\right)}^{\scriptstyle oooo}_{\scriptstyle \,1\,3}&=&
G_{\alpha\mu \nu\gamma\sigma\lambda}{\check{O}}^{42}\,\,\,\,\,\,,\,\,\,
{\left( \,G\cdot O \,\right)}^{\scriptstyle oooo}_{\scriptstyle \,1\,2}=
G_{\alpha\mu \beta\nu\sigma\lambda}{\check{O}}^{40}\,,\nonumber\\
\vspace*{0.3cm}
\displaystyle {\left( \,G\cdot O \,\right)}^{\scriptstyle oooo}_{\scriptstyle \,3\,4}&=&
G_{\mu \nu\gamma\sigma\rho\lambda}{\check{O}}^{43}\,\,\,\,\,\,,\,\,\,
{\left( \,G\cdot O \,\right)}^{\scriptstyle oooo}_{\scriptstyle \,2\,4}=
G_{\mu \beta\nu\sigma\rho\lambda}{\check{O}}^{41}\,,\nonumber\\
\vspace*{0.3cm}
\displaystyle {\left( \,G\cdot O \,\right)}^{\scriptstyle oooo}_{\scriptstyle \,2\,3}&=&
G_{\mu\beta \nu\gamma\sigma\lambda}{\check{O}}^{44}\,\,\,\,\,\,,\,\,\,
{\left( \,G\cdot O \,\right)}^{\scriptstyle oooo}=
G_{\mu \nu\sigma\lambda}{\check{O}}^{45}\,,
\end{eqnarray}
where we have assumed again the notation for the traces given in
eq.(\ref{eq:traces}) and the corresponding {\em operators} introduced here are,
\begin{eqnarray}
\check{O}^{1\, \mu \,\nu\,\sigma\,\lambda}_{\,\,\,i\,j\,k\,l} &=& e^{4} 
A_{\mu} \,A_{\nu}\,A_{\sigma}\,A_{\lambda}\,\delta_{ij}\,\delta_{jk} \,\delta_{kl}
\,\delta_{li}
-e^{3}\frac{g}{2\cw}\left[\, Z_{\mu} \,A_{\nu}\, A_{\sigma}\,A_{\lambda}\,
\delta_{jk}\,\delta_{kl}\,\delta_{li} 
({O'}_{L}+{O'}_{R})_{i\,j}\right.\nonumber\\ 
&+& A_{\mu} \,Z_{\nu}\,A_{\sigma}\, A_{\lambda}\, 
\delta_{ij}\,\delta_{kl}\,\delta_{li}({O'}_{L}+{O'}_{R})_{j\,k} +
A_{\mu} \,A_{\nu}\, Z_{\sigma}\,A_{\lambda}\, 
\delta_{ij}\,\delta_{jk}\,\delta_{li}({O'}_{L}+{O'}_{R})_{k\,l}\nonumber\\
&+&\left.A_{\mu} \,A_{\nu}\, A_{\sigma}\,Z_{\lambda}\,
\delta_{ij}\,\delta_{jk}\,\delta_{kl} 
({O'}_{L}+{O'}_{R})_{l\,i}\,\right]\,\,,\nonumber\\
\check{O}^{2\,(3)\, \mu \,\nu\,\sigma\,\lambda}_{\hspace*{0.6cm}i\,j\,k\,l} &=& e^{2} \frac{g^{2}}{2{\cw}^{2}} 
A_{\mu}\, A_{\nu}\,Z_{\sigma}\,Z_{\lambda} \,\delta_{ij}\,\delta_{jk} 
\left( {O'}_{L_{k\,l}} {O'}_{L\,(R)_{\,l\,i}}+
{O'}_{R_{k\,l}} {O'}_{R\,(L)_{\,l\,i}}\right)\,\,,\nonumber\\
\check{O}^{4\,(5)\, \mu \,\nu\,\sigma\,\lambda}_{\hspace*{0.6cm}i\,j\,k\,l} &=& e^{2} \frac{g^{2}}{2{\cw}^{2}} 
A_{\mu}\, A_{\sigma}\,Z_{\nu} \,Z_{\lambda}\,\delta_{ij}\,\delta_{kl} 
 \left( {O'}_{L_{j\,k}} {O'}_{L\,(R)_{\,l\,i}}+
{O'}_{R_{j\,k}} {O'}_{R\,(L)_{\,l\,i}}\right)\,\,,\nonumber\\
\check{O}^{6\,(7)\, \mu \,\nu\,\sigma\,\lambda}_{\hspace*{0.6cm}i\,j\,k\,l} &=& e^{2} \frac{g^{2}}{2{\cw}^{2}} 
A_{\nu}\, A_{\sigma}\,Z_{\mu} \,Z_{\lambda}\,\delta_{jk}\,\delta_{kl}  
\left( {O'}_{L_{i\,j}} {O'}_{L\,(R)_{l\,i}}+
{O'}_{R_{i\,j}} {O'}_{R\,(L)_{l\,i}}\right)\,\,,\nonumber\\
\check{O}^{8\,(9)\, \mu \,\nu\,\sigma\,\lambda}_{\hspace*{0.6cm}i\,j\,k\,l} &=& e^{2} \frac{g^{2}}{2{\cw}^{2}} 
A_{\mu}\, A_{\lambda}\,Z_{\nu} \,Z_{\sigma}\,\delta_{ij}\,\delta_{li} 
\left( {O'}_{L_{j\,k}} {O'}_{L\,(R)_{k\,l}}+
{O'}_{R_{j\,k}} {O'}_{R\,(L)_{k\,l}}\right)\,\,,\nonumber\\
\check{O}^{10\,(11)\, \mu \,\nu\,\sigma\,\lambda}_{\hspace*{0.9cm}i\,j\,k\,l} &=& e^{2} \frac{g^{2}}{2{\cw}^{2}} 
A_{\nu}\, A_{\lambda}\,Z_{\mu} \,Z_{\sigma}\,\delta_{jk}\,\delta_{li}  
\left( {O'}_{L_{i\,j}} {O'}_{L\,(R)_{k\,l}}+
{O'}_{R_{i\,j}} {O'}_{R\,(L)_{k\,l}}\right)\,\,,\nonumber\\
\check{O}^{12\,(13)\, \mu \,\nu\,\sigma\,\lambda}_{\hspace*{0.9cm}i\,j\,k\,l} &=& e^{2} \frac{g^{2}}{2{\cw}^{2}} 
A_{\sigma}\, A_{\lambda}\,Z_{\mu} \,Z_{\nu}\,\delta_{kl}\,\delta_{li} 
 \left( {O'}_{L_{i\,j}} {O'}_{L\,(R)_{j\,k}}+
{O'}_{R_{i\,j}} {O'}_{R\,(L)_{j\,k}}\right)\,\,,\nonumber\\
\check{O}^{14\,(15)\, \mu \,\nu\,\sigma\,\lambda}_{\hspace*{0.9cm}i\,j\,k\,l} 
&=& -e \frac{g^{3}}{2{\cw}^{3}} 
A_{\mu}\, Z_{\nu} \,Z_{\sigma}\,Z_{\lambda}\,\delta_{ij}\, 
\left({O'}_{L_{j\,k}} {O'}_{L\,(R)_{k\,l}} {O'}_{L_{l\,i}}+
{O'}_{R_{j\,k}}{O'}_{R\,(L)_{k\,l}} {O'}_{R_{l\,i}}\right)\,,\nonumber\\
\check{O}^{16\,(17)\, \mu \,\nu\,\sigma\,\lambda}_{\hspace*{0.9cm}i\,j\,k\,l} 
&=& -e \frac{g^{3}}{2{\cw}^{3}} 
A_{\mu}\, Z_{\nu} \,Z_{\sigma}\,Z_{\lambda}\, \delta_{ij}\,
\left({O'}_{L_{j\,k}} {O'}_{L\,(R)_{k\,l}} {O'}_{R_{l\,i}}+
{O'}_{R_{j\,k}}{O'}_{R\,(L)_{k\,l}} {O'}_{L_{l\,i}}\right)\,,\nonumber\\
\check{O}^{18\,(19)\, \mu \,\nu\,\sigma\,\lambda}_{\hspace*{0.9cm}i\,j\,k\,l} &=& 
-e \frac{g^{3}}{2{\cw}^{3}} 
A_{\nu}\, Z_{\mu} \,Z_{\sigma}\,Z_{\lambda}\, \delta_{jk}\,
\left( {O'}_{L_{i\,j}} {O'}_{L\,(R)_{k\,l}}{O'}_{L_{l\,i}}+
{O'}_{R_{i\,j}} {O'}_{R\,(L)_{k\,l}}{O'}_{R_{l\,i}}\right)\,,\nonumber\\
\check{O}^{20\,(21)\, \mu \,\nu\,\sigma\,\lambda}_{\hspace*{1.0cm}i\,j\,k\,l} &=& 
-e \frac{g^{3}}{2{\cw}^{3}} 
A_{\nu}\, Z_{\mu} \,Z_{\sigma}\,Z_{\lambda}\,\delta_{jk}\,
\left( {O'}_{L_{i\,j}} {O'}_{L\,(R)_{k\,l}}{O'}_{R_{l\,i}}+
{O'}_{R_{i\,j}} {O'}_{R\,(L)_{k\,l}}{O'}_{L_{l\,i}}\right)\,,\nonumber\\
\check{O}^{22\,(23)\, \mu \,\nu\,\sigma\,\lambda}_{\hspace*{0.9cm}i\,j\,k\,l} &=& 
-e \frac{g^{3}}{2{\cw}^{3}} 
A_{\sigma}\, Z_{\mu} \,Z_{\nu}\,Z_{\lambda}\, \delta_{kl}\,
\left( {O'}_{L_{i\,j}} {O'}_{L\,(R)_{j\,k}}{O'}_{L_{l\,i}}+
{O'}_{R_{i\,j}} {O'}_{R\,(L)_{j\,k}}{O'}_{R_{l\,i}}\right)\,,\nonumber\\
\check{O}^{24\,(25)\, \mu \,\nu\,\sigma\,\lambda}_{\hspace*{1.0cm}i\,j\,k\,l} &=& 
-e \frac{g^{3}}{2{\cw}^{3}} 
A_{\sigma}\, Z_{\mu} \,Z_{\nu}\,Z_{\lambda}\,\delta_{kl}\,
\left( {O'}_{L_{i\,j}} {O'}_{L\,(R)_{j\,k}}{O'}_{R_{l\,i}}+
{O'}_{R_{i\,j}} {O'}_{R\,(L)_{j\,k}}{O'}_{L_{l\,i}}\right)\,,\nonumber\\
\check{O}^{26\,(27)\, \mu \,\nu\,\sigma\,\lambda}_{\hspace*{0.9cm}i\,j\,k\,l} &=& 
-e \frac{g^{3}}{2{\cw}^{3}} 
A_{\lambda}\, Z_{\mu} \,Z_{\nu}\,Z_{\sigma}\, \delta_{li}\,
\left( {O'}_{L_{i\,j}} {O'}_{L\,(R)_{j\,k}}{O'}_{L_{k\,l}}+
{O'}_{R_{i\,j}} {O'}_{R\,(L)_{j\,k}}{O'}_{R_{k\,l}}\right)\,,\nonumber\\
\check{O}^{28\,(29)\, \mu \,\nu\,\sigma\,\lambda}_{\hspace*{1.0cm}i\,j\,k\,l} &=& 
-e \frac{g^{3}}{2{\cw}^{3}} 
A_{\lambda}\, Z_{\mu} \,Z_{\nu}\,Z_{\sigma}\,\delta_{li}\,
\left( {O'}_{L_{i\,j}} {O'}_{L\,(R)_{j\,k}}{O'}_{R_{k\,l}}+
{O'}_{R_{i\,j}} {O'}_{R\,(L)_{j\,k}}{O'}_{L_{k\,l}}\right)\,,\nonumber\\
\check{O}^{30\,(31)\, \mu \,\nu\,\sigma\,\lambda}_{\hspace*{0.9cm}i\,j\,k\,l} 
&=& \frac{g^{4}}{2\cw^{4}} 
Z_{\mu}\, Z_{\nu} \,Z_{\sigma}\,Z_{\lambda}\,\left( {O'}_{L_{i\,j}} 
{O'}_{L\,(R)_{j\,k}}{O'}_{L_{k\,l}} {O'}_{L_{l\,i}}+{O'}_{R_{i\,j}} 
{O'}_{R\,(L)_{j\,k}}{O'}_{R_{k\,l}} {O'}_{R_{l\,i}}\right),\nonumber\\ 
\check{O}^{32\,(33)\, \mu \,\nu\,\sigma\,\lambda}_{\hspace*{0.9cm}i\,j\,k\,l} &=& 
\frac{g^{4}}{2\cw^{4}} Z_{\mu}\, Z_{\nu} \,Z_{\sigma}\,Z_{\lambda}\,
\left(  {O'}_{L_{i\,j}} 
{O'}_{L_{j\,k}}{O'}_{R_{k\,l}} {O'}_{L\,(R)_{l\,i}}+{O'}_{R_{i\,j}} 
{O'}_{R_{j\,k}}{O'}_{L_{k\,l}} {O'}_{R\,(L)_{l\,i}}\right)\,,\nonumber\\
\check{O}^{34\,(35)\, \mu \,\nu\,\sigma\,\lambda}_{\hspace*{0.9cm}i\,j\,k\,l} 
&=& \frac{g^{4}}{2\cw^{4}} 
Z_{\mu}\, Z_{\nu} \,Z_{\sigma}\,Z_{\lambda}\,\left( {O'}_{L_{i\,j}} 
{O'}_{R_{j\,k}}{O'}_{R_{k\,l}} {O'}_{L\,(R)_{l\,i}}+{O'}_{R_{i\,j}} 
{O'}_{L_{j\,k}}{O'}_{L_{k\,l}} {O'}_{R\,(L)_{l\,i}}\right),\nonumber\\ 
\check{O}^{36\,(37)\, \mu \,\nu\,\sigma\,\lambda}_{\hspace*{0.9cm}i\,j\,k\,l} &=& 
\frac{g^{4}}{2\cw^{4}} Z_{\mu}\, Z_{\nu}Z_{\sigma}\, \,Z_{\lambda}\,
\left(  {O'}_{L_{i\,j}} 
{O'}_{L\,(R)_{j\,k}}{O'}_{L_{k\,l}} {O'}_{R_{l\,i}}+{O'}_{R_{i\,j}} 
{O'}_{R\,(L)_{j\,k}}{O'}_{R_{k\,l}} {O'}_{L_{l\,i}}\right)\,,\nonumber\\
\check{O}^{38\,(39)\, \mu \,\nu\,\sigma\,\lambda}_{\hspace*{0.9cm}i\,j\,k\,l} 
&=& \frac{g^{4}}{2\cw^{4}} 
Z_{\mu}\, Z_{\nu} \,Z_{\sigma}\,Z_{\lambda}\,\left( {O''}_{L_{i\,j}} 
{O''}_{L\,(R)_{j\,k}}{O''}_{L_{k\,l}} {O''}_{L_{l\,i}}+{O''}_{R_{i\,j}} 
{O''}_{R\,(L)_{j\,k}}{O''}_{R_{k\,l}} {O''}_{R_{l\,i}}\right),\nonumber\\ 
\check{O}^{40\,(41)\, \mu \,\nu\,\sigma\,\lambda}_{\hspace*{0.9cm}i\,j\,k\,l} &=& 
\frac{g^{4}}{2\cw^{4}} Z_{\mu}\, Z_{\nu} \,Z_{\sigma}\,Z_{\lambda}\,
\left(  {O''}_{L_{i\,j}} 
{O''}_{L_{j\,k}}{O''}_{R_{k\,l}} {O''}_{L\,(R)_{l\,i}}+{O''}_{R_{i\,j}} 
{O''}_{R_{j\,k}}{O''}_{L_{k\,l}} {O''}_{R\,(L)_{l\,i}}\right),\nonumber\\
\check{O}^{42\,(43)\, \mu \,\nu\,\sigma\,\lambda}_{\hspace*{0.9cm}i\,j\,k\,l} 
&=& \frac{g^{4}}{2\cw^{4}} 
Z_{\mu}\, Z_{\nu} \,Z_{\sigma}\,Z_{\lambda}\,\left( {O''}_{L_{i\,j}} 
{O''}_{R_{j\,k}}{O''}_{R_{k\,l}} {O''}_{L\,(R)_{l\,i}}+{O''}_{R_{i\,j}} 
{O''}_{L_{j\,k}}{O''}_{L_{k\,l}} {O''}_{R\,(L)_{l\,i}}\right),\nonumber\\ 
\check{O}^{44\,(45)\, \mu \,\nu\,\sigma\,\lambda}_{\hspace*{0.9cm}i\,j\,k\,l} &=& 
\frac{g^{4}}{2\cw^{4}} Z_{\mu}\, Z_{\nu} \,Z_{\sigma}\,Z_{\lambda}\,
\left(  {O''}_{L_{i\,j}} 
{O''}_{L\,(R)_{j\,k}}{O''}_{L_{k\,l}} {O''}_{R_{l\,i}}+{O''}_{R_{i\,j}} 
{O''}_{R\,(L)_{j\,k}}{O''}_{R_{k\,l}} {O''}_{L_{l\,i}}\right),\nonumber\\
\check{O}^{46\,(47)\, \mu \,\nu\,\sigma\,\lambda}_{\hspace*{0.9cm}i\,j\,k\,l} &=&
 -e \frac{g^{3}}{2{\cw}} 
W^{-}_{\mu} \,A_{\nu} \,Z_{\sigma}\,W^{+}_{\lambda}\,\delta_{jk}\,
\left( {O}_{L_{i\,j}} {O'}_{L\,(R)_{k\,l}}{O}^{+}_{L_{l\,i}}+
{O}_{R_{i\,j}} {O'}_{R\,(L)_{k\,l}}{O}^{+}_{R_{l\,i}}\right)\,,\nonumber\\ 
\check{O}^{48\,(49)\, \mu \,\nu\,\sigma\,\lambda}_{\hspace*{0.9cm}i\,j\,k\,l} &=&
 -e \frac{g^{3}}{2{\cw}} 
W^{-}_{\mu} \,A_{\nu} \,Z_{\sigma}\,W^{+}_{\lambda}\,\delta_{jk}\,
\left( {O}_{L_{i\,j}} {O'}_{L\,(R)_{k\,l}}{O}^{+}_{R_{l\,i}}+
{O}_{R_{i\,j}} {O'}_{R\,(L)_{k\,l}}{O}^{+}_{L_{l\,i}}\right)\,,\nonumber\\
\check{O}^{50\,(51)\, \mu \,\nu\,\sigma\,\lambda}_{\hspace*{0.9cm}i\,j\,k\,l}
 &=& -e \frac{g^{3}}{2{\cw}} 
W^{-}_{\mu} \,Z_{\nu} \,A_{\sigma}\,W^{+}_{\lambda}\,\delta_{kl}\,
\left( {O}_{L_{i\,j}} {O'}_{L\,(R)_{j\,k}}{O}^{+}_{L_{l\,i}}+
{O}_{R_{i\,j}} {O'}_{R\,(L)_{j\,k}}{O}^{+}_{R_{l\,i}}\right)\,,\nonumber\\
\check{O}^{52\,(53)\, \mu \,\nu\,\sigma\,\lambda}_{\hspace*{0.9cm}i\,j\,k\,l} 
&=& -e \frac{g^{3}}{2{\cw}} 
W^{-}_{\mu} \,Z_{\nu} \,A_{\sigma}\,W^{+}_{\lambda}\,\delta_{kl}\,
\left( {O}_{L_{i\,j}} {O'}_{L\,(R)_{j\,k}}{O}^{+}_{R_{l\,i}}+
{O}_{R_{i\,j}} {O'}_{R\,(L)_{j\,k}}{O}^{+}_{L_{l\,i}}
\right)\,\,,\nonumber\\
\check{O}^{54\,(55)\, \mu \,\nu\,\sigma\,\lambda}_{\hspace*{0.9cm}i\,j\,k\,l} 
&=& -e \frac{g^{3}}{2{\cw}} 
W^{-}_{\nu} \,Z_{\mu} \,A_{\sigma}\,W^{+}_{\lambda}\,\delta_{kl}\,
\left( {O''}_{L_{i\,j}} {O}_{L\,(R)_{j\,k}}{O}^{+}_{L_{l\,i}}+
{O''}_{R_{i\,j}} {O}_{R\,(L)_{j\,k}}{O}^{+}_{R_{l\,i}}\right)\,,\nonumber\\
\check{O}^{56\,(57)\, \mu \,\nu\,\sigma\,\lambda}_{\hspace*{0.9cm}i\,j\,k\,l} 
&=& -e \frac{g^{3}}{2{\cw}} 
W^{-}_{\nu} \,Z_{\mu} \,A_{\sigma}\,W^{+}_{\lambda}\,\delta_{kl}\,
\left( {O''}_{L_{i\,j}} {O}_{L\,(R)_{j\,k}}{O}^{+}_{R_{l\,i}}+
{O''}_{R_{i\,j}} {O}_{R\,(L)_{j\,k}}{O}^{+}_{L_{l\,i}}
\right)\,\,,\nonumber\\
\check{O}^{58\,(59)\, \mu \,\nu\,\sigma\,\lambda}_{\hspace*{0.9cm}i\,j\,k\,l} 
&=& \frac{e^{2}g^{2}}{2} W^{-}_{\mu} \,A_{\nu} \,A_{\sigma}\,W^{+}_{\lambda}
\,\delta_{jk}\,\delta_{kl}\, \left( {O}_{L_{i\,j}} {O}^{+}_{L\,(R)_{l\,i}}+
{O}_{R_{i\,j}} {O}^{+}_{R\,(L)_{l\,i}}\right)\,\,,\nonumber\\
\check{O}^{60\,(61)\, \mu \,\nu\,\sigma\,\lambda}_{\hspace*{0.9cm}i\,j\,k\,l} &=& 
\frac{g^{4} }{2\cw^{2}}
W^{-}_{\mu} \,Z_{\nu} \,Z_{\sigma}\,W^{+}_{\lambda}
\, \left({O}_{L_{i\,j}} 
{O'}_{L\,(R)_{j\,k}}{O'}_{L_{k\,l}} {O}^{+}_{L_{l\,i}}+{O}_{R_{i\,j}} 
{O'}_{R\,(L)_{j\,k}}{O'}_{R_{k\,l}} {O}^{+}_{R_{l\,i}}\right)\,,\nonumber\\
\check{O}^{62\,(63)\, \mu \,\nu\,\sigma\,\lambda}_{\hspace*{0.9cm}i\,j\,k\,l} &=& 
\frac{g^{4} }{2\cw^{2}}
W^{-}_{\mu} \,Z_{\nu} \,Z_{\sigma}\,W^{+}_{\lambda}
\, \left( {O}_{L_{i\,j}} 
{O'}_{L_{j\,k}}{O'}_{R_{k\,l}} {O}^{+}_{L\,(R)_{l\,i}}+{O}_{R_{i\,j}} 
{O'}_{R_{j\,k}}{O'}_{L_{k\,l}} {O}^{+}_{R\,(L)_{l\,i}}\right)\,,\nonumber\\
\check{O}^{64\,(65)\, \mu \,\nu\,\sigma\,\lambda}_{\hspace*{0.9cm}i\,j\,k\,l} &=& 
\frac{g^{4} }{2\cw^{2}}W^{-}_{\mu} \,Z_{\nu} \,Z_{\sigma}\,W^{+}_{\lambda}
\, \left({O}_{L_{i\,j}} 
{O'}_{R_{j\,k}}{O'}_{R_{k\,l}} {O}^{+}_{L\,(R)_{l\,i}}+{O}_{R_{i\,j}} 
{O'}_{L_{j\,k}}{O'}_{L_{k\,l}} {O}^{+}_{R\,(L)_{l\,i}}\right),\nonumber\\ 
\check{O}^{66\,(67)\, \mu \,\nu\,\sigma\,\lambda}_{\hspace*{0.9cm}i\,j\,k\,l} &=& 
\frac{g^{4} }{2\cw^{2}}W^{-}_{\mu} \,Z_{\nu} \,Z_{\sigma}\,W^{+}_{\lambda}
\, \left( {O}_{L_{i\,j}} 
{O'}_{L\,(R)_{j\,k}}{O'}_{L_{k\,l}} {O}^{+}_{R_{l\,i}}+{O}_{R_{i\,j}} 
{O'}_{R\,(L)_{j\,k}}{O'}_{R_{k\,l}} {O}^{+}_{L_{l\,i}}\right)\,,\nonumber\\
\check{O}^{68\,(69)\, \mu \,\nu\,\sigma\,\lambda}_{\hspace*{0.9cm}i\,j\,k\,l} &=& 
\frac{g^{4} }{2\cw^{2}}
W^{-}_{\nu} \,Z_{\mu} \,Z_{\sigma}\,W^{+}_{\lambda}
\, \left({O''}_{L_{i\,j}} 
{O}_{L\,(R)_{j\,k}}{O'}_{L_{k\,l}} {O}^{+}_{L_{l\,i}}+{O''}_{R_{i\,j}} 
{O}_{R\,(L)_{j\,k}}{O'}_{R_{k\,l}} {O}^{+}_{R_{l\,i}}\right)\,,\nonumber\\
\check{O}^{70\,(71)\, \mu \,\nu\,\sigma\,\lambda}_{\hspace*{0.9cm}i\,j\,k\,l} &=& 
\frac{g^{4} }{2\cw^{2}}
W^{-}_{\nu} \,Z_{\mu} \,Z_{\sigma}\,W^{+}_{\lambda}
\, \left( {O''}_{L_{i\,j}} 
{O}_{L_{j\,k}}{O'}_{R_{k\,l}} {O}^{+}_{L\,(R)_{l\,i}}+{O''}_{R_{i\,j}} 
{O}_{R_{j\,k}}{O'}_{L_{k\,l}} {O}^{+}_{R\,(L)_{l\,i}}\right)\,,\nonumber\\
\check{O}^{72\,(73)\, \mu \,\nu\,\sigma\,\lambda}_{\hspace*{0.9cm}i\,j\,k\,l} &=& 
\frac{g^{4} }{2\cw^{2}}W^{-}_{\nu} \,Z_{\mu} \,Z_{\sigma}\,W^{+}_{\lambda}
\, \left({O''}_{L_{i\,j}} 
{O}_{R_{j\,k}}{O'}_{R_{k\,l}} {O}^{+}_{L\,(R)_{l\,i}}+{O''}_{R_{i\,j}} 
{O}_{L_{j\,k}}{O'}_{L_{k\,l}} {O}^{+}_{R\,(L)_{l\,i}}\right),\nonumber\\ 
\check{O}^{74\,(75)\, \mu \,\nu\,\sigma\,\lambda}_{\hspace*{0.9cm}i\,j\,k\,l} &=& 
\frac{g^{4} }{2\cw^{2}}W^{-}_{\nu} \,Z_{\mu} \,Z_{\sigma}\,W^{+}_{\lambda}
\, \left( {O''}_{L_{i\,j}} 
{O}_{L\,(R)_{j\,k}}{O'}_{L_{k\,l}} {O}^{+}_{R_{l\,i}}+{O''}_{R_{i\,j}} 
{O}_{R\,(L)_{j\,k}}{O'}_{R_{k\,l}} {O}^{+}_{L_{l\,i}}\right)\,,\nonumber\\
\check{O}^{76\,(77)\, \mu \,\nu\,\sigma\,\lambda}_{\hspace*{0.9cm}i\,j\,k\,l} &=& 
\frac{g^{4} }{2\cw^{2}}W^{-}_{\sigma} \,Z_{\mu} \,Z_{\nu}\,W^{+}_{\lambda}
\, \left({O''}_{L_{i\,j}} 
{O''}_{L\,(R)_{j\,k}}{O}_{L_{k\,l}} {O}^{+}_{L_{l\,i}}+{O''}_{R_{i\,j}} 
{O''}_{R\,(L)_{j\,k}}{O}_{R_{k\,l}} {O}^{+}_{R_{l\,i}}\right)\,,\nonumber\\
\check{O}^{78\,(79)\, \mu \,\nu\,\sigma\,\lambda}_{\hspace*{0.9cm}i\,j\,k\,l} &=& 
\frac{g^{4} }{2\cw^{2}}W^{-}_{\sigma} \,Z_{\mu} \,Z_{\nu}\,W^{+}_{\lambda}
\, \left( {O''}_{L_{i\,j}} 
{O''}_{L_{j\,k}}{O}_{R_{k\,l}} {O}^{+}_{L\,(R)_{l\,i}}+{O''}_{R_{i\,j}} 
{O''}_{R_{j\,k}}{O}_{L_{k\,l}} {O}^{+}_{R\,(L)_{l\,i}}\right)\,,\nonumber\\
\check{O}^{80\,(81)\, \mu \,\nu\,\sigma\,\lambda}_{\hspace*{0.9cm}i\,j\,k\,l} &=& 
\frac{g^{4} }{2\cw^{2}}W^{-}_{\sigma} \,Z_{\mu} \,Z_{\nu}\,W^{+}_{\lambda}
\, \left({O''}_{L_{i\,j}} 
{O''}_{R_{j\,k}}{O}_{R_{k\,l}} {O}^{+}_{L\,(R)_{l\,i}}+{O''}_{R_{i\,j}} 
{O''}_{L_{j\,k}}{O}_{L_{k\,l}} {O}^{+}_{R\,(L)_{l\,i}}\right),\nonumber\\ 
\check{O}^{82\,(83)\, \mu \,\nu\,\sigma\,\lambda}_{\hspace*{0.9cm}i\,j\,k\,l} &=& 
\frac{g^{4} }{2\cw^{2}}W^{-}_{\sigma} \,Z_{\mu} \,Z_{\nu}\,W^{+}_{\lambda}
\, \left( {O''}_{L_{i\,j}} 
{O''}_{L\,(R)_{j\,k}}{O}_{L_{k\,l}} {O}^{+}_{R_{l\,i}}+{O''}_{R_{i\,j}} 
{O''}_{R\,(L)_{j\,k}}{O}_{R_{k\,l}} {O}^{+}_{L_{l\,i}}\right)\,,\nonumber\\
\check{O}^{84\,(85)\, \mu \,\nu\,\sigma\,\lambda}_{\hspace*{0.9cm}i\,j\,k\,l} &=& 
\frac{g^{4} }{2}
W^{-}_{\mu} \,W^{+}_{\nu} \,W^{-}_{\sigma}\,W^{+}_{\lambda}
\, \left({O}_{L_{i\,k}} 
{O}^{+}_{L\,(R)_{k\,j}}{O}_{L_{j\,l}} {O}^{+}_{L_{l\,i}}+{O}_{R_{i\,k}} 
{O}^{+}_{R\,(L)_{k\,j}}{O}_{R_{j\,l}} {O}^{+}_{R_{l\,i}}\right)\,,\nonumber\\
\check{O}^{86\,(87)\, \mu \,\nu\,\sigma\,\lambda}_{\hspace*{0.9cm}i\,j\,k\,l} &=& 
\frac{g^{4} }{2}W^{-}_{\mu} \,W^{+}_{\nu} \,W^{-}_{\sigma}\,
W^{+}_{\lambda}\, \left( {O}_{L_{i\,k}} 
{O}^{+}_{L_{k\,j}}{O}_{R_{j\,l}} {O}^{+}_{L\,(R)_{l\,i}}+{O}_{R_{i\,k}} 
{O}^{+}_{R_{k\,j}}{O}_{L_{j\,l}} {O}^{+}_{R\,(L)_{l\,i}}
\right)\,,\nonumber\\
\check{O}^{88\,(89)\, \mu \,\nu\,\sigma\,\lambda}_{\hspace*{0.9cm}i\,j\,k\,l} &=& 
\frac{g^{4} }{2}W^{-}_{\mu} \,W^{+}_{\nu} \,W^{-}_{\sigma}\,
W^{+}_{\lambda}\, \left({O}_{L_{i\,k}} 
{O}^{+}_{R_{k\,j}}{O}_{R_{j\,l}} {O}^{+}_{L\,(R)_{l\,i}}+{O}_{R_{i\,k}} 
{O}^{+}_{L_{k\,j}}{O}_{L_{j\,l}} {O}^{+}_{R\,(L)_{l\,i}}\right),\nonumber\\
\check{O}^{90\,(91)\, \mu \,\nu\,\sigma\,\lambda}_{\hspace*{0.9cm}i\,j\,k\,l} &=& 
\frac{g^{4} }{2}W^{-}_{\mu} \,W^{+}_{\nu} \,W^{-}_{\sigma}\,
W^{+}_{\lambda}\, \left({O}_{L_{i\,k}} 
{O}^{+}_{L\,(R)_{k\,j}}{O}_{L_{j\,l}} {O}^{+}_{R_{l\,i}}+{O}_{R_{i\,k}} 
{O}^{+}_{R\,(L)_{k\,j}}{O}_{R_{j\,l}} {O}^{+}_{L_{l\,i}}\right)\,,\nonumber\\
\end{eqnarray}

The definitions of the coupling matrices, 
$\hat{Q}_f \,, \hat{G}_f \,,\Sigma_{f}^{tb} \,, \Sigma_{f}^{bt} \,, 
\Sigma_{f}\,,O_{L,R}\,,O_{L,R}^{\,\prime}$ and
$O_{L,R}^{\,\prime\prime}$ of the above equations can be found in~\cite{GEISHA}.
 
Finally, we give explicitely in the following the expressions for the  $f_{i}$ and $g_{i}$
functions:
\begin{eqnarray}
f_{1}&=&-2c_{b}^{2}c_{t}^{2}\log \frac{2\tilde{m}_{t_{1}}^{2}+\tilde{m}_{b_{1}}^{2}}{3\mu_{o}^{2}}-
2s_{b}^{2}c_{t}^{2}\log \frac{2\tilde{m}_{t_{1}}^{2}+\tilde{m}_{b_{2}}^{2}}{3\mu_{o}^{2}}-
2c_{b}^{2}s_{t}^{2}\log \frac{2\tilde{m}_{t_{2}}^{2}+\tilde{m}_{b_{1}}^{2}}{3\mu_{o}^{2}}\nonumber\\
&-& 2s_{b}^{2} s_{t}^{2}\log \frac{2\tilde{m}_{t_{2}}^{2}+\tilde{m}_{b_{2}}^{2}}{3\mu_{o}^{2}}+
c_{b}^{2}c_{t}^{2}\log \frac{\tilde{m}_{t_{1}}^{2}+2\tilde{m}_{b_{1}}^{2}}{3\mu_{o}^{2}}+
s_{b}^{2}c_{t}^{2}\log \frac{\tilde{m}_{t_{1}}^{2}+2\tilde{m}_{b_{2}}^{2}}{3\mu_{o}^{2}}\nonumber\\
&+& c_{b}^{2}s_{t}^{2}\log \frac{\tilde{m}_{t_{2}}^{2}+2\tilde{m}_{b_{1}}^{2}}{3\mu_{o}^{2}}
+s_{b}^{2}s_{t}^{2}\log \frac{\tilde{m}_{t_{2}}^{2}+2\tilde{m}_{b_{2}}^{2}}{3\mu_{o}^{2}}\,,\nonumber\\
\nonumber\\
f_{2} &=& c_{b}^{2}c_{t}^{2}\left[\left(\frac{c_{t}^{2}}{2}-
\frac{2}{3}\sw^{2}\right) \log
\frac{2\tilde{m}_{t_{1}}^{2}+\tilde{m}_{b_{1}}^{2}}{3\mu_{o}^{2}}+
\left(-\frac{c_{b}^{2}}{2}+
\frac{1}{3}\sw^{2}\right) \log
\frac{\tilde{m}_{t_{1}}^{2}+2\tilde{m}_{b_{1}}^{2}}{3\mu_{o}^{2}} 
\right]\nonumber\\
&+& s_{b}^{2}s_{t}^{2}\left[\left(\frac{s_{t}^{2}}{2}-
\frac{2}{3}\sw^{2}\right) \log
\frac{2\tilde{m}_{t_{2}}^{2}+\tilde{m}_{b_{2}}^{2}}{3\mu_{o}^{2}}+
\left(-\frac{s_{b}^{2}}{2}+
\frac{1}{3}\sw^{2}\right) \log
\frac{\tilde{m}_{t_{2}}^{2}+2\tilde{m}_{b_{2}}^{2}}{3\mu_{o}^{2}}
\right]\nonumber\\
&+& s_{b}^{2}c_{t}^{2}\left[\left(\frac{c_{t}^{2}}{2}-
\frac{2}{3}\sw^{2}\right) \log
\frac{2\tilde{m}_{t_{1}}^{2}+\tilde{m}_{b_{2}}^{2}}{3\mu_{o}^{2}}+
\left(-\frac{s_{b}^{2}}{2}+
\frac{1}{3}\sw^{2}\right) \log
\frac{\tilde{m}_{t_{1}}^{2}+2\tilde{m}_{b_{2}}^{2}}{3\mu_{o}^{2}} 
\right]\nonumber\\
&+& c_{b}^{2}s_{t}^{2}\left[\left(\frac{s_{t}^{2}}{2}-
\frac{2}{3}\sw^{2}\right) \log
\frac{2\tilde{m}_{t_{2}}^{2}+\tilde{m}_{b_{1}}^{2}}{3\mu_{o}^{2}}+
\left(-\frac{c_{b}^{2}}{2}+
\frac{1}{3}\sw^{2}\right) \log
\frac{\tilde{m}_{t_{2}}^{2}+2\tilde{m}_{b_{1}}^{2}}{3\mu_{o}^{2}}
\right]\nonumber\\
&+& c_{b}^{2}c_{t}^{2}s_{t}^{2} \log 
\frac{\tilde{m}_{t_{1}}^{2}+\tilde{m}_{t_{2}}^{2}+\tilde{m}_{b_{1}}^{2}}
{3\mu_{o}^{2}}+
 s_{b}^{2}c_{t}^{2}s_{t}^{2} \log 
\frac{\tilde{m}_{t_{1}}^{2}+\tilde{m}_{t_{2}}^{2}+\tilde{m}_{b_{2}}^{2}}
{3\mu_{o}^{2}}\nonumber\\
&-& c_{b}^{2}s_{b}^{2}c_{t}^{2} \log 
\frac{\tilde{m}_{t_{1}}^{2}+\tilde{m}_{b_{1}}^{2}+\tilde{m}_{b_{2}}^{2}}
{3\mu_{o}^{2}}-
 c_{b}^{2}s_{b}^{2}s_{t}^{2} \log 
\frac{\tilde{m}_{t_{2}}^{2}+\tilde{m}_{b_{1}}^{2}+\tilde{m}_{b_{2}}^{2}}
{3\mu_{o}^{2}}\,,\nonumber\\
\nonumber\\
f_{3}&=& -\log \frac{2\mcu+\mnd}{3\mu_{o}^{2}} 
-\frac{1}{4} \log \frac{2\mcd+\mnt}{3\mu_{o}^{2}}
-\frac{1}{4} \log \frac{2\mcd+\mnc}{3\mu_{o}^{2}}\,,\nonumber\\
\nonumber\\
f_{4}&=& -(\sw^{2}-1) \log \frac{2\mcu+\mnd}{3\mu_{o}^{2}}
-\frac{1}{8} (2\sw^{2}-1)
\left( \log \frac{2\mcd+\mnt}{3\mu_{o}^{2}}+ \log 
\frac{2\mcd+\mnc}{3\mu_{o}^{2}}\right)\nonumber\\
&+& \frac{1}{4} \log \frac{\mcd+\mnt+\mnc}{3\mu_{o}^{2}}\,,
\end{eqnarray}
and,
\begin{eqnarray}
g_{1}&=& -\frac{2}{3}\left\{c_{b}^{\,2}c_{t}^{\,2}\left[\log
\frac{\tilde{m}_{t_{1}}^{2}+2\tilde{m}_{b_{1}}^{2}}{3\mu_{o}^{2}}+
4\log \frac{2\tilde{m}_{t_{1}}^{2}+\tilde{m}_{b_{1}}^{2}}{3\mu_{o}^{2}}-
\frac{8}{3}\log\frac{3\tilde{m}_{t_{1}}^{2}+\tilde{m}_{b_{1}}^{2}}{4\mu_{o}^{2}}
-\frac{2}{3}\log\frac{\tilde{m}_{t_{1}}^{2}+3\tilde{m}_{b_{1}}^{2}}{4\mu_{o}^{2}}
\right.\right.\nonumber\\
&+&\left.
\frac{4}{3}\log\frac{\tilde{m}_{t_{1}}^{2}+\tilde{m}_{b_{1}}^{2}}{2\mu_{o}^{2}}
\right]+s_{b}^{\,2}c_{t}^{\,2}\left[\log
\frac{\tilde{m}_{t_{1}}^{2}+2\tilde{m}_{b_{2}}^{2}}{3\mu_{o}^{2}}+
4\log\frac{2\tilde{m}_{t_{1}}^{2}+\tilde{m}_{b_{2}}^{2}}{3\mu_{o}^{2}}-
\frac{8}{3}\log\frac{3\tilde{m}_{t_{1}}^{2}+\tilde{m}_{b_{2}}^{2}}{4\mu_{o}^{2}}
\right.\nonumber\\
&-& \left.
\frac{2}{3}\log\frac{\tilde{m}_{t_{1}}^{2}+3\tilde{m}_{b_{2}}^{2}}{4\mu_{o}^{2}}+
\frac{4}{3}\log\frac{\tilde{m}_{t_{1}}^{2}+\tilde{m}_{b_{2}}^{2}}{2\mu_{o}^{2}}
\right]+ c_{b}^{2}s_{t}^{2} \left[\log
\frac{\tilde{m}_{t_{2}}^{2}+2\tilde{m}_{b_{1}}^{2}}{3\mu_{o}^{2}}+
4\log\frac{2\tilde{m}_{t_{2}}^{2}+\tilde{m}_{b_{1}}^{2}}{3\mu_{o}^{2}}\right.\nonumber\\
&-&\left.
\frac{8}{3}\log\frac{3\tilde{m}_{t_{2}}^{2}+\tilde{m}_{b_{1}}^{2}}{4\mu_{o}^{2}}
-\frac{2}{3}\log\frac{\tilde{m}_{t_{2}}^{2}+3\tilde{m}_{b_{1}}^{2}}{4\mu_{o}^{2}}+
\frac{4}{3}\log\frac{\tilde{m}_{t_{2}}^{2}+\tilde{m}_{b_{1}}^{2}}{2\mu_{o}^{2}}\right]
+s_{b}^{2}s_{t}^{2}\left[\log
\frac{\tilde{m}_{t_{2}}^{2}+2\tilde{m}_{b_{2}}^{2}}{3\mu_{o}^{2}}\right.\nonumber\\
&+&\left.\left.
4\log \frac{2\tilde{m}_{t_{2}}^{2}+\tilde{m}_{b_{2}}^{2}}{3\mu_{o}^{2}}-
\frac{8}{3}\log\frac{3\tilde{m}_{t_{2}}^{2}+\tilde{m}_{b_{2}}^{2}}{4\mu_{o}^{2}}
-\frac{2}{3}\log\frac{\tilde{m}_{t_{2}}^{2}+3\tilde{m}_{b_{2}}^{2}}{4\mu_{o}^{2}}
+\frac{4}{3}\log\frac{\tilde{m}_{t_{2}}^{2}+\tilde{m}_{b_{2}}^{2}}{2\mu_{o}^{2}}
\right]\right\}
\nonumber\\
g_{2}&=& -\frac{2}{3}\left\{c_{b}^{\,2}c_{t}^{\,2}\left[-\frac{1}{2}\log
\frac{\tilde{m}_{t_{1}}^{2}+\tilde{m}_{b_{1}}^{2}}{2\mu_{o}^{2}}-
\log \frac{\tilde{m}_{t_{1}}^{2}+2\tilde{m}_{b_{1}}^{2}}{3\mu_{o}^{2}}+
2\log\frac{2\tilde{m}_{t_{1}}^{2}+\tilde{m}_{b_{1}}^{2}}{3\mu_{o}^{2}}
-\frac{8}{3}\log\frac{3\tilde{m}_{t_{1}}^{2}+\tilde{m}_{b_{1}}^{2}}{4\mu_{o}^{2}}
\right.\right.\nonumber\\
&-&\left.
\frac{2}{3}\log\frac{\tilde{m}_{t_{1}}^{2}+3\tilde{m}_{b_{1}}^{2}}{4\mu_{o}^{2}}
+\frac{4}{3}\log\frac{\tilde{m}_{t_{1}}^{2}+\tilde{m}_{b_{1}}^{2}}{2\mu_{o}^{2}}
\right]+s_{b}^{\,2}c_{t}^{\,2}\left[\frac{1}{2}\log
\frac{\tilde{m}_{t_{1}}^{2}+\tilde{m}_{b_{2}}^{2}}{2\mu_{o}^{2}}+
\log\frac{\tilde{m}_{t_{1}}^{2}+2\tilde{m}_{b_{2}}^{2}}{3\mu_{o}^{2}}\right.\nonumber\\
&-&2\log\frac{2\tilde{m}_{t_{1}}^{2}+\tilde{m}_{b_{2}}^{2}}{3\mu_{o}^{2}}
+\left.
\frac{8}{3}\log\frac{3\tilde{m}_{t_{1}}^{2}+\tilde{m}_{b_{2}}^{2}}{4\mu_{o}^{2}}+
\frac{2}{3}\log\frac{\tilde{m}_{t_{1}}^{2}+3\tilde{m}_{b_{2}}^{2}}{4\mu_{o}^{2}}-
\frac{4}{3}\log\frac{\tilde{m}_{t_{1}}^{2}+\tilde{m}_{b_{2}}^{2}}{2\mu_{o}^{2}}
\right]\nonumber\\
&+&c_{b}^{2}s_{t}^{2} \left[-\frac{1}{2}\log
\frac{\tilde{m}_{t_{2}}^{2}+\tilde{m}_{b_{1}}^{2}}{\mu_{o}^{2}}-
\log\frac{\tilde{m}_{t_{2}}^{2}+2\tilde{m}_{b_{1}}^{2}}{3\mu_{o}^{2}}\right.
+2\log\frac{2\tilde{m}_{t_{2}}^{2}+\tilde{m}_{b_{1}}^{2}}{3\mu_{o}^{2}}
-\frac{8}{3}\log\frac{3\tilde{m}_{t_{2}}^{2}+\tilde{m}_{b_{1}}^{2}}{4\mu_{o}^{2}}\nonumber\\
&-&\left.\frac{2}{3}\log\frac{\tilde{m}_{t_{2}}^{2}+
3\tilde{m}_{b_{1}}^{2}}{4\mu_{o}^{2}}+
\frac{4}{3}\log\frac{\tilde{m}_{t_{2}}^{2}+\tilde{m}_{b_{1}}^{2}}{2\mu_{o}^{2}}\right]
+s_{b}^{2}s_{t}^{2}\left[-\frac{1}{2}\log
\frac{\tilde{m}_{t_{2}}^{2}+\tilde{m}_{b_{2}}^{2}}{2\mu_{o}^{2}}
-\log \frac{\tilde{m}_{t_{2}}^{2}+2\tilde{m}_{b_{2}}^{2}}{3\mu_{o}^{2}}\right.\nonumber\\
&+&\left.\left.
2\log \frac{2\tilde{m}_{t_{2}}^{2}+\tilde{m}_{b_{2}}^{2}}{3\mu_{o}^{2}}-
\frac{8}{3}\log\frac{3\tilde{m}_{t_{2}}^{2}+\tilde{m}_{b_{2}}^{2}}{4\mu_{o}^{2}}
-\frac{2}{3}\log\frac{\tilde{m}_{t_{2}}^{2}+3\tilde{m}_{b_{2}}^{2}}{4\mu_{o}^{2}}
+\frac{4}{3}\log\frac{\tilde{m}_{t_{2}}^{2}+\tilde{m}_{b_{2}}^{2}}{2\mu_{o}^{2}}
\right]\right\}
\nonumber\\
g_{3}&=&c_{t}^{\,2}s_{t}^{\,2}\left[4\log  
\frac{\tilde{m}_{t_{1}}^{2}+\tilde{m}_{t_{2}}^{2}}{2\mu_{o}^{2}}-2\log
\frac{2\tilde{m}_{t_{1}}^{2}+\tilde{m}_{t_{2}}^{2}}{3\mu_{o}^{2}}-2\log
\frac{\tilde{m}_{t_{1}}^{2}+2\tilde{m}_{t_{2}}^{2}}{3\mu_{o}^{2}}\right]\nonumber\\
&+&c_{b}^{\,2}c_{t}^{\,2}\left[\frac{2}{3}(-3c_{t}^{2}+4\sw^{2})\log
\frac{2\tilde{m}_{t_{1}}^{2}+\tilde{m}_{b_{1}}^{2}}{3\mu_{o}^{2}}+
\frac{1}{3}(-3c_{b}^{2}+2\sw^{2})\log
\frac{\tilde{m}_{t_{1}}^{2}+2\tilde{m}_{b_{1}}^{2}}{3\mu_{o}^{2}}\right.\nonumber\\
&+&\frac{1}{9}(-6c_{b}^{2}-3c_{t}^{2}+8\sw^{2})\log
\frac{\tilde{m}_{t_{1}}^{2}+\tilde{m}_{b_{1}}^{2}}{2\mu_{o}^{2}}
+\frac{4}{9}\,(3c_{t}^{2}-4\sw^{2})\log
\frac{3\tilde{m}_{t_{1}}^{2}+\tilde{m}_{b_{1}}^{2}}{4\mu_{o}^{2}}\nonumber\\
&+&\left.\frac{4}{9}(3c_{b}^{2}-2\sw^{2})\log
\frac{\tilde{m}_{t_{1}}^{2}+3\tilde{m}_{b_{1}}^{2}}{4\mu_{o}^{2}}\right]
+s_{b}^{2}c_{t}^{2}\left[\frac{2}{3} (-3c_{t}^{2}+4\sw^{2})\log
\frac{2\tilde{m}_{t_{1}}^{2}+\tilde{m}_{b_{2}}^{2}}{3\mu_{o}^{2}}\right.\nonumber\\
&+&\frac{1}{3} (-3s_{b}^{2}+2\sw^{2})\log
\frac{\tilde{m}_{t_{1}}^{2}+2\tilde{m}_{b_{2}}^{2}}{3\mu_{o}^{2}}+
\frac{1}{9} (-6s_{b}^{2}-3c_{t}^{2}+8\sw^{2})\log
\frac{\tilde{m}_{t_{1}}^{2}+\tilde{m}_{b_{2}}^{2}}{2\mu_{o}^{2}}\nonumber\\
&+&\left.\frac{4}{9} (3c_{t}^{2}-4\sw^{2})\log
\frac{3\tilde{m}_{t_{1}}^{2}+\tilde{m}_{b_{2}}^{2}}{4\mu_{o}^{2}}+
\frac{2}{9} (3s_{b}^{2}-2\sw^{2})\log
\frac{\tilde{m}_{t_{1}}^{2}+3\tilde{m}_{b_{2}}^{2}}{4\mu_{o}^{2}}\right]\nonumber\\
&+&c_{b}^{2}s_{t}^{2}\left[\frac{2}{3} (-3s_{t}^{2}+4\sw^{2})\log
\frac{2\tilde{m}_{t_{2}}^{2}+\tilde{m}_{b_{1}}^{2}}{3\mu_{o}^{2}}+
\frac{1}{3} (-3c_{b}^{2}+2\sw^{2})\log
\frac{\tilde{m}_{t_{2}}^{2}+2\tilde{m}_{b_{1}}^{2}}{3\mu_{o}^{2}}\right.\nonumber\\
&+&\frac{1}{9} (-6c_{b}^{2}-3s_{t}^{2}+8\sw^{2})\log
\frac{\tilde{m}_{t_{2}}^{2}+\tilde{m}_{b_{1}}^{2}}{2\mu_{o}^{2}}
+\frac{4}{9} (3s_{t}^{2}-4\sw^{2})\log
\frac{3\tilde{m}_{t_{2}}^{2}+\tilde{m}_{b_{1}}^{2}}{4\mu_{o}^{2}}\nonumber\\
&+&\left.\frac{2}{9} (3c_{b}^{2}-2\sw^{2})\log
\frac{\tilde{m}_{t_{2}}^{2}+3\tilde{m}_{b_{1}}^{2}}{4\mu_{o}^{2}}\right]+
s_{b}^{2}s_{t}^{2}\left[\frac{2}{3} (-3s_{t}^{2}+4\sw^{2})\log
\frac{2\tilde{m}_{t_{2}}^{2}+\tilde{m}_{b_{2}}^{2}}{3\mu_{o}^{2}}\right.\nonumber\\
&+&\frac{1}{3} (-3s_{b}^{2}+2\sw^{2})\log
\frac{\tilde{m}_{t_{2}}^{2}+2\tilde{m}_{b_{2}}^{2}}{3\mu_{o}^{2}}+
\frac{1}{9} (-6s_{b}^{2}-3s_{t}^{2}+8\sw^{2})\log
\frac{\tilde{m}_{t_{2}}^{2}+\tilde{m}_{b_{2}}^{2}}{2\mu_{o}^{2}}\nonumber\\
&+&\frac{4}{9} (3s_{t}^{2}-4\sw^{2})\log
\frac{3\tilde{m}_{t_{2}}^{2}+\tilde{m}_{b_{2}}^{2}}{4\mu_{o}^{2}}+
\left.\frac{2}{9} (3s_{b}^{2}-2\sw^{2})\log
\frac{\tilde{m}_{t_{2}}^{2}+3\tilde{m}_{b_{2}}^{2}}{4\mu_{o}^{2}}\right]\nonumber\\
&+&c_{b}^{\,2}s_{b}^{\,2}\left[2\log  
\frac{\tilde{m}_{b_{1}}^{2}+\tilde{m}_{b_{2}}^{2}}{2\mu_{o}^{2}}-\log
\frac{2\tilde{m}_{b_{1}}^{2}+\tilde{m}_{b_{2}}^{2}}{3\mu_{o}^{2}}-\log
\frac{\tilde{m}_{b_{1}}^{2}+2\tilde{m}_{b_{2}}^{2}}{3\mu_{o}^{2}}\right]\nonumber\\
&+&c_{b}^{\,2}c_{t}^{\,2}s_{t}^{\,2}\left[-4\log
\frac{\tilde{m}_{t_{1}}^{2}+\tilde{m}_{t_{2}}^{2}+\tilde{m}_{b_{1}}^{2}}{3\mu_{o}^{2}}
+\frac{4}{3}\log
\frac{\tilde{m}_{t_{1}}^{2}+2\tilde{m}_{t_{2}}^{2}+\tilde{m}_{b_{1}}^{2}}{4\mu_{o}^{2}}
+\frac{4}{3}\log
\frac{2\tilde{m}_{t_{1}}^{2}+\tilde{m}_{t_{2}}^{2}+\tilde{m}_{b_{1}}^{2}}{4\mu_{o}^{2}}
\right.\nonumber\\
&-&\left.\frac{2}{3}\log\frac{\tilde{m}_{t_{1}}^{2}+\tilde{m}_{t_{2}}^{2}+2\tilde{m}_{b_{1}}^{2}}
{4\mu_{o}^{2}}\right]+
s_{b}^{\,2}c_{t}^{\,2}s_{t}^{\,2}\left[-4\log
\frac{\tilde{m}_{t_{1}}^{2}+\tilde{m}_{t_{2}}^{2}+\tilde{m}_{b_{2}}^{2}}{3\mu_{o}^{2}}
\right.+\frac{4}{3}\log
\frac{2\tilde{m}_{t_{1}}^{2}+\tilde{m}_{t_{2}}^{2}+\tilde{m}_{b_{2}}^{2}}{4\mu_{o}^{2}}
\nonumber\\
&+&\frac{4}{3}\log
\frac{\tilde{m}_{t_{1}}^{2}+2\tilde{m}_{t_{2}}^{2}+\tilde{m}_{b_{2}}^{2}}{4\mu_{o}^{2}}
-\left.\frac{2}{3}\log\frac{\tilde{m}_{t_{1}}^{2}+\tilde{m}_{t_{2}}^{2}+2\tilde{m}_{b_{2}}^{2}}
{4\mu_{o}^{2}}\right]+s_{b}^{\,2}c_{b}^{\,2}c_{t}^{\,2}\left[-2\log
\frac{\tilde{m}_{t_{1}}^{2}+\tilde{m}_{b_{1}}^{2}+\tilde{m}_{b_{2}}^{2}}{3\mu_{o}^{2}}
\right.\nonumber\\
&-&\frac{4}{3}\log
\frac{2\tilde{m}_{t_{1}}^{2}+\tilde{m}_{b_{1}}^{2}+\tilde{m}_{b_{2}}^{2}}{4\mu_{o}^{2}}
+\frac{2}{3}\log
\frac{\tilde{m}_{t_{1}}^{2}+2\tilde{m}_{b_{1}}^{2}+\tilde{m}_{b_{2}}^{2}}{4\mu_{o}^{2}}
+\left.\frac{2}{3}\log
\frac{\tilde{m}_{t_{1}}^{2}+\tilde{m}_{b_{1}}^{2}+2\tilde{m}_{b_{2}}^{2}}{4\mu_{o}^{2}}
\right]\nonumber\\
&+&s_{b}^{\,2}c_{b}^{\,2}s_{t}^{\,2}\left[-2\log
\frac{\tilde{m}_{t_{2}}^{2}+\tilde{m}_{b_{1}}^{2}+\tilde{m}_{b_{2}}^{2}}{3\mu_{o}^{2}}
\right.+\frac{2}{3}\log
\frac{\tilde{m}_{t_{2}}^{2}+2\tilde{m}_{b_{1}}^{2}+\tilde{m}_{b_{2}}^{2}}{4\mu_{o}^{2}}
+\frac{2}{3}\log
\frac{\tilde{m}_{t_{2}}^{2}+\tilde{m}_{b_{1}}^{2}+2\tilde{m}_{b_{2}}^{2}}{4\mu_{o}^{2}}
\nonumber\\
&-&\left.\left.\frac{4}{3}\log
\frac{2\tilde{m}_{t_{2}}^{2}+\tilde{m}_{b_{1}}^{2}+\tilde{m}_{b_{2}}^{2}}{4\mu_{o}^{2}}
\right]\right\}\nonumber\\
\nonumber\\
g_{4}&=&c_{b}^{\,2}c_{t}^{\,2}\left[\frac{1}{6}(-3c_{t}^{2}+8\sw^{2})\log
\frac{2\tilde{m}_{t_{1}}^{2}+\tilde{m}_{b_{1}}^{2}}{3\mu_{o}^{2}}+
\frac{1}{6}(3c_{b}^{2}-4\sw^{2})\log
\frac{\tilde{m}_{t_{1}}^{2}+2\tilde{m}_{b_{1}}^{2}}{3\mu_{o}^{2}}\right.\nonumber\\
&+&\frac{1}{9}(-6c_{b}^{2}-3c_{t}^{2}+5\sw^{2})\log
\frac{\tilde{m}_{t_{1}}^{2}+\tilde{m}_{b_{1}}^{2}}{2\mu_{o}^{2}}
+\frac{4}{9}\,(3c_{t}^{2}-4\sw^{2})\log
\frac{3\tilde{m}_{t_{1}}^{2}+\tilde{m}_{b_{1}}^{2}}{4\mu_{o}^{2}}\nonumber\\
&+&\left.\frac{2}{9}(3c_{b}^{2}-2\sw^{2})\log
\frac{\tilde{m}_{t_{1}}^{2}+3\tilde{m}_{b_{1}}^{2}}{4\mu_{o}^{2}}\right]
+s_{b}^{2}c_{t}^{2}\left[\frac{1}{6} (-3c_{t}^{2}+8\sw^{2})\log
\frac{2\tilde{m}_{t_{1}}^{2}+\tilde{m}_{b_{2}}^{2}}{3\mu_{o}^{2}}\right.\nonumber\\
&+&\frac{1}{6} (3s_{b}^{2}-4\sw^{2})\log
\frac{\tilde{m}_{t_{1}}^{2}+2\tilde{m}_{b_{2}}^{2}}{3\mu_{o}^{2}}+
\frac{1}{9} (-6s_{b}^{2}-3c_{t}^{2}+5\sw^{2})\log
\frac{\tilde{m}_{t_{1}}^{2}+\tilde{m}_{b_{2}}^{2}}{2\mu_{o}^{2}}\nonumber\\
&+&\left.\frac{4}{9} (3c_{t}^{2}-4\sw^{2})\log
\frac{3\tilde{m}_{t_{1}}^{2}+\tilde{m}_{b_{2}}^{2}}{4\mu_{o}^{2}}+
\frac{2}{9} (3s_{b}^{2}-2\sw^{2})\log
\frac{\tilde{m}_{t_{1}}^{2}+3\tilde{m}_{b_{2}}^{2}}{4\mu_{o}^{2}}\right]\nonumber\\
&+&c_{b}^{2}s_{t}^{2}\left[\frac{1}{6} (-3s_{t}^{2}+8\sw^{2})\log
\frac{2\tilde{m}_{t_{2}}^{2}+\tilde{m}_{b_{1}}^{2}}{3\mu_{o}^{2}}+
\frac{1}{6} (3c_{b}^{2}-4\sw^{2})\log
\frac{\tilde{m}_{t_{2}}^{2}+2\tilde{m}_{b_{1}}^{2}}{3\mu_{o}^{2}}\right.\nonumber\\
&+&\frac{1}{9} (-6c_{b}^{2}-3s_{t}^{2}+5\sw^{2})\log
\frac{\tilde{m}_{t_{2}}^{2}+\tilde{m}_{b_{1}}^{2}}{2\mu_{o}^{2}}
+\frac{4}{9} (3s_{t}^{2}-4\sw^{2})\log
\frac{3\tilde{m}_{t_{2}}^{2}+\tilde{m}_{b_{1}}^{2}}{4\mu_{o}^{2}}\nonumber\\
&+&\left.\frac{2}{9} (3c_{b}^{2}-2\sw^{2})\log
\frac{\tilde{m}_{t_{2}}^{2}+3\tilde{m}_{b_{1}}^{2}}{4\mu_{o}^{2}}\right]+
s_{b}^{2}s_{t}^{2}\left[\frac{1}{6} (-3s_{t}^{2}+8\sw^{2})\log
\frac{2\tilde{m}_{t_{2}}^{2}+\tilde{m}_{b_{2}}^{2}}{3\mu_{o}^{2}}\right.\nonumber\\
&+&\frac{1}{6} (3s_{b}^{2}-4\sw^{2})\log
\frac{\tilde{m}_{t_{2}}^{2}+2\tilde{m}_{b_{2}}^{2}}{3\mu_{o}^{2}}+
\frac{1}{9} (-6s_{b}^{2}-3s_{t}^{2}+5\sw^{2})\log
\frac{\tilde{m}_{t_{2}}^{2}+\tilde{m}_{b_{2}}^{2}}{2\mu_{o}^{2}}\nonumber\\
&+&\frac{4}{9} (3s_{t}^{2}-4\sw^{2})\log
\frac{3\tilde{m}_{t_{2}}^{2}+\tilde{m}_{b_{2}}^{2}}{4\mu_{o}^{2}}+
\left.\frac{2}{9} (3s_{b}^{2}-2\sw^{2})\log
\frac{\tilde{m}_{t_{2}}^{2}+3\tilde{m}_{b_{2}}^{2}}{4\mu_{o}^{2}}\right]\nonumber\\
&+&c_{b}^{\,2}c_{t}^{\,2}s_{t}^{\,2}\left[-\log
\frac{\tilde{m}_{t_{1}}^{2}+\tilde{m}_{t_{2}}^{2}+\tilde{m}_{b_{1}}^{2}}{3\mu_{o}^{2}}
+\frac{4}{3}\log
\frac{2\tilde{m}_{t_{1}}^{2}+\tilde{m}_{t_{2}}^{2}+\tilde{m}_{b_{1}}^{2}}{4\mu_{o}^{2}}
+\frac{4}{3}\log
\frac{\tilde{m}_{t_{1}}^{2}+2\tilde{m}_{t_{2}}^{2}+\tilde{m}_{b_{1}}^{2}}{4\mu_{o}^{2}}
\right.\nonumber\\
&-&\left.\frac{2}{3}\log\frac{\tilde{m}_{t_{1}}^{2}+\tilde{m}_{t_{2}}^{2}+2\tilde{m}_{b_{1}}^{2}}
{4\mu_{o}^{2}}\right]+
s_{b}^{\,2}c_{t}^{\,2}s_{t}^{\,2}\left[-\log
\frac{\tilde{m}_{t_{1}}^{2}+\tilde{m}_{t_{2}}^{2}+\tilde{m}_{b_{2}}^{2}}{3\mu_{o}^{2}}
\right.+\frac{4}{3}\log
\frac{2\tilde{m}_{t_{1}}^{2}+\tilde{m}_{t_{2}}^{2}+\tilde{m}_{b_{2}}^{2}}{4\mu_{o}^{2}}
\nonumber\\
&+&\frac{4}{3}\log
\frac{\tilde{m}_{t_{1}}^{2}+2\tilde{m}_{t_{2}}^{2}+\tilde{m}_{b_{2}}^{2}}{4\mu_{o}^{2}}
-\left.\frac{2}{3}\log\frac{\tilde{m}_{t_{1}}^{2}+\tilde{m}_{t_{2}}^{2}+2\tilde{m}_{b_{2}}^{2}}
{4\mu_{o}^{2}}\right]+s_{b}^{\,2}c_{b}^{\,2}c_{t}^{\,2}\left[\log
\frac{\tilde{m}_{t_{1}}^{2}+\tilde{m}_{b_{1}}^{2}+\tilde{m}_{b_{2}}^{2}}{3\mu_{o}^{2}}
\right.\nonumber\\
&-&\frac{4}{3}\log
\frac{2\tilde{m}_{t_{1}}^{2}+\tilde{m}_{b_{1}}^{2}+\tilde{m}_{b_{2}}^{2}}{4\mu_{o}^{2}}
+\frac{2}{3}\log
\frac{\tilde{m}_{t_{1}}^{2}+2\tilde{m}_{b_{1}}^{2}+\tilde{m}_{b_{2}}^{2}}{4\mu_{o}^{2}}
+\left.\frac{2}{3}\log
\frac{\tilde{m}_{t_{1}}^{2}+\tilde{m}_{b_{1}}^{2}+2\tilde{m}_{b_{2}}^{2}}{4\mu_{o}^{2}}
\right]\nonumber\\
&+&s_{b}^{\,2}c_{b}^{\,2}s_{t}^{\,2}\left[\log
\frac{\tilde{m}_{t_{2}}^{2}+\tilde{m}_{b_{1}}^{2}+\tilde{m}_{b_{2}}^{2}}{3\mu_{o}^{2}}
\right.+\frac{2}{3}\log
\frac{\tilde{m}_{t_{2}}^{2}+2\tilde{m}_{b_{1}}^{2}+\tilde{m}_{b_{2}}^{2}}{4\mu_{o}^{2}}
+\frac{2}{3}\log
\frac{\tilde{m}_{t_{2}}^{2}+\tilde{m}_{b_{1}}^{2}+2\tilde{m}_{b_{2}}^{2}}{4\mu_{o}^{2}}
\nonumber\\
&-&\left.\left.\frac{4}{3}\log
\frac{2\tilde{m}_{t_{2}}^{2}+\tilde{m}_{b_{1}}^{2}+\tilde{m}_{b_{2}}^{2}}{4\mu_{o}^{2}}
\right]\right\}\nonumber\\
\nonumber\\
g_{5}&=& \frac{3}{2}s_{t}^{2}c_{t}^{4}\log \frac{\tilde{m}_{t_{1}}^{2}}{\mu_{o}^{2}}
+\frac{3}{2}s_{t}^{4}c_{t}^{2}\log \frac{\tilde{m}_{t_{2}}^{2}}{\mu_{o}^{2}}+
\frac{3}{2}c_{b}^{4}s_{b}^{2}\log
\frac{\tilde{m}_{b_{1}}^{2}}{\mu_{o}^{2}}+
\frac{3}{2}c_{b}^{2}s_{b}^{4}\log
\frac{\tilde{m}_{b_{2}}^{2}}{\mu_{o}^{2}}\nonumber\\
&-&c_{b}^{2}c_{t}^{2}\left[\left(\frac{3}{2}s_{t}^{2}c_{t}^{2}+6
{\left(\frac{c_{t}^{2}}{2}-\frac{2}{3}\sw^{2}\right)}^{2}\right)\log
\frac{2\tilde{m}_{t_{1}}^{2}+\tilde{m}_{b_{1}}^{2}}{3\mu_{o}^{2}}-
4{\left(\frac{c_{t}^{2}}{2}-\frac{2}{3}\sw^{2}\right)}^{2}\log
\frac{3\tilde{m}_{t_{1}}^{2}+\tilde{m}_{b_{1}}^{2}}{4\mu_{o}^{2}}\right.\nonumber\\
&+&\left(\frac{3}{2}s_{b}^{2}c_{b}^{2}+{\left(-\frac{c_{b}^{2}}{2}+
\frac{1}{3}\sw^{2}\right)}^{2}\right) \log
\frac{\tilde{m}_{t_{1}}^{2}+2\tilde{m}_{b_{1}}^{2}}{3\mu_{o}^{2}}-
4{\left(-\frac{c_{b}^{2}}{2}+\frac{1}{3}\sw^{2}\right)}^{2}\log
\frac{\tilde{m}_{t_{1}}^{2}+3\tilde{m}_{b_{1}}^{2}}{4\mu_{o}^{2}}\nonumber\\
&-&\left.\frac{1}{9}(3c_{t}^{2}-4\sw^{2})(-3c_{b}^{2}+2\sw^{2}) \log
\frac{\tilde{m}_{t_{1}}^{2}+\tilde{m}_{b_{1}}^{2}}{2\mu_{o}^{2}}\right]
-s_{b}^{2}s_{t}^{2}\left[-4{\left(\frac{s_{t}^{2}}{2}-\frac{2}{3}\sw^{2}\right)}^{2}\log
\frac{3\tilde{m}_{t_{2}}^{2}+\tilde{m}_{b_{2}}^{2}}{4\mu_{o}^{2}}\right.\nonumber\\
&+&\left(\frac{3}{2}s_{t}^{2}c_{t}^{2}+6
{\left(\frac{s_{t}^{2}}{2}-\frac{2}{3}\sw^{2}\right)}^{2}\right)\log
\frac{2\tilde{m}_{t_{2}}^{2}+\tilde{m}_{b_{2}}^{2}}{3\mu_{o}^{2}}+
\left(\frac{3}{2}s_{b}^{2}c_{b}^{2}+6{\left(-\frac{s_{b}^{2}}{2}+
\frac{1}{3}\sw^{2}\right)}^{2}\right) \log
\frac{\tilde{m}_{t_{2}}^{2}+2\tilde{m}_{b_{2}}^{2}}{3\mu_{o}^{2}}\nonumber\\
&-&4{\left(-\frac{s_{b}^{2}}{2}+\frac{1}{3}\sw^{2}\right)}^{2}\log
\frac{\tilde{m}_{t_{2}}^{2}+3\tilde{m}_{b_{2}}^{2}}{4\mu_{o}^{2}}-
\left.\frac{1}{9}(3s_{t}^{2}-4\sw^{2})(-3s_{b}^{2}+2\sw^{2}) \log
\frac{\tilde{m}_{t_{2}}^{2}+\tilde{m}_{b_{2}}^{2}}{2\mu_{o}^{2}} 
\right]\nonumber\\
&-&s_{b}^{2}c_{t}^{2}\left[\left(\frac{3}{2}s_{t}^{2}c_{t}^{2}+6
{\left(\frac{c_{t}^{2}}{2}-\frac{2}{3}\sw^{2}\right)}^{2}\right)\log
\frac{2\tilde{m}_{t_{1}}^{2}+\tilde{m}_{b_{2}}^{2}}{3\mu_{o}^{2}}-
4{\left(\frac{c_{t}^{2}}{2}-\frac{2}{3}\sw^{2}\right)}^{2}\log
\frac{3\tilde{m}_{t_{1}}^{2}+\tilde{m}_{b_{2}}^{2}}{4\mu_{o}^{2}}\right.\nonumber\\
&+&\left(\frac{3}{2}s_{b}^{2}c_{b}^{2}+6{\left(-\frac{s_{b}^{2}}{2}+
\frac{1}{3}\sw^{2}\right)}^{2}\right) \log
\frac{\tilde{m}_{t_{1}}^{2}+2\tilde{m}_{b_{2}}^{2}}{3\mu_{o}^{2}}-
4{\left(-\frac{s_{b}^{2}}{2}+\frac{1}{3}\sw^{2}\right)}^{2}\log
\frac{\tilde{m}_{t_{1}}^{2}+3\tilde{m}_{b_{2}}^{2}}{4\mu_{o}^{2}}\nonumber\\
&-&\left.\frac{1}{9}(3c_{t}^{2}-4\sw^{2})(-3s_{b}^{2}+2\sw^{2}) \log
\frac{\tilde{m}_{t_{1}}^{2}+\tilde{m}_{b_{2}}^{2}}{2\mu_{o}^{2}} 
\right]-c_{b}^{2}s_{t}^{2}\left[-
4{\left(\frac{s_{t}^{2}}{2}-\frac{2}{3}\sw^{2}\right)}^{2}\log
\frac{3\tilde{m}_{t_{2}}^{2}+\tilde{m}_{b_{1}}^{2}}{4\mu_{o}^{2}}\right.\nonumber\\
&+&\left(\frac{3}{2}s_{t}^{2}c_{t}^{2}+6
{\left(\frac{s_{t}^{2}}{2}-\frac{2}{3}\sw^{2}\right)}^{2}\right)\log
\frac{2\tilde{m}_{t_{2}}^{2}+\tilde{m}_{b_{1}}^{2}}{3\mu_{o}^{2}}
+\left(\frac{3}{2}s_{b}^{2}c_{b}^{2}+6{\left(-\frac{c_{b}^{2}}{2}+
\frac{1}{3}\sw^{2}\right)}^{2}\right) \log
\frac{\tilde{m}_{t_{2}}^{2}+2\tilde{m}_{b_{1}}^{2}}{3\mu_{o}^{2}}\nonumber\\
&-&4{\left(-\frac{c_{b}^{2}}{2}+\frac{1}{3}\sw^{2}\right)}^{2}\log
\frac{\tilde{m}_{t_{2}}^{2}+3\tilde{m}_{b_{1}}^{2}}{4\mu_{o}^{2}}-
\left.\frac{1}{9}(3s_{t}^{2}-4\sw^{2})(-3c_{b}^{2}+2\sw^{2}) \log
\frac{\tilde{m}_{t_{2}}^{2}+\tilde{m}_{b_{1}}^{2}}{2\mu_{o}^{2}} 
\right]\nonumber\\
&-&s_{t}^{2}c_{t}^{2}\left[-(3-8\sw^{2})\log
\frac{\tilde{m}_{t_{1}}^{2}+\tilde{m}_{t_{2}}^{2}}{2\mu_{o}^{2}}+
\frac{1}{2}(9c_{t}^{2}-8\sw^{2})\log
\frac{2\tilde{m}_{t_{1}}^{2}+\tilde{m}_{t_{2}}^{2}}{3\mu_{o}^{2}}\right.\nonumber\\
&+&\left.\frac{1}{2}(9s_{t}^{2}-8\sw^{2})\log
\frac{\tilde{m}_{t_{1}}^{2}+2\tilde{m}_{t_{2}}^{2}}{3\mu_{o}^{2}}\right]-
s_{b}^{2}c_{b}^{2}\left[-(3-4\sw^{2})\log
\frac{\tilde{m}_{b_{1}}^{2}+\tilde{m}_{b_{2}}^{2}}{2\mu_{o}^{2}}\right.\nonumber\\
&+&\left.\frac{1}{2}(9c_{b}^{2}-4\sw^{2})\log
\frac{2\tilde{m}_{b_{1}}^{2}+\tilde{m}_{b_{2}}^{2}}{3\mu_{o}^{2}}+
\frac{1}{2}(9s_{b}^{2}-4\sw^{2})\log
\frac{\tilde{m}_{b_{1}}^{2}+2\tilde{m}_{b_{2}}^{2}}{3\mu_{o}^{2}}\right]\nonumber\\
&+& 4c_{b}^{2}c_{t}^{2}s_{t}^{2}\left(-\frac{c_{b}^{2}}{2}+
\frac{1}{3}\sw^{2}\right)\log 
\frac{\tilde{m}_{t_{1}}^{2}+\tilde{m}_{t_{2}}^{2}+2\tilde{m}_{b_{1}}^{2}}
{4\mu_{o}^{2}}+4 s_{b}^{2}c_{t}^{2}s_{t}^{2}
\left(-\frac{s_{b}^{2}}{2}+\frac{1}{3} \sw^{2} \right) \log
\frac{\tilde{m}_{t_{1}}^{2}+\tilde{m}_{t_{2}}^{2}+2\tilde{m}_{b_{2}}^{2}}
{4\mu_{o}^{2}}\nonumber\\
&-&4c_{b}^{2}c_{t}^{2}s_{b}^{2}\left(\frac{c_{t}^{2}}{2}-
\frac{2}{3}\sw^{2}\right)\log 
\frac{2\tilde{m}_{t_{1}}^{2}+\tilde{m}_{b_{1}}^{2}+\tilde{m}_{b_{2}}^{2}}
{4\mu_{o}^{2}}-4s_{b}^{2}c_{b}^{2}s_{t}^{2}
\left(\frac{s_{t}^{2}}{2}-\frac{2}{3} \sw^{2} \right) \log
\frac{2\tilde{m}_{t_{2}}^{2}+\tilde{m}_{b_{1}}^{2}+\tilde{m}_{b_{2}}^{2}}
{4\mu_{o}^{2}}\nonumber\\
&+&\frac{1}{3}c_{b}^{2}s_{t}^{2}c_{t}^{2}\left[(9c_{t}^{2}-8\sw^{2})\log 
\frac{2\tilde{m}_{t_{1}}^{2}+\tilde{m}_{t_{2}}^{2}+\tilde{m}_{b_{1}}^{2}}
{4\mu_{o}^{2}}+(9s_{t}^{2}-8\sw^{2})\log 
\frac{\tilde{m}_{t_{1}}^{2}+2\tilde{m}_{t_{2}}^{2}+\tilde{m}_{b_{1}}^{2}}
{4\mu_{o}^{2}}\right.\nonumber\\
&-&\left. 3(3-8\sw^{2})\log
\frac{\tilde{m}_{t_{1}}^{2}+\tilde{m}_{t_{2}}^{2}+\tilde{m}_{b_{1}}^{2}}
{3\mu_{o}^{2}}\right]+
\frac{1}{3}s_{b}^{2}s_{t}^{2}c_{t}^{2}\left[(9c_{t}^{2}-8\sw^{2})\log 
\frac{2\tilde{m}_{t_{1}}^{2}+\tilde{m}_{t_{2}}^{2}+\tilde{m}_{b_{2}}^{2}}
{4\mu_{o}^{2}}\right.\nonumber\\
&+&\left.(9s_{t}^{2}-8\sw^{2})\log 
\frac{\tilde{m}_{t_{1}}^{2}+2\tilde{m}_{t_{2}}^{2}+\tilde{m}_{b_{2}}^{2}}
{4\mu_{o}^{2}}-3(3-8\sw^{2})\log
\frac{\tilde{m}_{t_{1}}^{2}+\tilde{m}_{t_{2}}^{2}+\tilde{m}_{b_{2}}^{2}}
{3\mu_{o}^{2}}\right]\nonumber\\
&+& \frac{1}{3}c_{b}^{2}s_{b}^{2}c_{t}^{2}\left[(9c_{b}^{2}-4\sw^{2})\log 
\frac{\tilde{m}_{t_{1}}^{2}+2\tilde{m}_{b_{1}}^{2}+\tilde{m}_{b_{2}}^{2}}
{4\mu_{o}^{2}}+(9s_{b}^{2}-4\sw^{2})\log 
\frac{\tilde{m}_{t_{1}}^{2}+\tilde{m}_{b_{1}}^{2}+2\tilde{m}_{b_{2}}^{2}}
{4\mu_{o}^{2}}\right.\nonumber\\
&-&\left.3(3-4\sw^{2})\log
\frac{\tilde{m}_{t_{1}}^{2}+\tilde{m}_{b_{1}}^{2}+\tilde{m}_{b_{2}}^{2}}
{3\mu_{o}^{2}}\right]
+\frac{1}{3}c_{b}^{2}s_{b}^{2}s_{t}^{2}\left[(9c_{b}^{2}-4\sw^{2})\log 
\frac{\tilde{m}_{t_{2}}^{2}+2\tilde{m}_{b_{1}}^{2}+\tilde{m}_{b_{2}}^{2}}
{4\mu_{o}^{2}}\right.\nonumber\\
&+&\left.(9s_{b}^{2}-4\sw^{2})\log 
\frac{\tilde{m}_{t_{2}}^{2}+\tilde{m}_{b_{1}}^{2}+2\tilde{m}_{b_{2}}^{2}}
{4\mu_{o}^{2}}-3(3-4\sw^{2})\log
\frac{\tilde{m}_{t_{2}}^{2}+\tilde{m}_{b_{1}}^{2}+\tilde{m}_{b_{2}}^{2}}
{3\mu_{o}^{2}}\right]\nonumber\\
&-& 4c_{b}^{2}c_{t}^{2}s_{b}^{2}s_{t}^{2}\log
\frac{\tilde{m}_{t_{1}}^{2}+\tilde{m}_{t_{2}}^{2}+\tilde{m}_{b_{1}}^{2}+
\tilde{m}_{b_{2}}^{2}}{4\mu_{o}^{2}}\,,\nonumber\\
\nonumber\\
g_{6}&=& c_{b}^{2}c_{t}^{2}\left[\frac{\sw^{4}}{3}\log
\frac{\tilde{m}_{t_{1}}^{2}+\tilde{m}_{b_{1}}^{2}}{2\mu_{o}^{2}}+
\frac{\sw^{2}}{3}(3c_{t}^{2}-4\sw^{2})\log
\frac{2\tilde{m}_{t_{1}}^{2}+\tilde{m}_{b_{1}}^{2}}{3\mu_{o}^{2}}\right.\nonumber\\
&+&2\sw^{2}\left(-\frac{c_{b}^{2}}{2}+\frac{\sw^{2}}{3}\right)\log
\frac{\tilde{m}_{t_{1}}^{2}+2\tilde{m}_{b_{1}}^{2}}{3\mu_{o}^{2}}+
\frac{1}{9}(3c_{t}^{2}-4\sw^{2})(-3c_{b}^{2}+2\sw^{2}) \log
\frac{\tilde{m}_{t_{1}}^{2}+\tilde{m}_{b_{1}}^{2}}{2\mu_{o}^{2}}\nonumber\\
&+&\left.4{\left(\frac{c_{t}^{2}}{2}-\frac{2}{3}\sw^{2}\right)}^{2}\log
\frac{3\tilde{m}_{t_{1}}^{2}+\tilde{m}_{b_{1}}^{2}}{4\mu_{o}^{2}}
+4{\left(-\frac{c_{b}^{2}}{2}+\frac{\sw^{2}}{3}\right)}^{2}\log
\frac{\tilde{m}_{t_{1}}^{2}+3\tilde{m}_{b_{1}}^{2}}{4\mu_{o}^{2}}\right]\nonumber\\
&+& c_{b}^{2}s_{t}^{2}\left[\frac{\sw^{4}}{3}\log
\frac{\tilde{m}_{t_{2}}^{2}+\tilde{m}_{b_{1}}^{2}}{2\mu_{o}^{2}}+
\frac{\sw^{2}}{3}(3s_{t}^{2}-4\sw^{2})\log
\frac{2\tilde{m}_{t_{2}}^{2}+\tilde{m}_{b_{1}}^{2}}{3\mu_{o}^{2}}\right.\nonumber\\
&+&2\sw^{2}\left(-\frac{c_{b}^{2}}{2}+\frac{\sw^{2}}{3}\right)\log
\frac{\tilde{m}_{t_{2}}^{2}+2\tilde{m}_{b_{1}}^{2}}{3\mu_{o}^{2}}+
\frac{1}{9}(3s_{t}^{2}-4\sw^{2})(-3c_{b}^{2}+2\sw^{2}) \log
\frac{\tilde{m}_{t_{2}}^{2}+\tilde{m}_{b_{1}}^{2}}{2\mu_{o}^{2}}\nonumber\\
&+&\left.4{\left(\frac{s_{t}^{2}}{2}-\frac{2}{3}\sw^{2}\right)}^{2}\log
\frac{3\tilde{m}_{t_{2}}^{2}+\tilde{m}_{b_{1}}^{2}}{4\mu_{o}^{2}}
+4{\left(-\frac{c_{b}^{2}}{2}+\frac{\sw^{2}}{3}\right)}^{2}\log
\frac{\tilde{m}_{t_{2}}^{2}+3\tilde{m}_{b_{1}}^{2}}{4\mu_{o}^{2}}\right]\nonumber\\
&+& s_{b}^{2}c_{t}^{2}\left[\frac{\sw^{4}}{3}\log
\frac{\tilde{m}_{t_{1}}^{2}+\tilde{m}_{b_{2}}^{2}}{2\mu_{o}^{2}}+
\frac{\sw^{2}}{3}(3c_{t}^{2}-4\sw^{2})\log
\frac{2\tilde{m}_{t_{1}}^{2}+\tilde{m}_{b_{2}}^{2}}{3\mu_{o}^{2}}\right.\nonumber\\
&+&2\sw^{2}\left(-\frac{s_{b}^{2}}{2}+\frac{\sw^{2}}{3}\right)\log
\frac{\tilde{m}_{t_{1}}^{2}+2\tilde{m}_{b_{2}}^{2}}{3\mu_{o}^{2}}+
\frac{1}{9}(3c_{t}^{2}-4\sw^{2})(-3s_{b}^{2}+2\sw^{2}) \log
\frac{\tilde{m}_{t_{1}}^{2}+\tilde{m}_{b_{2}}^{2}}{2\mu_{o}^{2}}\nonumber\\
&+&\left.4{\left(\frac{c_{t}^{2}}{2}-\frac{2}{3}\sw^{2}\right)}^{2}\log
\frac{3\tilde{m}_{t_{1}}^{2}+\tilde{m}_{b_{2}}^{2}}{4\mu_{o}^{2}}
+4{\left(-\frac{s_{b}^{2}}{2}+\frac{\sw^{2}}{3}\right)}^{2}\log
\frac{\tilde{m}_{t_{1}}^{2}+3\tilde{m}_{b_{2}}^{2}}{4\mu_{o}^{2}}\right]\nonumber\\
&+& s_{b}^{2}s_{t}^{2}\left[\frac{\sw^{4}}{3}\log
\frac{\tilde{m}_{t_{2}}^{2}+\tilde{m}_{b_{2}}^{2}}{2\mu_{o}^{2}}+
\frac{\sw^{2}}{3}(3s_{t}^{2}-4\sw^{2})\log
\frac{2\tilde{m}_{t_{2}}^{2}+\tilde{m}_{b_{2}}^{2}}{3\mu_{o}^{2}}\right.\nonumber\\
&+&2\sw^{2}\left(-\frac{s_{b}^{2}}{2}+\frac{\sw^{2}}{3}\right)\log
\frac{\tilde{m}_{t_{2}}^{2}+2\tilde{m}_{b_{2}}^{2}}{3\mu_{o}^{2}}+
\frac{1}{9}(3s_{t}^{2}-4\sw^{2})(-3s_{b}^{2}+2\sw^{2}) \log
\frac{\tilde{m}_{t_{2}}^{2}+\tilde{m}_{b_{2}}^{2}}{2\mu_{o}^{2}}\nonumber\\
&+&\left.4{\left(\frac{s_{t}^{2}}{2}-\frac{2}{3}\sw^{2}\right)}^{2}\log
\frac{3\tilde{m}_{t_{2}}^{2}+\tilde{m}_{b_{2}}^{2}}{4\mu_{o}^{2}}
+4{\left(-\frac{s_{b}^{2}}{2}+\frac{\sw^{2}}{3}\right)}^{2}\log
\frac{\tilde{m}_{t_{2}}^{2}+3\tilde{m}_{b_{2}}^{2}}{4\mu_{o}^{2}}\right]\nonumber\\
&+& 4c_{b}^{2}c_{t}^{2}s_{t}^{2}\left(-\frac{c_{b}^{2}}{2}+
\frac{\sw^{2}}{3}\right)\log 
\frac{\tilde{m}_{t_{1}}^{2}+\tilde{m}_{t_{2}}^{2}+2\tilde{m}_{b_{1}}^{2}}
{4\mu_{o}^{2}}+4 s_{b}^{2}c_{t}^{2}s_{t}^{2}
\left(-\frac{s_{b}^{2}}{2}+\frac{\sw^{2}}{3} \right) \log
\frac{\tilde{m}_{t_{1}}^{2}+\tilde{m}_{t_{2}}^{2}+2\tilde{m}_{b_{2}}^{2}}
{4\mu_{o}^{2}}\nonumber\\
&-&4c_{b}^{2}c_{t}^{2}s_{b}^{2}\left(\frac{c_{t}^{2}}{2}-
\frac{2}{3}\sw^{2}\right)\log 
\frac{2\tilde{m}_{t_{1}}^{2}+\tilde{m}_{b_{1}}^{2}+\tilde{m}_{b_{2}}^{2}}
{4\mu_{o}^{2}}-4s_{b}^{2}c_{b}^{2}s_{t}^{2}
\left(\frac{s_{t}^{2}}{2}-\frac{2}{3} \sw^{2} \right) \log
\frac{2\tilde{m}_{t_{2}}^{2}+\tilde{m}_{b_{1}}^{2}+\tilde{m}_{b_{2}}^{2}}
{4\mu_{o}^{2}}\nonumber\\
&+& \frac{1}{3}c_{b}^{2}s_{t}^{2}c_{t}^{2}\left[6\sw^{2}\log
\frac{\tilde{m}_{t_{1}}^{2}+\tilde{m}_{t_{2}}^{2}+\tilde{m}_{b_{1}}^{2}}{3\mu_{o}^{2}}
+(9c_{t}^{2}-8\sw^{2})\log 
\frac{2\tilde{m}_{t_{1}}^{2}+\tilde{m}_{t_{2}}^{2}+\tilde{m}_{b_{1}}^{2}}
{4\mu_{o}^{2}}\right.\nonumber\\
&+&\left.(9s_{t}^{2}-8\sw^{2})\log 
\frac{\tilde{m}_{t_{1}}^{2}+2\tilde{m}_{t_{2}}^{2}+\tilde{m}_{b_{1}}^{2}}
{4\mu_{o}^{2}}\right]+
\frac{1}{3}s_{b}^{2}s_{t}^{2}c_{t}^{2}\left[6\sw^{2}\log
\frac{\tilde{m}_{t_{1}}^{2}+\tilde{m}_{t_{2}}^{2}+\tilde{m}_{b_{2}}^{2}}{3\mu_{o}^{2}}
\right.\nonumber\\
&+&\left.(9c_{t}^{2}-8\sw^{2})\log 
\frac{2\tilde{m}_{t_{1}}^{2}+\tilde{m}_{t_{2}}^{2}+\tilde{m}_{b_{2}}^{2}}
{4\mu_{o}^{2}}+(9s_{t}^{2}-8\sw^{2})\log 
\frac{\tilde{m}_{t_{1}}^{2}+2\tilde{m}_{t_{2}}^{2}+\tilde{m}_{b_{2}}^{2}}
{4\mu_{o}^{2}}\right]\nonumber\\
&+& \frac{1}{3}c_{b}^{2}s_{b}^{2}c_{t}^{2}\left[-6\sw^{2}\log
\frac{\tilde{m}_{t_{1}}^{2}+\tilde{m}_{b_{1}}^{2}+\tilde{m}_{b_{2}}^{2}}{3\mu_{o}^{2}}
+(9c_{b}^{2}-4\sw^{2})\log 
\frac{\tilde{m}_{t_{1}}^{2}+2\tilde{m}_{b_{1}}^{2}+\tilde{m}_{b_{2}}^{2}}
{4\mu_{o}^{2}}\right.\nonumber\\
&+&\left.(9s_{b}^{2}-4\sw^{2})\log 
\frac{\tilde{m}_{t_{1}}^{2}+\tilde{m}_{b_{1}}^{2}+2\tilde{m}_{b_{2}}^{2}}
{4\mu_{o}^{2}}\right]+\frac{1}{3}c_{b}^{2}s_{b}^{2}s_{t}^{2}\left[-6\sw^{2}\log
\frac{\tilde{m}_{t_{2}}^{2}+\tilde{m}_{b_{1}}^{2}+\tilde{m}_{b_{2}}^{2}}{3\mu_{o}^{2}}
\right.\nonumber\\
&+&\left.(9c_{b}^{2}-4\sw^{2})\log 
\frac{\tilde{m}_{t_{2}}^{2}+2\tilde{m}_{b_{1}}^{2}+\tilde{m}_{b_{2}}^{2}}
{4\mu_{o}^{2}}+(9s_{b}^{2}-4\sw^{2})\log 
\frac{\tilde{m}_{t_{2}}^{2}+\tilde{m}_{b_{1}}^{2}+2\tilde{m}_{b_{2}}^{2}}
{4\mu_{o}^{2}}\right]\nonumber\\
&-& 4c_{b}^{2}c_{t}^{2}s_{b}^{2}s_{t}^{2}\log
\frac{\tilde{m}_{t_{1}}^{2}+\tilde{m}_{t_{2}}^{2}+\tilde{m}_{b_{1}}^{2}+
\tilde{m}_{b_{2}}^{2}}{4\mu_{o}^{2}}\,,\nonumber\\
\nonumber\\
g_{7}&=& \frac{3}{2}c_{t}^{\,4}\log
\frac{\tilde{m}_{t_{1}}^{2}}{\mu_{o}^{2}}+
\frac{3}{2}s_{t}^{\,4}\log
\frac{\tilde{m}_{t_{2}}^{2}}{\mu_{o}^{2}}+
\frac{3}{2}c_{b}^{\,4}\log
\frac{\tilde{m}_{b_{1}}^{2}}{\mu_{o}^{2}}+
\frac{3}{2}s_{b}^{\,4}\log
\frac{\tilde{m}_{b_{2}}^{2}}{\mu_{o}^{2}}+3s_{t}^{2}c_{t}^{2} \log \frac 
{\tilde{m}_{t_{1}}^{2}+\tilde{m}_{t_{2}}^{2}}{2\mu_{o}^{2}}\nonumber\\
&+&c_{t}^{4}c_{b}^{4} \log
\frac{\tilde{m}_{t_{1}}^{2}+\tilde{m}_{b_{1}}^{2}}{2\mu_{o}^{2}}+
s_{t}^{4}c_{b}^{4} \log
\frac{\tilde{m}_{t_{2}}^{2}+\tilde{m}_{b_{1}}^{2}}{2\mu_{o}^{2}}-
3c_{b}^{2}c_{t}^{4} \log
\frac{2\tilde{m}_{t_{1}}^{2}+\tilde{m}_{b_{1}}^{2}}{3\mu_{o}^{2}}-
3s_{t}^{4}c_{b}^{2} \log
\frac{2\tilde{m}_{t_{2}}^{2}+\tilde{m}_{b_{1}}^{2}}{3\mu_{o}^{2}}\nonumber\\
&-& 3c_{t}^{2}c_{b}^{4} \log
\frac{\tilde{m}_{t_{1}}^{2}+2\tilde{m}_{b_{1}}^{2}}{3\mu_{o}^{2}}-
3s_{t}^{2}c_{b}^{4} \log
\frac{\tilde{m}_{t_{2}}^{2}+2\tilde{m}_{b_{1}}^{2}}{3\mu_{o}^{2}}
-6c_{b}^{2}s_{t}^{2}c_{t}^{2}\log 
\frac{\tilde{m}_{t_{1}}^{2}+\tilde{m}_{t_{2}}^{2}+\tilde{m}_{b_{1}}^{2}}
{3\mu_{o}^{2}}\nonumber\\
&+&2c_{b}^{4}s_{t}^{2}c_{t}^{2}\log 
\frac{\tilde{m}_{t_{1}}^{2}+\tilde{m}_{t_{2}}^{2}+2\tilde{m}_{b_{1}}^{2}}{4\mu_{o}^{2}}+
c_{t}^{4}s_{b}^{4} \log
\frac{\tilde{m}_{t_{1}}^{2}+\tilde{m}_{b_{2}}^{2}}{2\mu_{o}^{2}}-
3c_{t}^{4}s_{b}^{2} \log
\frac{2\tilde{m}_{t_{1}}^{2}+\tilde{m}_{b_{2}}^{2}}{3\mu_{o}^{2}}\nonumber\\ 
&+&s_{t}^{4}s_{b}^{4} \log
\frac{\tilde{m}_{t_{2}}^{2}+\tilde{m}_{b_{2}}^{2}}{2\mu_{o}^{2}}
-3s_{t}^{4}s_{b}^{2} \log
\frac{2\tilde{m}_{t_{2}}^{2}+\tilde{m}_{b_{2}}^{2}}{3\mu_{o}^{2}}-
6c_{t}^{2}s_{t}^{2}s_{b}^{2}\log 
\frac{\tilde{m}_{t_{1}}^{2}+\tilde{m}_{t_{2}}^{2}+\tilde{m}_{b_{2}}^{2}}{3\mu_{o}^{2}}\nonumber\\
&+&3c_{b}^{2}s_{b}^{2} \log
\frac{\tilde{m}_{b_{1}}^{2}+\tilde{m}_{b_{2}}^{2}}{2\mu_{o}^{2}}-
3c_{t}^{2}s_{b}^{4} \log
\frac{\tilde{m}_{t_{1}}^{2}+2\tilde{m}_{b_{2}}^{2}}{3\mu_{o}^{2}}-
3s_{b}^{4}s_{t}^{2} \log
\frac{\tilde{m}_{t_{2}}^{2}+2\tilde{m}_{b_{2}}^{2}}{3\mu_{o}^{2}}\nonumber\\
&-&6c_{b}^{2}s_{b}^{2}c_{t}^{2}\log 
\frac{\tilde{m}_{t_{1}}^{2}+\tilde{m}_{b_{1}}^{2}+\tilde{m}_{b_{2}}^{2}}
{3\mu_{o}^{2}}-
6c_{b}^{2}s_{b}^{2}s_{t}^{2}\log 
\frac{\tilde{m}_{t_{2}}^{2}+\tilde{m}_{b_{1}}^{2}+\tilde{m}_{b_{2}}^{2}}{3\mu_{o}^{2}}\nonumber\\
&+&2c_{b}^{2}s_{b}^{2}c_{t}^{4}\log 
\frac{2\tilde{m}_{t_{1}}^{2}+\tilde{m}_{b_{1}}^{2}+\tilde{m}_{b_{2}}^{2}}{4\mu_{o}^{2}}
+2c_{b}^{2}s_{b}^{2}s_{t}^{4}\log 
\frac{2\tilde{m}_{t_{2}}^{2}+\tilde{m}_{b_{1}}^{2}+\tilde{m}_{b_{2}}^{2}}
{4\mu_{o}^{2}}\nonumber\\
&+& 2c_{t}^{2}s_{t}^{2}s_{b}^{4}\log 
\frac{\tilde{m}_{t_{1}}^{2}+\tilde{m}_{t_{2}}^{2}+2\tilde{m}_{b_{2}}^{2}}{4\mu_{o}^{2}}+
4c_{b}^{2}s_{b}^{2}c_{t}^{2}s_{t}^{2}\log 
\frac{\tilde{m}_{t_{1}}^{2}+\tilde{m}_{t_{2}}^{2}+\tilde{m}_{b_{1}}^{2}
+\tilde{m}_{b_{2}}^{2}}{4\mu_{o}^{2}}\,,\nonumber\\
\nonumber\\
g_{8}&=& c_{t}^{4}c_{b}^{4} \log
\frac{\tilde{m}_{t_{1}}^{2}+\tilde{m}_{b_{1}}^{2}}{2\mu_{o}^{2}}+
s_{t}^{4}c_{b}^{4} \log
\frac{\tilde{m}_{t_{2}}^{2}+\tilde{m}_{b_{1}}^{2}}{2\mu_{o}^{2}}+
s_{b}^{4}c_{t}^{4} \log
\frac{\tilde{m}_{t_{1}}^{2}+\tilde{m}_{b_{2}}^{2}}{2\mu_{o}^{2}}+
s_{t}^{4}s_{b}^{4} \log
\frac{\tilde{m}_{t_{2}}^{2}+\tilde{m}_{b_{2}}^{2}}{4\mu_{o}^{2}}\nonumber\\
&+&2c_{b}^{4}s_{t}^{2}c_{t}^{2}\log 
\frac{\tilde{m}_{t_{1}}^{2}+\tilde{m}_{t_{2}}^{2}+2\tilde{m}_{b_{1}}^{2}}{4\mu_{o}^{2}}+
2c_{t}^{2}s_{t}^{2}s_{b}^{4}\log 
\frac{\tilde{m}_{t_{1}}^{2}+\tilde{m}_{t_{2}}^{2}+2\tilde{m}_{b_{2}}^{2}}{4\mu_{o}^{2}}\nonumber\\
&+&2c_{b}^{2}s_{b}^{2}c_{t}^{4}\log 
\frac{2\tilde{m}_{t_{1}}^{2}+\tilde{m}_{b_{1}}^{2}+\tilde{m}_{b_{2}}^{2}}
{4\mu_{o}^{2}}+
2c_{b}^{2}s_{b}^{2}s_{t}^{4}\log 
\frac{2\tilde{m}_{t_{2}}^{2}+\tilde{m}_{b_{1}}^{2}+\tilde{m}_{b_{2}}^{2}}{4\mu_{o}^{2}}\nonumber\\
&+&4c_{b}^{2}s_{b}^{2}c_{t}^{2}s_{t}^{2}\log 
\frac{\tilde{m}_{t_{1}}^{2}+\tilde{m}_{t_{2}}^{2}+\tilde{m}_{b_{1}}^{2}
+\tilde{m}_{b_{2}}^{2}}{4\mu_{o}^{2}}\,,\nonumber\\
g_{9}&=& \log \frac{3\mcu+\mnd}{4\mu_{o}^{2}} +\frac{1}{4} \log \frac{3\mcd+\mnt}{4\mu_{o}^{2}}
+\frac{1}{4} \log \frac{3\mcd+\mnc}{4\mu_{o}^{2}}\nonumber\\
\nonumber\\
g_{10}&=& \cw^{2}\log \frac{3\mcu+\mnd}{4\mu_{o}^{2}}
+\frac{1}{4}\log \frac{2\mcd+\mnt+\mnc}{4\mu_{o}^{2}}\nonumber\\
&-& \frac{1}{8}(2\sw^{2}-1) \left( \log \frac{3\mcd+\mnt}{4\mu_{o}^{2}}
+ \log \frac{3\mcd+\mnc}{4\mu_{o}^{2}}\right)\,,\nonumber\\
\nonumber\\
g_{11}&=& -4\cw^{4}\log \frac{3\mcu+\mnd}{4\mu_{o}^{2}}-
\frac{1}{4}{(2\sw^{2}-1)}^{2} \left( \log \frac{3\mcd+\mnt}{4\mu_{o}^{2}}
+ \log \frac{3\mcd+\mnc}{4\mu_{o}^{2}}\right)\nonumber\\
&+&(2\sw^{2}-1) \log \frac{2\mcd+\mnt+\mnc}{4\mu_{o}^{2}}
-\frac{1}{4} \log \frac{\mcd+2\mnt+\mnc}{4\mu_{o}^{2}}\nonumber\\
&-&\frac{1}{4} \log \frac{\mcd+\mnt+2\mnc}{4\mu_{o}^{2}}\nonumber\\
\nonumber\\
g_{12}&=& -8 \log \frac{\mcu+\mnd}{2\mu_{o}^{2}} - 
\frac{1}{2}\log \frac{\mcd+\mnt}{2\mu_{o}^{2}}
-\frac{1}{2}\log \frac{\mcd+\mnc}{2\mu_{o}^{2}}\nonumber\\
&-&\log \frac{2\mcd+\mnt+\mnc}{4\mu_{o}^{2}}\,.\nonumber\\
\end{eqnarray}

\section*{Appendix C.}
\vspace{0.4cm}
\setcounter{equation}{0}
\renewcommand{\theequation}{C.\arabic{equation}}

This appendix is devoted to present the exact results to one loop for the  
three and four point sfermions contributions as well as the three point
{\it inos} contributions which are denoted in the text by
${\Delta {\Gamma}_{\mu \,\nu \,\sigma\hspace*{0.4cm}\tilde{q}}^{A W^+ W^-}}$,
${\Delta {\Gamma}_{\mu \,\nu \,\sigma\hspace*{0.4cm}\tilde{q}}^{Z W^+ W^-}}$,
and ${\Delta {\Gamma}_{\mu \,\nu \,\sigma\,\lambda
\hspace*{0.4cm}\tilde{q}}^{A A W^+ W^-}}$, 
${\Delta {\Gamma}_{\mu \,\nu \,\sigma\,\lambda
\hspace*{0.4cm}\tilde{q}}^{A Z W^+ W^-}}$,
${\Delta {\Gamma}_{\mu \,\nu \,\sigma\,\lambda
\hspace*{0.4cm}\tilde{q}}^{Z Z W^+ W^-}}$,
$\,{\Delta {\Gamma}_{\mu \,\nu \,\sigma\,\lambda
\hspace*{0.4cm}\tilde{q}}^{W^+ W^- W^+ W^-}}$ and 
${\Delta {\Gamma}_{\,\mu \,\nu \,\sigma\hspace*{0.4cm}{\sg}}^{A W^+ W^-}}$,
${\Delta {\Gamma}_{\,\mu \,\nu \,\sigma\hspace*{0.4cm}{\sg}}^{Z W^+ W^-}}$
respectively. The exact formulae for the four point {\it inos} contributions
have also been computed by us and are available upon request, but they are not
shown here due to their extreme length. The momenta
assignements for the external gauge bosons are $\,V^{\mu}_{1}(p)\, 
V^{\nu}_{2}(k)\,V^{\sigma}_{3}(r)$ in the three point functions,
${\Delta {\Gamma}_{\,\mu \,\nu \,\sigma}^{V_{1}V_{2}V_{3}}}$, and
$V^{\mu}_{1}(p)\, V^{\nu}_{2}(q)\,V^{\sigma}_{3}(r)\,V^{\lambda}_{4}(t)$ in
the four point functions, 
${\Delta {\Gamma}_{\mu \,\nu \,\sigma\,\lambda}^{V_{1}V_{2}V_{3}V_{4}}}$, and
the convention is with all the external momenta outgoing. For shortness we omit
here the arguments in the Feynman integrals since the notation we have chosen
for them is self-explanatory. Thus, for instance, 
$$J^{a\,b\,c\,d}_{\mu\,\nu\,\sigma\,\lambda}\equiv
J^{a\,b\,c\,d}_{\mu\,\nu\,\sigma\,\lambda}(p,k,r,t,\tilde{m}_{f_a}, 
\tilde{m}_{f_b},\tilde{m}_{f_c})\,\,,\,\,
J^{a\,b\,c\,d}_{\lambda\,\nu\,\sigma\,\mu}\equiv
J^{a\,b\,c\,d}_{\lambda\,\nu\,\sigma\,\mu}(t,k,r,p,\tilde{m}_{f_a}, 
\tilde{m}_{f_b},\tilde{m}_{f_c})\,,$$ and so on.

$\bullet$ Three point sfermions contributions.
\begin{eqnarray}
\label{eq:AWWexact}
\displaystyle &&{\Delta {\Gamma}_{\mu \,\nu \,\sigma\hspace*{0.4cm}\tilde{q}}
^{A W^+ W^-}}=
-eg^{2} N_{c} \frac{\pi^{2}}{2}\, \sum_{\tilde{f}}\left\{ 
\sum_{a,b}\left[ \,(\widehat{Q}_{f})_{a\, b}
(\Sigma_{f})_{b\, a} T^{a\,b}_{\mu}\,g_{\nu \sigma}+\frac{1}{3}\right.
\left((\Sigma_{f}^{t\,b})_{a\, b} (\Sigma_{f}^{b\,t})_{b\, a}
T^{a\,b}_{\sigma}\,g_{\mu \sigma}\right.\right.\nonumber\\
&&+\left.\left.
(\Sigma_{f}^{b\,t})_{a\, b} (\Sigma_{f}^{t\,b})_{b\, a}
T^{a\,b}_{\nu}\,g_{\mu \nu}\right)\right]
-\frac{1}{3} \sum_{a,b,c}\, \left[\,(\widehat{Q}_{f})_{a\, b}
(\Sigma_{f}^{t\,b})_{b\, c} (\Sigma_{f}^{b\,t})_{c\, a}
T^{a\,b\,c}_{\mu\,\nu\,\sigma}\right.\nonumber\\ 
&&+ (\widehat{Q}_{f})_{a\, b}
(\Sigma_{f}^{b\,t})_{b\, c} (\Sigma_{f}^{t\,b})_{c\, a}
T^{a\,b\,c}_{\mu\,\sigma\,\nu}+
(\Sigma_{f}^{t\,b})_{a\, b} (\widehat{Q}_{f})_{b\, c}
 (\Sigma_{f}^{b\,t})_{c\, a}T^{a\,b\,c}_{\nu\,\mu\,\sigma}
 +(\Sigma_{f}^{b\,t})_{a\, b}(\widehat{Q}_{f})_{b\, c}
 (\Sigma_{f}^{t\,b})_{c\, a}T^{a\,b\,c}_{\sigma\,\mu\,\nu}\nonumber\\
 \nonumber\\
&&+(\Sigma_{f}^{t\,b})_{a\, b}(\Sigma_{f}^{b\,t})_{b\, c}
(\widehat{Q}_{f})_{c\,a}T^{a\,b\,c}_{\nu\,\sigma\,\mu}
 +\left.\left.(\Sigma_{f}^{b\,t})_{a\, b} 
(\Sigma_{f}^{t\,b})_{a\, b} (\widehat{Q}_{f})_{a\, b}\,
T^{a\,b\, c}_{\sigma\,\nu \,\mu}\,\right]\,\right\}\,\,,\\
\nonumber\\
\label{eq:gammaZWWexact} 
\displaystyle&& {\Delta {\Gamma}_{\mu \,\nu \,\sigma\hspace*{0.4cm}\tilde{q}}
^{Z W^+ W^-}}=
-g^{3} N_{c} \frac{\pi^{2}}{2\,\cw}\, \sum_{\tilde{f}}\left\{ 
\sum_{a,b} \left[\,(\widehat{G}_{f})_{a\, b}
(\Sigma_{f})_{a\, b} T^{a\,b}_{\mu}\,g_{\nu \sigma}\right.\right.-
\frac{1}{3}\sw^{2}\left((\Sigma_{f}^{t\,b})_{a\, b} (\Sigma_{f}^{b\,t})_{b\, a}
T^{a\,b}_{\sigma}\,g_{\mu \sigma}\right.\nonumber\\
&&+\left.\left.
(\Sigma_{f}^{b\,t})_{a\, b} (\Sigma_{f}^{t\,b})_{b\, a}
T^{a\,b}_{\nu}\,g_{\mu \nu}\right)\right]
-\frac{1}{3} \sum_{a,b,c}\,\left[(\widehat{G}_{f})_{a\, b}
(\Sigma_{f}^{t\,b})_{b\, c} (\Sigma_{f}^{b\,t})_{c\, a}\,
T^{a\,b\,c}_{\mu\,\nu\,\sigma}\right.\nonumber\\ 
&& +(\widehat{G}_{f})_{a\, b}
(\Sigma_{f}^{b\,t})_{b\, c} (\Sigma_{f}^{t\,b})_{c\, a}
T^{a\,b\,c}_{\sigma\,\mu\,\nu}
+ (\Sigma_{f}^{t\,b})_{a\, b} (\widehat{G}_{f})_{b\, c}
 (\Sigma_{f}^{b\,t})_{c\, a}T^{a\,b\,c}_{\nu\,\mu\,\sigma}
 +(\Sigma_{f}^{b\,t})_{a\, b}(\widehat{G}_{f})_{b\, c}
 (\Sigma_{f}^{t\,b})_{c\, a}T^{a\,b\,c}_{\sigma\,\mu\,\nu}\nonumber\\
 \nonumber\\
 &&+ (\Sigma_{f}^{t\,b})_{a\, b} 
 (\Sigma_{f}^{b\,t})_{b\, c}(\widehat{G}_{f})_{c\, a}
T^{a\,b\,c}_{\nu\,\sigma\,\mu} 
 +\left.\left.(\Sigma_{f}^{b\,t})_{a\, b} 
(\Sigma_{f}^{t\,b})_{b\, c} (\widehat{G}_{f})_{c\, a}\right]
\, T^{a\,b\, c}_{\sigma\,\nu \,\mu}\,\right\}\,\,.
\end{eqnarray}

$\bullet$ Four point sfermions contributions.
\begin{eqnarray}
\label{eq:AAWWexact}
\displaystyle && {\Delta {\Gamma}_{\mu \,\nu \,\sigma\,\lambda
\hspace*{0.4cm}\tilde{q}}^{A A W^+ W^-}}=
e^{2} g^{2} {\pi^{2}} N_{c} \, \sum_{\tilde{f}}\left\{ \frac{1}{2}
\sum_{a,b} \left[ \,\left((\widehat{Q}^{2}_{f})_{a\, b}
(\Sigma_{f})_{b\, a}+(\Sigma_{f})_{a\, b}(\widehat{Q}^{2}_{f})_{b\, a}\right)
 \,g_{\mu \nu}\,g_{\sigma \lambda}J^{a\,b}_{p+k}\right.\right.\nonumber\\
&& + \frac{1}{9}\left(
(\Sigma_{f}^{t\,b})_{a\, b}(\Sigma_{f}^{b\,t})_{b\, a}
\,g_{\mu \sigma}\,g_{\nu \lambda}\,J^{a\,b}_{p+r}+
(\Sigma_{f}^{b\,t})_{a\, b}(\Sigma_{f}^{t\,b})_{b\, a}
\,g_{\mu \lambda}\,g_{\nu \sigma}\,J^{a\,b}_{p+t}\right)\left.\right]\nonumber\\
&& - \sum_{a,b,c}\, \left[ 
(\widehat{Q}_{f})_{a\, b} (\widehat{Q}_{f})_{b\, c}(\Sigma_{f})_{c\, a}\,
J^{a\,b\, c}_{\mu \,\nu} \,g_{\sigma \lambda}+
(\Sigma_{f}^{t\,b})_{a\, b} (\Sigma_{f}^{b\,t})_{b\, c}
(\widehat{Q}^{2}_{f})_{c\, a}\,J^{a\,b\, c}_{\lambda\,\sigma}
\,g_{\mu\nu}\right.\nonumber\\ 
&&+ (\Sigma_{f}^{b\,t})_{a\, b} (\Sigma_{f}^{t\,b})_{b\, c}
(\widehat{Q}^{2}_{f})_{c\, a}J^{a\,b\, c}_{\sigma\,\lambda} \,g_{\mu\nu}
+\frac{1}{3} \left( (\widehat{Q}_{f})_{a\, b}
(\Sigma_{f}^{b\,t})_{b\, c} (\Sigma_{f}^{t\,b})_{c\, a}
J^{a\,b\, c}_{\mu \,\lambda} \,g_{\nu \sigma}\right.\nonumber\\
&&+ (\Sigma_{f}^{t\,b})_{a\, b} (\widehat{Q}_{f})_{b\, c}
 (\Sigma_{f}^{b\,t})_{c\, a}\,J^{a\,b\, c}_{\sigma \,\nu} 
 g_{\mu \lambda}+ (\Sigma_{f}^{b\,t})_{a\, b} 
 (\widehat{Q}_{f})_{b\, c}(\Sigma_{f}^{t\,b})_{c\, a}
J^{a\,b\, c}_{\lambda\,\nu} g_{\mu \sigma}
\left.\left. +(\widehat{Q}_{f})_{a\, b}(\Sigma_{f}^{t\,b})_{b\, c} 
 (\Sigma_{f}^{b\,t})_{c\, a}\,
J^{a\,b\,c}_{\mu \,\sigma} g_{\nu \lambda}\right)\right]\nonumber\\
&&+\frac{1}{4} \sum_{a,b,c,d}\, \left[ (\widehat{Q}_{f})_{a\, b} 
(\widehat{Q}_{f})_{b\, c}
\left((\Sigma_{f}^{t\,b})_{c\, d} (\Sigma_{f}^{b\,t})_{d\, a}
\, J^{a\,b\,c\,d}_{\mu \,\nu\, \sigma\,\lambda}+
(\Sigma_{f}^{b\,t})_{c\, d} (\Sigma_{f}^{t\,b})_{d\, a}
\, J^{a\,b\,c\,d}_{\mu \,\nu\, \lambda\,\sigma}\right)\right.\nonumber\\
&&+(\widehat{Q}_{f})_{a\, b}(\Sigma_{f}^{t\,b})_{b\, c}(\widehat{Q}_{f})_{c\, d}
(\Sigma_{f}^{b\,t})_{d\, a}
\, J^{a\,b\,c\,d}_{\mu \,\sigma\,\nu\, \lambda}+
(\widehat{Q}_{f})_{a\, b}(\Sigma_{f}^{b\,t})_{b\, c}
(\widehat{Q}_{f})_{c\, d} (\Sigma_{f}^{t\,b})_{d\, a}
\, J^{a\,b\,c\,d}_{\mu \,\lambda\,\nu\, \sigma}\nonumber\\
&&+(\widehat{Q}_{f})_{a\, b}\left((\Sigma_{f}^{t\,b})_{b\, c} (\Sigma_{f}^{b\,t})_{c\,d}
\, J^{a\,b\,c\,d}_{\mu \, \sigma\,\lambda\,\nu}+
(\Sigma_{f}^{b\,t})_{b\, c} (\Sigma_{f}^{t\,b})_{c\, d}
\, J^{a\,b\,c\,d}_{\mu \,\lambda\, \sigma\,\nu}\right)
(\widehat{Q}_{f})_{d\, a}\nonumber\\
&&+(\Sigma_{f}^{t\,b})_{a\, b}
(\widehat{Q}_{f})_{b\, c} (\Sigma_{f}^{b\,t})_{c\, d}(\widehat{Q}_{f})_{d\, a}
\, J^{a\,b\,c\,d}_{\sigma\,\nu \,\lambda\, \mu}+
(\Sigma_{f}^{b\,t})_{a\, b}(\widehat{Q}_{f})_{b\, c} (\Sigma_{f}^{t\,b})_{c\, d}
(\widehat{Q}_{f})_{d\, a}
\, J^{a\,b\,c\,d}_{\lambda\,\nu \,\sigma\, \mu}\nonumber\\
&& \left((\Sigma_{f}^{t\,b})_{a\, b} (\Sigma_{f}^{b\,t})_{b\,c}
\, J^{a\,b\,c\,d}_{\sigma\,\lambda\,\mu \, \nu}+
(\Sigma_{f}^{b\,t})_{a\, b} (\Sigma_{f}^{t\,b})_{b\, c}
\, J^{a\,b\,c\,d}_{\lambda\,\sigma\,\mu \, \nu}\right)
(\widehat{Q}_{f})_{c\, d} (\widehat{Q}_{f})_{d\, a}\nonumber\\
&& \left.\left.+(\Sigma_{f}^{t\,b})_{a\, b}
(\widehat{Q}_{f})_{b\, c}(\widehat{Q}_{f})_{c\, d}
 (\Sigma_{f}^{b\,t})_{d\,a}\, J^{a\,b\,c\,d}_{\sigma\, \nu\,\mu \,\lambda}
 + (\Sigma_{f}^{b\,t})_{a\, b}
(\widehat{Q}_{f})_{b\, c}(\widehat{Q}_{f})_{c\, d}
 (\Sigma_{f}^{t\,b})_{d\,a}
\, J^{a\,b\,c\,d}_{\lambda\, \nu\,\mu \,\sigma} \right]\,\right\} 
\end{eqnarray}
\begin{eqnarray}
\label{eq:AZWWexact}
\displaystyle && {\Delta {\Gamma}_{\mu \,\nu \,\sigma\,\lambda
\hspace*{0.4cm}\tilde{q}}^{A Z W^+ W^-}}=
\frac{e g^{3}}{\cw} \frac{\pi^{2}}{2} N_{c}\, \sum_{\tilde{f}}\left\{ 
\sum_{a,b} \left[ \left( \,(\widehat{Q}_{f}\,\widehat{G}_{f})_{a\, b}
(\Sigma_{f})_{b\, a}\,J^{a\,b}_{p+k}
+ (\Sigma_{f})_{a\, b} (\widehat{Q}_{f}\,\widehat{G}_{f})_{b\, a}\,
J^{a\,b}_{r+t}\right)\,g_{\mu \nu}\,g_{\sigma \lambda}\right.\right.\nonumber\\
&&-\frac{\sw^{2}}{18}\left( \left(
(\Sigma_{f}^{t\,b})_{a\, b}(\Sigma_{f}^{b\,t})_{b\, a}\,J^{a\,b}_{p+r}+
(\Sigma_{f}^{b\,t})_{a\, b}(\Sigma_{f}^{t\,b})_{b\, a}\,J^{a\,b}_{k+r}\right)
\,g_{\mu \sigma}\,g_{\nu \lambda}\right.\nonumber\\
&&+\left.\left.\left((\Sigma_{f}^{t\,b})_{a\, b}(\Sigma_{f}^{b\,t})_{b\, a}
\,J^{a\,b}_{k+r}+
(\Sigma_{f}^{b\,t})_{a\, b}(\Sigma_{f}^{t\,b})_{b\, a}\,J^{a\,b}_{p+t}
\right)\,g_{\mu \lambda}\,g_{\nu \sigma}\right)\right]\nonumber\\
&& -\sum_{a,b,c}\, \left[ \left(
(\widehat{Q}_{f})_{a\, b} (\widehat{G}_{f})_{b\, c}\,
J^{a\,b\, c}_{\mu \,\nu} \,g_{\sigma \lambda}+
(\widehat{G}_{f})_{a\, b} (\widehat{Q}_{f})_{b\, c}\,
J^{a\,b\, c}_{\nu \,\mu} \,g_{\sigma \lambda}\right)
(\Sigma_{f})_{c\, a}\right.\nonumber\\
&&+2\,((\Sigma_{f}^{t\,b})_{a\, b} 
(\Sigma_{f}^{b\,t})_{b\, c}\,J^{a\,b\, c}_{\sigma\, \lambda} \,g_{\mu \nu}+
(\Sigma_{f}^{b\,t})_{a\, b}\,J^{a\,b\, c}_{\lambda\,\sigma} \,g_{\nu \mu}) 
(\Sigma_{f}^{t\,b})_{b\, c})(\widehat{Q}_{f}\,
\widehat{G}_{f})_{b\, a}\nonumber\\ 
&& -\frac{\sw^{2}}{3} \left( (\widehat{Q}_{f})_{a\, b}
(\Sigma_{f}^{b\,t})_{b\, c} (\Sigma_{f}^{t\,b})_{c\, a}
\,J^{a\,b\, c}_{\mu\, \lambda} \,g_{\nu \sigma}
+ (\Sigma_{f}^{b\,t})_{a\, b} (\widehat{Q}_{f})_{b\, c}
 (\Sigma_{f}^{t\,b})_{c\, a}\,J^{a\,b\, c}_{ \lambda\,\mu} \,g_{\nu \sigma}
 \right.\nonumber\\
&& + \left. (\widehat{Q}_{f})_{a\, b}
(\Sigma_{f}^{t\,b})_{b\, c} (\Sigma_{f}^{b\,t})_{c\, a}\,
J^{a\,b\, c}_{\mu\, \sigma} \,g_{\nu\lambda }
+ (\Sigma_{f}^{t\,b})_{a\, b} (\widehat{Q}_{f})_{b\, c}
 (\Sigma_{f}^{b\,t})_{c\, a}\,J^{a\,b\, c}_{\sigma\, \mu} 
 \,g_{\nu\lambda }\right)\nonumber\\
&&+\frac{1}{3}\left( (\widehat{G}_{f})_{a\, b}
(\Sigma_{f}^{t\,b})_{b\, c} (\Sigma_{f}^{b\,t})_{c\, a}\,
J^{a\,b\, c}_{\nu\, \sigma} \,g_{\mu\lambda }
+ (\Sigma_{f}^{t\,b})_{a\, b} (\widehat{G}_{f})_{b\, c}
 (\Sigma_{f}^{b\,t})_{c\, a}\,J^{a\,b\, c}_{\sigma\,\nu} \,g_{\mu\lambda} 
 \right.\nonumber\\ 
&&+ \left.\left.(\widehat{G}_{f})_{a\, b}
(\Sigma_{f}^{b\,t})_{b\, c} (\Sigma_{f}^{t\,b})_{c\, a}\,
\,J^{a\,b\, c}_{\nu\, \lambda} \,g_{\mu \sigma}
+ (\Sigma_{f}^{b\,t})_{a\, b} (\widehat{G}_{f})_{b\, c}
 (\Sigma_{f}^{t\,b})_{c\, a}\,J^{a\,b\, c}_{\lambda\,\nu} \,g_{\mu \sigma}
 \right)\right]\nonumber\\
&&+\frac{1}{4} \sum_{a,b,c,d}\, \left[ \left((\widehat{Q}_{f})_{a\, b} 
(\widehat{G}_{f})_{b\, c}\, J^{a\,b\,c\,d}_{\mu \,\nu\, \sigma\,\lambda}
+(\widehat{G}_{f})_{a\, b} 
(\widehat{Q}_{f})_{b\, c}\,J^{a\,b\,c\,d}_{\nu \,\mu\, \sigma\,\lambda}\right)\,
(\Sigma_{f}^{t\,b})_{c\, d} (\Sigma_{f}^{b\,t})_{d\, a}\right.\nonumber\\
&&+\left((\widehat{Q}_{f})_{a\, b} 
(\widehat{G}_{f})_{b\, c}\, J^{a\,b\,c\,d}_{\mu \,\nu\,\lambda\, \sigma}
+(\widehat{G}_{f})_{a\, b} 
(\widehat{Q}_{f})_{b\, c}\,J^{a\,b\,c\,d}_{\nu \,\mu\, \lambda\,\sigma}\right)\,
(\Sigma_{f}^{b\,t})_{c\, d} (\Sigma_{f}^{t\,b})_{d\, a}\nonumber\\
&& +(\widehat{Q}_{f})_{a\, b}
(\Sigma_{f}^{t\,b})_{b\, c}(\widehat{G}_{f})_{c\, d}
(\Sigma_{f}^{b\,t})_{d\, a}\, J^{a\,b\,c\,d}_{\mu \,\sigma\,\nu\,\lambda}+
(\widehat{G}_{f})_{a\, b}(\Sigma_{f}^{t\,b})_{b\,c}(\widehat{Q}_{f})_{c\, d}
(\Sigma_{f}^{b\,t})_{d\, a}\, J^{a\,b\,c\,d}_{\nu
\,\sigma\,\mu\,\lambda}\nonumber\\
&&+(\widehat{Q}_{f})_{a\, b}(\Sigma_{f}^{b\,t})_{b\, c}(\widehat{G}_{f})_{c\, d}
(\Sigma_{f}^{t\,b})_{d\, a}\, J^{a\,b\,c\,d}_{\mu \,\lambda\,\nu\,\sigma}
+(\widehat{G}_{f})_{a\, b}
(\Sigma_{f}^{b\,t})_{b\, c}(\widehat{Q}_{f})_{c\, d}
(\Sigma_{f}^{t\,b})_{d\, a}\, J^{a\,b\,c\,d}_{\nu
\,\lambda\,\mu\,\sigma}\nonumber\\
&&+(\widehat{Q}_{f})_{a\,b}\left((\Sigma_{f}^{t\,b})_{b\, c} (\Sigma_{f}^{b\,t})_{c\,d}
\, J^{a\,b\,c\,d}_{\mu \,\sigma\,\lambda\,\nu}+
(\Sigma_{f}^{b\,t})_{b\, c} (\Sigma_{f}^{t\,b})_{c\, d}
\, J^{a\,b\,c\,d}_{\mu \,\lambda\,\sigma\,\nu}\right)
(\widehat{G}_{f})_{d\, a}\nonumber\\
&&+(\widehat{G}_{f})_{a\, b}
\left((\Sigma_{f}^{t\,b})_{b\, c} (\Sigma_{f}^{b\,t})_{c\,d}\,
J^{a\,b\,c\,d}_{\nu \,\sigma\,\lambda\,\mu}+
(\Sigma_{f}^{b\,t})_{b\, c} (\Sigma_{f}^{t\,b})_{c\, d}\,
J^{a\,b\,c\,d}_{\nu \,\lambda\,\sigma\,\mu}\right)
(\widehat{Q}_{f})_{d\, a}\nonumber\\
&& +(\Sigma_{f}^{t\,b})_{a\, b}
(\widehat{Q}_{f})_{b\, c} (\Sigma_{f}^{b\,t})_{c\, d}(\widehat{G}_{f})_{d\, a}\,
J^{a\,b\,c\,d}_{\sigma\,\mu \,\lambda\,\nu}+
(\Sigma_{f}^{t\,b})_{a\, b}(\widehat{G}_{f})_{b\, c}
 (\Sigma_{f}^{b\,t})_{c\, d}(\widehat{Q}_{f})_{d\, a}\,
J^{a\,b\,c\,d}_{\sigma\,\nu \,\lambda\,\mu} \nonumber\\
&& +(\Sigma_{f}^{b\,t})_{a\, b}
(\widehat{Q}_{f})_{b\, c} (\Sigma_{f}^{t\,b})_{c\, d}(\widehat{G}_{f})_{d\, a}\,
J^{a\,b\,c\,d}_{\lambda\,\mu \,\sigma\,\nu}+
(\Sigma_{f}^{b\,t})_{a\, b}(\widehat{G}_{f})_{b\, c}
 (\Sigma_{f}^{t\,b})_{c\, d}(\widehat{Q}_{f})_{d\, a}\,
J^{a\,b\,c\,d}_{\lambda\,\nu\,\sigma \,\mu} \nonumber\\
&&+\left((\Sigma_{f}^{t\,b})_{a\, b} (\Sigma_{f}^{b\,t})_{b\,c}\,
J^{a\,b\,c\,d}_{\sigma\,\lambda\,\nu \,\mu}+
(\Sigma_{f}^{b\,t})_{a\, b} (\Sigma_{f}^{t\,b})_{b\, c}\,
J^{a\,b\,c\,d}_{\lambda\,\sigma\,\nu \,\mu}\right)\,
(\widehat{Q}_{f})_{c\, d} (\widehat{G}_{f})_{d\, a}\nonumber\\
&&+\left((\Sigma_{f}^{t\,b})_{a\, b} (\Sigma_{f}^{b\,t})_{b\,c}\,
J^{a\,b\,c\,d}_{\sigma\,\lambda\,\mu \,\nu}+
(\Sigma_{f}^{b\,t})_{a\, b} (\Sigma_{f}^{t\,b})_{b\, c}\,
J^{a\,b\,c\,d}_{\lambda\,\sigma\,\mu \,\nu}\right)\,
 (\widehat{G}_{f})_{c\, d} (\widehat{Q}_{f})_{d\, a}\nonumber\\
&& +(\Sigma_{f}^{t\,b})_{a\, b}
\left( (\widehat{Q}_{f})_{c\, d} (\widehat{G}_{f})_{d\, a}\,
J^{a\,b\,c\,d}_{\sigma\,\mu \,\nu\,\lambda}+
 (\widehat{G}_{f})_{c\, d} (\widehat{Q}_{f})_{d\, a}\,
J^{a\,b\,c\,d}_{\sigma\,\nu \,\mu\,\lambda }\right)
(\Sigma_{f}^{b\,t})_{d\,a}\nonumber\\
&&\left.\left.+ (\Sigma_{f}^{b\,t})_{a\, b}
\left( (\widehat{Q}_{f})_{c\, d} (\widehat{G}_{f})_{d\, a}\,
J^{a\,b\,c\,d}_{\lambda\,\mu \,\nu\,\sigma} +
 (\widehat{G}_{f})_{c\, d} (\widehat{Q}_{f})_{d\, a}\,
J^{a\,b\,c\,d}_{\lambda\,\nu \,\mu\,\sigma} \right)
(\Sigma_{f}^{t\,b})_{d\,a}\right]\,\right\} 
\end{eqnarray}
\begin{eqnarray}
\label{eq:ZZWWexact}
\displaystyle && {\Delta {\Gamma}_{\mu \,\nu \,\sigma\,\lambda
\hspace*{0.4cm}\tilde{q}}^{Z Z W^+ W^-}}=
\frac{g^{4}}{\cw^{2}} {\pi^{2}} N_{c}  \, \sum_{\tilde{f}}\left\{ \frac{1}{2}
\sum_{a,b} \left[ \,\left((\widehat{G}^{2}_{f})_{a\, b}
(\Sigma_{f})_{b\, a}+(\Sigma_{f})_{a\, b}(\widehat{G}^{2}_{f})_{b\, a}\right)
 \,g_{\mu \nu}\,g_{\sigma \lambda}J^{a\,b}_{p+k}\right.\right.\nonumber\\
&& + \frac{\sw^{4}}{9}\left(
(\Sigma_{f}^{t\,b})_{a\, b}(\Sigma_{f}^{b\,t})_{b\, a}
\,g_{\mu \sigma}\,g_{\nu \lambda}\,J^{a\,b}_{p+r}+
(\Sigma_{f}^{b\,t})_{a\, b}(\Sigma_{f}^{t\,b})_{b\, a}
\,g_{\mu \lambda}\,g_{\nu \sigma}\,J^{a\,b}_{p+t}\right)\left.\right]\nonumber\\
&& - \sum_{a,b,c}\, \left[ 
(\widehat{G}_{f})_{a\, b} (\widehat{G}_{f})_{b\, c}(\Sigma_{f})_{c\, a}\,
J^{a\,b\, c}_{\mu \,\nu} \,g_{\sigma \lambda}+
(\Sigma_{f}^{t\,b})_{a\, b} (\Sigma_{f}^{b\,t})_{b\, c}
(\widehat{G}^{2}_{f})_{c\, a}\,J^{a\,b\, c}_{\lambda\,\sigma}
\,g_{\mu\nu}\right.\nonumber\\ 
&&+ (\Sigma_{f}^{b\,t})_{a\, b} (\Sigma_{f}^{t\,b})_{b\, c}
(\widehat{G}^{2}_{f})_{c\, a}J^{a\,b\, c}_{\sigma\,\lambda} \,g_{\mu\nu}
-\frac{\sw^{2}}{3} \left( (\widehat{G}_{f})_{a\, b}
(\Sigma_{f}^{b\,t})_{b\, c} (\Sigma_{f}^{t\,b})_{c\, a}
J^{a\,b\, c}_{\mu \,\lambda} \,g_{\nu \sigma}\right.\nonumber\\
&&+ (\Sigma_{f}^{t\,b})_{a\, b} (\widehat{G}_{f})_{b\, c}
 (\Sigma_{f}^{b\,t})_{c\, a}\,J^{a\,b\, c}_{\sigma \,\nu} 
 g_{\mu \lambda}+ (\Sigma_{f}^{b\,t})_{a\, b} 
 (\widehat{G}_{f})_{b\, c}(\Sigma_{f}^{t\,b})_{c\, a}
J^{a\,b\, c}_{\lambda\,\nu} g_{\mu \sigma}
\left.\left. +(\widehat{G}_{f})_{a\, b}(\Sigma_{f}^{t\,b})_{b\, c} 
 (\Sigma_{f}^{b\,t})_{c\, a}\,
J^{a\,b\,c}_{\mu \,\sigma} g_{\nu \lambda}\right)\right]\nonumber\\
&&+\frac{1}{4} \sum_{a,b,c,d}\, \left[ (\widehat{G}_{f})_{a\, b} 
(\widehat{G}_{f})_{b\, c}
\left((\Sigma_{f}^{t\,b})_{c\, d} (\Sigma_{f}^{b\,t})_{d\, a}
\, J^{a\,b\,c\,d}_{\mu \,\nu\, \sigma\,\lambda}+
(\Sigma_{f}^{b\,t})_{c\, d} (\Sigma_{f}^{t\,b})_{d\, a}
\, J^{a\,b\,c\,d}_{\mu \,\nu\, \lambda\,\sigma}\right)\right.\nonumber\\
&&+(\widehat{G}_{f})_{a\, b}(\Sigma_{f}^{t\,b})_{b\, c}(\widehat{G}_{f})_{c\, d}
(\Sigma_{f}^{b\,t})_{d\, a}
\, J^{a\,b\,c\,d}_{\mu \,\sigma\,\nu\, \lambda}+
(\widehat{G}_{f})_{a\, b}(\Sigma_{f}^{b\,t})_{b\, c}
(\widehat{G}_{f})_{c\, d} (\Sigma_{f}^{t\,b})_{d\, a}
\, J^{a\,b\,c\,d}_{\mu \,\lambda\,\nu\, \sigma}\nonumber\\
&&+(\widehat{G}_{f})_{a\, b}\left((\Sigma_{f}^{t\,b})_{b\, c} (\Sigma_{f}^{b\,t})_{c\,d}
\, J^{a\,b\,c\,d}_{\mu \, \sigma\,\lambda\,\nu}+
(\Sigma_{f}^{b\,t})_{b\, c} (\Sigma_{f}^{t\,b})_{c\, d}
\, J^{a\,b\,c\,d}_{\mu \,\lambda\, \sigma\,\nu}\right)
(\widehat{G}_{f})_{d\, a}\nonumber\\
&&+(\Sigma_{f}^{t\,b})_{a\, b}
(\widehat{G}_{f})_{b\, c} (\Sigma_{f}^{b\,t})_{c\, d}(\widehat{G}_{f})_{d\, a}
\, J^{a\,b\,c\,d}_{\sigma\,\nu \,\lambda\, \mu}+
(\Sigma_{f}^{b\,t})_{a\, b}(\widehat{G}_{f})_{b\, c} (\Sigma_{f}^{t\,b})_{c\, d}
(\widehat{G}_{f})_{d\, a}
\, J^{a\,b\,c\,d}_{\lambda\,\nu \,\sigma\, \mu}\nonumber\\
&& \left((\Sigma_{f}^{t\,b})_{a\, b} (\Sigma_{f}^{b\,t})_{b\,c}
\, J^{a\,b\,c\,d}_{\sigma\,\lambda\,\mu \, \nu}+
(\Sigma_{f}^{b\,t})_{a\, b} (\Sigma_{f}^{t\,b})_{b\, c}
\, J^{a\,b\,c\,d}_{\lambda\,\sigma\,\mu \, \nu}\right)
(\widehat{G}_{f})_{c\, d} (\widehat{G}_{f})_{d\, a}\nonumber\\
&& \left.\left.+(\Sigma_{f}^{t\,b})_{a\, b}
(\widehat{G}_{f})_{b\, c}(\widehat{G}_{f})_{c\, d}
 (\Sigma_{f}^{b\,t})_{d\,a}\, J^{a\,b\,c\,d}_{\sigma\, \nu\,\mu \,\lambda}
 + (\Sigma_{f}^{b\,t})_{a\, b}
(\widehat{G}_{f})_{b\, c}(\widehat{G}_{f})_{c\, d}
 (\Sigma_{f}^{t\,b})_{d\,a}
\, J^{a\,b\,c\,d}_{\lambda\, \nu\,\mu \,\sigma} \right]\,\right\}\\
\nonumber\\ 
\label{eq:WWWWexact}
&& \displaystyle{\Delta {\Gamma}_{\mu \,\nu \,\sigma\,\lambda
\hspace*{0.4cm}\tilde{q}}^{ W^+ W^- W^+ W^-}} =
g^{4} {\pi^{2}} N_{c} \, \sum_{\tilde{f}}\left\{ \frac{1}{2}
\sum_{a,b} \left[ \,(\Sigma_{f})_{a\, b} (\Sigma_{f})_{b\, a}
\,g_{\mu \lambda}\,g_{\nu \sigma}\right]\right.\,J^{a\,b}_{p+k}\nonumber\\
&&- \sum_{a,b,c}\, \left[ \left(
 (\Sigma_{f}^{t\,b})_{a\, b} (\Sigma_{f}^{b\,t})_{b\, c}\,
J^{a\,b\, c}_{\mu \,\nu} \,g_{\sigma \lambda} +
+ (\Sigma_{f}^{b\,t})_{a\, b} (\Sigma_{f}^{t\,b})_{b\, c}\,
J^{a\,b\, c}_{\nu \,\mu} \,g_{\sigma \lambda}\right)
(\Sigma_{f})_{c\, a}\right]\nonumber\\
&&+ \frac{1}{4} \sum_{a,b,c,d}\, \left[\,
(\Sigma_{f}^{t\,b})_{a\, b} (\Sigma_{f}^{t\,b})_{b\, c}
(\Sigma_{f}^{b\,t})_{c\, d} (\Sigma_{f}^{b\,t})_{d\, a}\,
 J^{a\,b\,c\,d}_{\mu \,\sigma\,\nu\, \lambda}+
(\Sigma_{f}^{b\,t})_{a\, b} (\Sigma_{f}^{b\,t})_{b\, c}
(\Sigma_{f}^{t\,b})_{c\, d} (\Sigma_{f}^{t\,b})_{d\, a}\,
 J^{a\,b\,c\,d}_{\lambda\,\nu \,\sigma\,\mu}\right.\nonumber\\
&& +(\Sigma_{f}^{t\,b})_{a\, b} (\Sigma_{f}^{b\,t})_{b\, c}
\left((\Sigma_{f}^{t\,b})_{c\, d} (\Sigma_{f}^{b\,t})_{d\,a}\,
J^{a\,b\,c\,d}_{\mu\,\nu \,\sigma\,\lambda}+
(\Sigma_{f}^{b\,t})_{c\, d} (\Sigma_{f}^{t\,b})_{d\, a}\,
J^{a\,b\,c\,d}_{\mu\,\nu \,\lambda\,\sigma}\right)\nonumber\\
&& \left.\left.+(\Sigma_{f}^{b\,t})_{a\, b}
(\Sigma_{f}^{t\,b})_{b\, c}
\left((\Sigma_{f}^{t\,b})_{c\, d} (\Sigma_{f}^{b\,t})_{d\,a}\,
J^{a\,b\,c\,d}_{\nu\,\mu \,\sigma\,\lambda}+
(\Sigma_{f}^{b\,t})_{c\, d} (\Sigma_{f}^{t\,b})_{d\, a}\,
J^{a\,b\,c\,d}_{\nu\,\mu \,\lambda\,\sigma}\right)\right]\,\right\} 
\end{eqnarray}

$\bullet$ Three point {\it inos} contributions.
\begin{eqnarray}
\label{eq:effAWW}
\displaystyle \Delta{\Gamma}_{\mu \,\nu \,\sigma\hspace*{0.4cm}\sg}^{AW^+W^-}
&=& -\frac{eg^2}{8{\pi}^2} \sum_{i=1}^{4} \sum_{j,k=1}^{2}\delta_{jk}\,\left\{
 (O_{L_{ij}}O_{L_{ki}}^{+}+O_{R_{ij}}O_{R_{ki}}^{+})\right.
\left[\,{\cal T}_{\,\sigma\,\mu \,\nu}^{\,i\,j\,k}
+\tilde{M}_{j}^{+} \tilde{M}_{k}^{+}\,
{\cal I}_{\,\sigma\,\mu \,\nu}^{\,i\,j\,k}\,\right]\nonumber\\
&+&(O_{L_{ij}}O_{R_{ki}}^{+}+O_{R_{ij}}O_{L_{ki}}^{+})\,\left[\,
\tilde{M}_{i}^{o} \tilde{M}_{k}^{+}\,
{\cal P}_{\,\sigma\,\mu \,\nu}^{\,i\,j\,k}
\left.+\tilde{M}_{i}^{o} \tilde{M}_{j}^{+}\,
{\cal J}_{\,\sigma\,\mu \,\nu}^{\,i\,j\,k}\,\right]\,\right\}\,,\\
\label{eq:effZWW}
\displaystyle \Delta{\Gamma}_{\mu \,\nu \,\sigma\hspace*{0.4cm}\sg}^{ZW^+W^-}
&=& \frac{g^3}{8{\pi}^2}\frac{1}{\cw} \left\{\sum_{i=1}^{4} \sum_{j,k=1}^{2}
\right.\left[\,(O_{L_{ij}}O'_{L_{jk}}O_{L_{ki}}^{+}+
O_{R_{ij}}O'_{R_{jk}}O_{R_{ki}}^{+})
{\cal T}_{\,\sigma\,\mu \,\nu}^{\,i\,j\,k}\right.\nonumber\\
&+&(O_{L_{ij}}O'_{R_{jk}}O_{R_{ki}}^{+}+
O_{R_{ij}}O'_{L_{jk}}O_{L_{ki}}^{+})\,\tilde{M}_{i}^{o} \tilde{M}_{j}^{+}\,
{\cal J}_{\,\sigma\,\mu \,\nu}^{\,i\,j\,k}\nonumber\\
&+&(O_{L_{ij}}O'_{R_{jk}}O_{L_{ki}}^{+}+
O_{R_{ij}}O'_{L_{jk}}O_{R_{ki}}^{+})\,
\tilde{M}_{j}^{+} \tilde{M}_{k}^{+}\,
{\cal I}_{\,\sigma\,\mu \,\nu}^{\,i\,j\,k}\nonumber\\
&+&\left.(O_{L_{ij}}O'_{L_{jk}}O_{R_{ki}}^{+}+
O_{R_{ij}}O'_{R_{jk}}O_{L_{ki}}^{+})\,\tilde{M}_{i}^{o} \tilde{M}_{k}^{+}\,
{\cal P}_{\,\sigma\,\mu \,\nu}^{\,i\,j\,k}\,\right]\nonumber\\
&& \displaystyle \hspace*{0.4cm}+\sum_{i,j=1}^{4} \sum_{k=1}^{2}
\left[\,(O''_{L_{ij}}O_{L_{jk}}O_{L_{ki}}^{+}+
O''_{R_{ij}}O_{R_{jk}}O_{R_{ki}}^{+})\,
{\cal T}_{\,\mu \,\sigma\,\nu}^{\,i\,j\,k}\right.\nonumber\\
&+&(O''_{L_{ij}}O_{R_{jk}}O_{R_{ki}}^{+}+
O''_{R_{ij}}O_{L_{jk}}O_{R_{ki}}^{+})\,
\tilde{M}_{i}^{o} \tilde{M}_{j}^{o}\
{\cal J}_{\,\mu \,\sigma\,\nu}^{\,i\,j\,k}\nonumber\\
&+&(O''_{L_{ij}}O_{R_{jk}}O_{L_{ki}}^{+}+
O''_{R_{ij}}O_{L_{jk}}O_{R_{ki}}^{+}) \,\tilde{M}_{j}^{o} \tilde{M}_{k}^{+}\,
{\cal I}_{\,\mu \,\sigma\,\nu}^{\,i\,j\,k}\nonumber\\
&+&\left.\left.(O''_{L_{ij}}O_{L_{jk}}O_{R_{ki}}^{+}+
O''_{R_{ij}}O_{R_{jk}}O_{L_{ki}}^{+})\,\tilde{M}_{i}^{o} \tilde{M}_{k}^{+}\,
{\cal P}_{\,\mu \,\sigma\,\nu \,}^{\,i\,j\,k}\,\right]\,\right\}
\end{eqnarray}

The integrals appearing in the above formula are given in terms of the 
standard one-loop integrals in appendix A
and we refer once more to~\cite{GEISHA} in order to find the 
explicit expressions of the coupling matrices.

\begin{center}
\epsfig{file=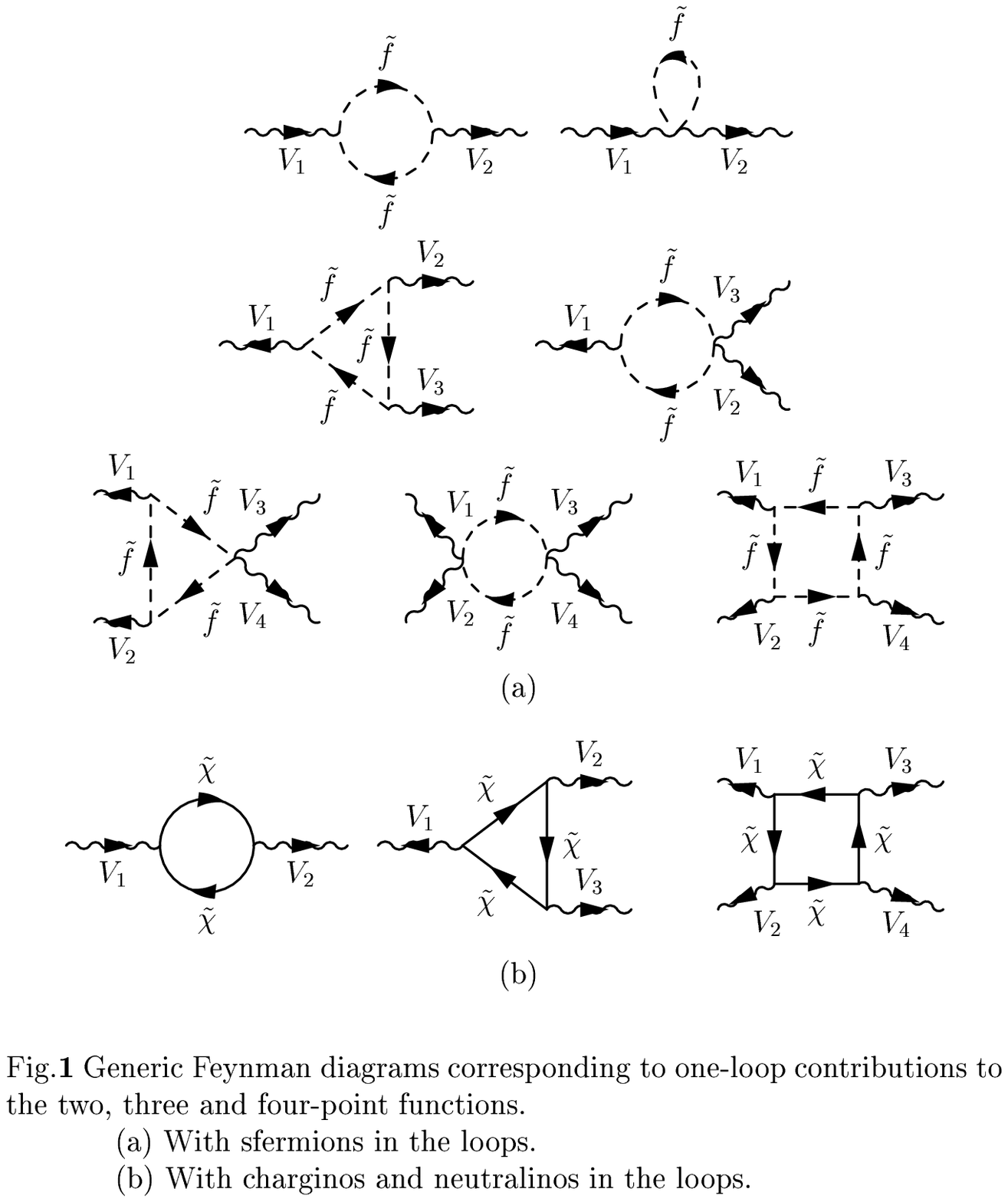,width=20cm}
\end{center}
\end{document}